\PassOptionsToPackage{final}{graphicx}
\documentclass[preprint,12pt,number]{elsarticle}

\usepackage{color}

\usepackage[utf8]{inputenc}
\usepackage{amssymb}
\usepackage{amsmath} 
\usepackage{amsfonts} 
\usepackage{dcolumn}
\usepackage{comment}
\usepackage{bm} 
\usepackage{hyperref} 
\usepackage{mathptmx}
\usepackage{textcomp}
\usepackage{mwe}
\usepackage{algorithm}
\usepackage{algpseudocode}
\usepackage{tabularx}
\usepackage{float}
\usepackage[caption = false]{subfig}
\usepackage{pdflscape}
\usepackage{booktabs}
\usepackage{array}
\usepackage{rotating}
\usepackage{subfig}
\usepackage{tikz}
\usepackage{subcaption}
\usepackage{textcomp}
\usepackage{siunitx}
\usepackage{xspace}
\usepackage{xcolor}
\usepackage{mathrsfs}
\usepackage{soul}

\usepackage[mathscr]{eucal}
\usepackage{tikz}
\usetikzlibrary{arrows.meta, positioning, shapes.geometric}
\usepackage[margin=1in]{geometry}

\newlength{\bigfboxsep}
\newcommand{\dd}{\mathrm{d}}

\newcommand{\bs}[1]{\boldsymbol{#1}}

\newcommand{\bse}{\bs{e}}

\newcommand{\bsX}{\bs{X}}

\newcommand{\bstheta}{\bs{\theta}}

\journal{}

\begin{document}

\begin{frontmatter}

\title{Fully differentiable framework for inverse identification of geometry and material parameters with application to determining stress-free configuration of soft tissues
}

\author[aff1]{Hyunoh Bae}
\author[aff1]{Tianyi Hu}
\author[aff2]{Adrian Buganza Tepole}
\author[aff1]{Hector Gomez\corref{cor1}}

\cortext[cor1]{Corresponding author. Email: hectorgomez@purdue.edu}

\affiliation[aff1]{%
    organization={School of Mechanical Engineering, Purdue University},
    addressline={585 Purdue Mall},
    city={West Lafayette},
    state={Indiana},
    postcode={47906},
    country={USA}
}

\affiliation[aff2]{%
    organization={Department of Mechanical Engineering, Columbia University},
    addressline={116th Street and Broadway},
    city={New York},
    state={NY},
    postcode={10027},
    country={USA}
}

\begin{abstract}
Medical images of soft tissues typically depict loaded configurations rather than true unloaded (stress-free) states, which can bias biomechanical simulations and inverse material identification. A gradient-based inverse finite element framework is presented that jointly reconstructs an effective unloaded reference geometry and estimates hyperelastic material parameters from two or more observed deformed configurations. The formulation is fully differentiable and leverages exact end-to-end gradients to enable unified, simultaneous optimization of geometry and constitutive parameters. The objective function combines a nodal-position misfit with a deformation-gradient-based mismatch term, improving robustness under large deformations. Benchmark studies quantify the influence of loading diversity and observation count and demonstrate reduced sensitivity to poor material initialization. Finally, application to an MRI-derived breast model shows accurate recovery of the unloaded configuration and constitutive parameters from multiple gravity-loaded states. The framework provides a unified and scalable tool for inverse biomechanics with potential applications in personalized modeling, elastography, and surgical planning.
\end{abstract}


\end{frontmatter}

\section{Introduction}
\label{Intro}

Medical images of soft tissues rarely represent a stress-free configuration. Contact loads and body forces—especially gravity—induce substantial deformation in compliant tissues such as the breast, abdominal organs, and the cardiovascular system \cite{Businaro_Studer_Pajic_Buchler_2015, Rajagopal_Lee_Chung_Warren_2007}. In addition, growth and remodeling generate intrinsic residual stresses \cite{yousefi_new_2026, Lee2021TheGO, vavourakis_inverse_2016, bols_computational_2013,hajhashemkhani_novel_2020}. Consequently, geometries extracted from magnetic resonance imaging (MRI), computed tomography (CT), or related imaging methods correspond to mechanically loaded and prestressed states. Treating such geometries as stress-free leads to biased predictions of deformation, strain, and material parameters in biomechanical simulation and diagnosis. For instance, in myocardial modeling, stiffness estimation from partially loaded configurations systematically overestimates material properties \cite{nikou_effects_2016}.

Unfortunately, while the in vivo geometry is easily available, and loading and boundary conditions can be reasonably well approximated \cite{vavourakis_inverse_2016,eiben_surface_2016}, the current stress field over such geometry is unknown and cannot be measured directly. Similarly, mechanical properties often cannot be measured directly \cite{smoljkic_comparison_2018}. While there exist noninvasive techniques such as magnetic resonance (MR) elastography or ultrasonic elastography, these methods are also limited in their ability to identify the stress-free configuration. Instead, they provide linearized estimates of the stiffness at the current deformation state \cite{capilnasiu_magnetic_2019,crutison_combined_2022}. Yet, the loaded geometry and boundary conditions indirectly contain information about the current state of stress, the material property field, and the stress-free configuration, through the momentum balance equations.
Early inverse strategies to reconstruct the stress-free geometry typically assumed known material parameters and focused only on recovering the unloaded geometry. 
One approach is direct formulations that cast the reverse elasticity problem as a modified equilibrium system, in which the reference configuration is treated as the unknown and solved in a single inverse-design step. A key limitation of these formulations is that they require problem-specific reformulations, making them less straightforward to integrate with standard forward finite element solvers, even though they may be mathematically well-posed \cite{gee_computational_2010, rajagopal_determining_2007, peirlinck_modular_2018,barnafi_reconstructing_2024,jilberto_identification_2025}. To improve implementation flexibility, iterative fixed-point strategies were later introduced, in which the reference configuration is updated incrementally through repeated forward analyses until the predicted loaded state matches the observation. These methods are easier to couple with existing solvers but may exhibit slow convergence or stability issues for large deformations \cite{bols_computational_2013, carter_biomechanical_2009, mira_biomechanical_2018, rausch_augmented_2017, sellier_iterative_2011}.

In parallel, a large body of inverse material characterization work has focused on identifying heterogeneous material properties when the reference or unloaded configuration is already known. For example, inverse homogenization and finite element updating approaches have been used to recover material parameters of heterogeneous materials \cite{klinge_inverse_2015,li_integrated_2026,hajhashemkhani_inverse_2018,goenezen_nonlinear_2011}.
More recent data-driven approaches have also identified three-dimensional heterogeneous modulus fields from measured displacement or strain fields \cite{liu_resolving_2024}. In image-based biomechanical modeling, heterogeneous regions such as tumors, inclusions, or anatomical subdomains may be obtainable from MRI segmentation or mesh labeling in the observed configuration \cite{guo_image_2005,schnabel_validation_2003,sivaramakrishna_3d_2005,amjad_computationally_2020}. In such cases, the inverse problem can be formulated as identifying the material parameters assigned to known regions, rather than reconstructing an arbitrary spatially varying material field. 
More recently, joint identification of unloaded geometry and material parameters has been investigated with inverse finite element frameworks \cite{parikh_biomechanical_2023, hajhashemkhani_identification_2021}, albeit for relatively simplified scenarios. For example, in~\cite{parikh_biomechanical_2023}, the optimization framework is restricted to a very reduced set of parameters for geometry and material properties. 
In ~\cite{hajhashemkhani_identification_2021}, the optimization requires analytically derived sensitivity matrices. In practice, this restricts the problem to simple geometries or a subset of geometric parameters (e.g. surface geometry only) and homogeneous or low-dimensional material parameters. 

Here we develop a unified gradient-based inverse framework that jointly optimizes the unloaded configuration and hyperelastic material parameters. The framework first treats the homogeneous case before extending the formulation to region-wise heterogeneous materials. The unloaded geometry is parameterized as a deformation field over an arbitrarily selected reference domain. The weak form of linear momentum balance is recast in such a way that, even though the reference domain stays fixed throughout the computation, the stresses and final deformed state are computed based on an intermediate, stress-free configuration which is the target of the optimization. Crucially, the formulation, implemented into the finite element package JAX-FEM, maintains end-to-end differentiability of the nonlinear forward problem with respect to the intermediate geometry and material properties, enabling the use of gradient-based optimizers. Numerical results demonstrate reliable reconstruction of both unloaded geometry and material properties in a variety of examples, including estimation of unloaded configuration for breast geometries from gravity-loaded states, an important piece of breast cancer surgical planning.

\section{Method}
\label{}

\subsection{Problem Formulation}
\label{sec:problem_formulation}

\begin{figure}[H]
    \centering
    \includegraphics[width=0.95\linewidth]{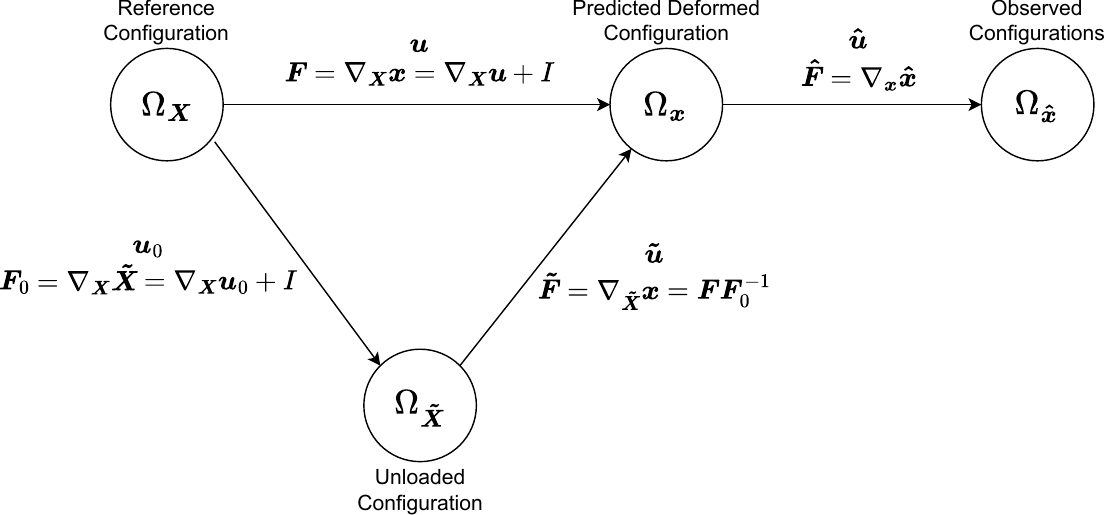}
    \caption{Configuration mapping in the inverse unloading framework. 
The body is described through four configurations: the arbitrary reference configuration $\Omega_{\boldsymbol{X}}$, the unloaded configuration $\Omega_{\tilde{\boldsymbol{X}}}$, the predicted loaded configuration $\Omega_{\boldsymbol{x}}$, and the observed configuration $\Omega_{\hat{\boldsymbol{x}}}$. The arrows indicate the mappings between configurations through the corresponding displacement fields and deformation gradients.}
    \label{fig:intermediate_config}
\end{figure}

We consider an elastic body that can occupy several configurations in $\mathbb{R}^3$, as illustrated in Fig.~\ref{fig:intermediate_config}.
Let $\Omega_{\boldsymbol{X}} \subset \mathbb{R}^3$ denote an arbitrary reference domain describing the body in its reference configuration, which is not necessarily stress-free. A material point in this configuration is identified by its position vector $\boldsymbol{X} \in \Omega_{\boldsymbol{X}}$. The unloaded (stress-free) configuration is denoted by $\Omega_{\tilde{\boldsymbol{X}}} \subset \mathbb{R}^3$, with material point coordinates $\tilde{\boldsymbol{X}} \in \Omega_{\tilde{\boldsymbol{X}}}$. Upon application of loads, the body deforms into a predicted configuration $\Omega_{\boldsymbol{x}} \subset \mathbb{R}^3$, described by spatial coordinates $\boldsymbol{x} \in \Omega_{\boldsymbol{x}}$. The observed configuration obtained from imaging data is denoted by $\Omega_{\hat{\boldsymbol{x}}} \subset \mathbb{R}^3$, with coordinates $\hat{\boldsymbol{x}} \in \Omega_{\hat{\boldsymbol{x}}}$.

Having defined the four configurations, we introduce deformation fields between them. The deformation from the stress-free configuration to the predicted loaded configuration is described by the displacement field $\tilde{\boldsymbol{u}}$ 
\begin{equation}
    \boldsymbol{x} = \tilde{\boldsymbol{X}} + \tilde{\boldsymbol{u}},
\end{equation}
with the corresponding deformation gradient with respect to the coordinates $\tilde{\boldsymbol{X}}$, 
\begin{equation}
    \tilde{\boldsymbol{F}} = \nabla_{\tilde{\boldsymbol{X}}}\boldsymbol{x} = \nabla_{\tilde{\boldsymbol{X}}}\tilde{\boldsymbol{u}} + \boldsymbol{I}.
\end{equation}
The deformation field from the reference configuration to the predicted loaded configuration is similarly described by a displacement field, in this case denoted $\boldsymbol{u}$, with the deformation gradient with respect to the $\boldsymbol{X}$ coordinates,
\begin{equation}
    \boldsymbol{F} = \nabla_{\boldsymbol{X}}\boldsymbol{x} = \nabla_{\boldsymbol{X}}\boldsymbol{u} + \boldsymbol{I}.
\end{equation}
We further introduce a displacement field $\boldsymbol{u}_0$ mapping the reference configuration coordinate to the unloaded configuration coordinate,
\begin{equation}
    \tilde{\boldsymbol{X}} = \boldsymbol{X} + \boldsymbol{u}_0,
\end{equation}
with associated deformation gradient 
\begin{equation}
    \boldsymbol{F}_0 = \nabla_{\boldsymbol{X}}\tilde{\boldsymbol{X}} = \nabla_{\boldsymbol{X}}\boldsymbol{u}_0 + \boldsymbol{I}.
\end{equation}
We remark that these displacement fields carry the assumption that deformations between configurations are compatible. In other words, the underlying assumption is the existence of a continuous displacement field such that the gradient is well-defined. Moreover, each deformation is invertible, which is specified by the determinants being positive, $J_0=\det\boldsymbol{F}_0>0$, $\tilde{J}=\det\tilde{\boldsymbol{F}}>0$, and $J=\det\boldsymbol{F}>0$, over the entire domains in which they are defined. It follows that the three deformation gradients are not independent; they are related by the chain rule,
\begin{equation}
    \tilde{\boldsymbol{F}} = \boldsymbol{F}\,\boldsymbol{F}_0^{-1}.
    \label{F_split}
\end{equation}

The observed configuration $\Omega_{\hat{\boldsymbol{x}}}$ is the \emph{experimentally} measured state of the body and is therefore in mechanical equilibrium. It is assumed that the loading and boundary conditions are known for $\Omega_{\hat{\boldsymbol{x}}}$, with boundary decomposed into disjoint parts such that $   \Gamma_t^{\hat{\boldsymbol{x}}} \cup \Gamma_u^{\hat{\boldsymbol{x}}} = \partial\Omega_{\hat{\boldsymbol{x}}}$, $\Gamma_t^{\hat{\boldsymbol{x}}} \cap \Gamma_u^{\hat{\boldsymbol{x}}} = \varnothing$,
where $\Gamma_u^{\hat{\boldsymbol{x}}}$ denotes the part of the boundary with prescribed displacements and $\Gamma_t^{\hat{\boldsymbol{x}}}$ the part with prescribed tractions $\bar{\mathbf{t}}$. For the purpose of numerical simulation, it is convenient to express the corresponding boundary conditions in the reference configuration $\Omega_{\boldsymbol{X}}$. The boundary $\partial\Omega_{\boldsymbol{X}}$ is accordingly decomposed into disjoint parts $\Gamma_t^{\boldsymbol{X}}$ and $\Gamma_u^{\boldsymbol{X}}$, with prescribed displacements $\boldsymbol{u} = \bar{\boldsymbol{u}}$ on $\Gamma_u^{\boldsymbol{X}}$ and prescribed nominal tractions $\boldsymbol{T} = \bar{\boldsymbol{T}}$ on $\Gamma_t^{\boldsymbol{X}}$.  Dirichlet boundary conditions in the unloaded configuration are $\boldsymbol{u}_0 = \bar{\boldsymbol{u}}_0$ on $\Gamma_u^{\tilde{\boldsymbol{X}}}$, with $\Gamma_u^{\tilde{\boldsymbol{X}}}$ encompassing the same material points as $\Gamma_u^{\boldsymbol{X}}$ but in the unloaded configuration $\Omega_{\tilde{\boldsymbol{X}}}$.

To compare the predicted and observed configurations, we introduce an auxiliary displacement field $\hat{\boldsymbol{u}}$ that maps the predicted configuration $\Omega_{\boldsymbol{x}}$ to the observed configuration $\Omega_{\hat{\boldsymbol{x}}}$.
The corresponding relative deformation gradient with respect to the predicted configuration is defined as

\begin{equation}
\hat{\boldsymbol{F}}
= \nabla_{\boldsymbol{x}} \hat{\boldsymbol{x}}.
\end{equation}

The displacement field $\hat{\boldsymbol{u}}$ and the tensor $\hat{\boldsymbol{F}}$ do not represent a physical deformation caused by loading. Instead, they quantify the geometric mismatch between the predicted and observed configurations, and will play a central role in the objective function introduced in Sec.~\ref{sec:inverse_problem}. The four configurations, the associated deformation fields, together with the assumption of non-vanishing determinants, establish a one-to-one correspondence between the various coordinates $\boldsymbol{X},\,\tilde{\boldsymbol{X}},\,\boldsymbol{x},\,\hat{\boldsymbol{x}}$.

The constitutive response is defined with respect to the unloaded configuration $\Omega_{\tilde{\boldsymbol{X}}}$ through a strain-energy density function $W(\tilde{\boldsymbol{F}})$. The associated first Piola-Kirchhoff stress is 
\begin{equation}
\tilde{\boldsymbol{P}}=\frac{\partial W(\tilde{\boldsymbol{F}})}{\partial \tilde{\boldsymbol{F}}}.
\end{equation}

Using the Piola transformation, we define the first Piola-Kirchhoff stress field on $\Omega_{\boldsymbol{X}}$ as
\begin{equation}
\boldsymbol{P}=J_0\,\tilde{\boldsymbol{P}}\,\boldsymbol{F}_0^{-T}.
\label{eq:Piola_pullback_P}
\end{equation}

In this work, a compressible neo-Hookean material parameterized by the shear modulus $\mu$ and bulk modulus $\kappa$ is employed,
with strain-energy density
\begin{equation}
W(\tilde{\boldsymbol{F}})=\frac{\mu}{2}(\tilde{I}_1-3)-\mu\ln\tilde{J}+\frac{\kappa}{2}(\ln\tilde{J})^2,
\end{equation}
where $\tilde{I}_1=\mathrm{tr}(\tilde{\boldsymbol{C}})$ and $\tilde{\boldsymbol{C}}=\tilde{\boldsymbol{F}}^{T}\tilde{\boldsymbol{F}}$. The first Piola-Kirchhoff stress in the unloaded configuration for the neo-Hookean strain energy is
\begin{equation}
\tilde{\boldsymbol{P}} = \mu\tilde{\boldsymbol{F}} + (\kappa\ln\tilde{J} - \mu)\tilde{\boldsymbol{F}}^{-T}
\label{P_tilde_neoHooke}
\end{equation}

The weak form of the momentum balance equation, written on the reference domain $\Omega_{\boldsymbol{X}}$, is to find $\boldsymbol{u}$ that satisfies the Dirichlet boundary conditions on $\Gamma_u^{\boldsymbol{X}}$ such that for all admissible variations $\delta\boldsymbol{u}$ that vanish on $\Gamma_u^{\boldsymbol{X}}$, it holds that
\begin{equation}
\int_{\Omega_{\boldsymbol{X}}}
    \boldsymbol{P}
    :
    \nabla_{\boldsymbol{X}} \delta \boldsymbol{u}
    \, \dd V_0
=
\int_{\Omega_{\boldsymbol{X}}}
    \rho \, \boldsymbol{b} \cdot \delta\boldsymbol{u} \, J_0 \, \dd V_0
+
\int_{\Gamma_t^{\boldsymbol{X}}}
    \bar{\boldsymbol{T}}
    \cdot
    \delta\boldsymbol{u} \dd A_0,
\label{inter_weak_form}
\end{equation}
where $\rho$ is the mass density defined in the unloaded configuration $\Omega_{\tilde{\boldsymbol{X}}}$, 
$\boldsymbol{b}$ is body force per unit mass, and $\bar{\boldsymbol{T}}$ are the prescribed nominal tractions defined on $\Gamma_t^{\boldsymbol{X}}$. Note that, if the known tractions $\bar{t}$ are defined in the corresponding boundary of the predicted configuration $\Gamma_t^{\boldsymbol{x}}$, the nominal traction is $\bar{\boldsymbol{T}}=\boldsymbol{\bar{t}}J\sqrt{ \boldsymbol{N}_0\cdot \boldsymbol{C}^{-1} \cdot \boldsymbol{N}_0}$, with $\boldsymbol{N}_0$ the unit outward normal to $\Omega_{\boldsymbol{X}}$. Eq. (\ref{inter_weak_form}) uses the notation $\boldsymbol{A}:\boldsymbol{B}$ for the tensor dot product which, in indicial notation, is expressed as $\boldsymbol{A}:\boldsymbol{B}=A_{ij}B_{ij}$. Here and in what follows, we assume that repeated indices indicate summation. For completeness, the standard dot product between vectors used in Eq. (\ref{inter_weak_form}) is expressed in indicial notation as $\boldsymbol{u}\cdot\boldsymbol{v}=u_iv_i$.

\subsection{Finite Element Discretization}
\label{sec:fem_discretization}

The optimization problem, described in detail in the following subsection, requires the discretization of the weak form in Eq.~\eqref{inter_weak_form}. Let $\mathcal{T}_h$ denote a conforming partition of $\Omega_{\boldsymbol{X}}$ into finite elements with nodal shape functions $N_a$, $a=1,\ldots,n_{sh}$, where $n_{sh}$ is the global number of nodes. The various configurations are interpolated using the nodal shape functions and corresponding nodal coordinates,
\begin{equation}
\boldsymbol{X}_h = N_a\,\boldsymbol{X}_a,
\qquad
\tilde{\boldsymbol{X}}_h =  N_a\,\tilde{\boldsymbol{X}}_a,
\qquad
\boldsymbol{x}_h = N_a\,\boldsymbol{x}_a,
\qquad
\hat{\boldsymbol{x}}_h = N_a\,\hat{\boldsymbol{x}}_a.
\end{equation}
The displacement field between the reference and predicted deformed configurations, and the corresponding variation field appearing in Eq.~\eqref{inter_weak_form}, are approximated as
\begin{equation}
\boldsymbol{u}_h 
= \boldsymbol{x}_h - \boldsymbol{X}_h
= N_a\,\boldsymbol{u}_a,
\qquad
\delta\boldsymbol{u}_h = N_a\,\delta\boldsymbol{u}_a.
\end{equation}
The remaining displacement fields $\boldsymbol{u}_0$, $\tilde{\boldsymbol{u}}$, and $\hat{\boldsymbol{u}}$ admit analogous approximations $\boldsymbol{u}_{0h}$, $\tilde{\boldsymbol{u}}_h$, and $\hat{\boldsymbol{u}}_h$ in terms of the nodal shape functions and the corresponding nodal values.

The displacement fields are not arbitrary but must satisfy the Dirichlet boundary conditions specified above. For the field $\boldsymbol{u}_h$, enforcing $\boldsymbol{u} = \bar{\boldsymbol{u}}$ on $\Gamma_u^{\boldsymbol{X}}$ is achieved by prescribing the corresponding nodal values, i.e., $\boldsymbol{u}_a = \bar{\boldsymbol{u}}(\boldsymbol{X}_a)$ for all nodes $a$ on $\Gamma_u^{\boldsymbol{X}}$. Collecting all nodal displacement values into the array $\boldsymbol{U} \in \mathbb{R}^{n_{sh} \times 3}$, we partition it as
\begin{equation}
    \boldsymbol{U} = [\boldsymbol{U}_D,\; \boldsymbol{U}_F],
\end{equation}
where $\boldsymbol{U}_D$ collects the nodal values prescribed by the Dirichlet boundary conditions, and $\boldsymbol{U}_F$ collects the \emph{free} nodal values, i.e., the degrees of freedom to be determined by enforcing the balance of linear momentum, Eq.~\eqref{inter_weak_form}. The Dirichlet boundary conditions propagate to the field $\boldsymbol{u}_{0h}$ connecting the reference and stress-free configurations, and the corresponding nodal array is partitioned analogously as
\begin{equation}
    \boldsymbol{U}_0 = [\boldsymbol{U}_{0D},\; \boldsymbol{U}_{0F}].
\end{equation}

With these definitions, we introduce the discrete residuals. For a node $a$ not on the Dirichlet boundary, the external (force) residual is
\begin{equation}
(\boldsymbol{R}_{\mathrm{ext}})_a
=
\int_{\Omega_{\boldsymbol{X}_h}}
    \rho\, N_a\, \boldsymbol{b}\, J_0 \, \dd V_0
+
\int_{\Gamma_{t,h}^{\boldsymbol{X}}}
    N_a\, \bar{\boldsymbol{T}} \, \dd A_0,
\label{R_ext}
\end{equation}
where $\Omega_{\boldsymbol{X}_h}$ and $\Gamma_{t,h}^{\boldsymbol{X}}$ are the discretizations of $\Omega_{\boldsymbol{X}}$ and $\Gamma_{t}^{\boldsymbol{X}}$, respectively. The external residual depends only on the known body force $\boldsymbol{b}$ and prescribed traction $\bar{\boldsymbol{T}}$. The internal (stress) residual for free node $a$ is
\begin{equation}
(\boldsymbol{R}_{\mathrm{int}})_a
=
\int_{\Omega_{\boldsymbol{X}_h}}
    \boldsymbol{P}_h\, \nabla_{\boldsymbol{X}} N_a
    \, \dd V_0,
\label{R_int}
\end{equation}
where we have used the identity $\boldsymbol{P}:\nabla_{\boldsymbol{X}}(N_a\,\delta\boldsymbol{u}_a) 
= (\boldsymbol{P}\,\nabla_{\boldsymbol{X}} N_a)\cdot\delta\boldsymbol{u}_a$ to reduce the double contraction to a matrix-vector product, yielding a vector residual in $\mathbb{R}^3$ for each node $a$.

Computing $\boldsymbol{P}_h$ in Eq.~\eqref{R_int} requires the following steps. Given a point $\boldsymbol{X}_h$, the discrete deformation gradient $\boldsymbol{F}_h$ is computed from the displacement field $\boldsymbol{u}_h$, which is entirely encoded in $\boldsymbol{U}$. The discrete deformation gradient $\boldsymbol{F}_{0h}$ is computed from the displacement field $\boldsymbol{u}_{0h}$, entirely encoded in $\boldsymbol{U}_0$. Given $\boldsymbol{F}_h$ and $\boldsymbol{F}_{0h}$, the 
deformation gradient relative to the stress-free configuration follows from Eq.~\eqref{F_split} as $\tilde{\boldsymbol{F}}_h = \boldsymbol{F}_h\,\boldsymbol{F}_{0h}^{-1}$. 
The discrete first Piola-Kirchhoff stress $\tilde{\boldsymbol{P}}_h$ is then evaluated from Eq.~\eqref{P_tilde_neoHooke}, and pulled back to $\boldsymbol{P}_h$ using Eq.~\eqref{eq:Piola_pullback_P}.

The material parameters are discretized in the same finite element space. In particular, the shear modulus and bulk modulus are assigned at the nodes and interpolated within each element using the nodal shape functions,
\begin{equation}
    \mu_h = N_a \mu_a,
    \qquad
    \kappa_h = N_a \kappa_a,
\end{equation}
where $\mu_a$ and $\kappa_a$ are the nodal material parameters. The constitutive response at each quadrature point is therefore evaluated using the interpolated values $\mu_h$ and $\kappa_h$. In the heterogeneous examples considered below, the spatial locations of the material regions are assumed to be known from segmentation or mesh labeling. Each node is assigned a material-region label. For a problem with $n_{\mathrm{in}}$ mechanically distinct inclusions, the domain $\Omega_{\tilde{\boldsymbol{X}}}$ is decomposed into the matrix region $\Omega_{\mathrm{mat}}$ and inclusion regions $\Omega_{\mathrm{in}_j}$, $j=1,\ldots,n_{\mathrm{in}}$. The unknown material parameters associated with the matrix are denoted by $\mu_{\mathrm{mat}}$ and $\kappa_{\mathrm{mat}}$, while those associated with the $j$-th inclusion are denoted by $\mu_{\mathrm{in}_j}$ and $\kappa_{\mathrm{in}_j}$. The nodal material parameters are then defined region-wise as
\begin{equation}
\mu_a =
\begin{cases}
\mu_{\mathrm{mat}}, & \tilde{\boldsymbol{X}}_a \in \Omega_{\mathrm{mat}},\\
\mu_{\mathrm{in}_j}, & \tilde{\boldsymbol{X}}_a \in \Omega_{\mathrm{in}_j}, \quad j=1,\ldots,n_{\mathrm{in}},
\end{cases}
\qquad
\kappa_a =
\begin{cases}
\kappa_{\mathrm{mat}}, & \tilde{\boldsymbol{X}}_a \in \Omega_{\mathrm{mat}},\\
\kappa_{\mathrm{in}_j}, & \tilde{\boldsymbol{X}}_a \in \Omega_{\mathrm{in}_j}, \quad j=1,\ldots,n_{\mathrm{in}}.
\end{cases}
\label{eq:regionwise_material_assignment}
\end{equation}

Accordingly, the vectors of material parameters that need to be optimized are denoted by
\begin{equation} \boldsymbol{\kappa} =
\left[
\kappa_{\mathrm{mat}},
\kappa_{\mathrm{in}_1},
\kappa_{\mathrm{in}_2},
\ldots,
\kappa_{\mathrm{in}_{n_{\mathrm{in}}}}
\right]^{T},\\
\boldsymbol{\mu} =
\left[
\mu_{\mathrm{mat}},
\mu_{\mathrm{in}_1},
\mu_{\mathrm{in}_2},
\ldots,
\mu_{\mathrm{in}_{n_{\mathrm{in}}}}
\right]^{T}.
\end{equation}

We note that in practice, the residuals are assembled element by element rather than node by node. The nodal viewpoint adopted here is chosen for notational clarity in identifying the role of each quantity. The internal residual $\boldsymbol{R}_{\mathrm{int}}$ depends on the displacement fields $\boldsymbol{U}$ and $\boldsymbol{U}_0$ as well as on the material parameter vectors $\boldsymbol{\mu}$ and $\boldsymbol{\kappa}$ through the constitutive response. In contrast, the external residual $\boldsymbol{R}_{\mathrm{ext}}$ depends on the applied loading, consisting of body forces $\boldsymbol{b}$, prescribed tractions $\bar{\boldsymbol{T}}$, and the displacement field $\boldsymbol{U}_0$. The dependence on the body force  $\boldsymbol{b}$ arises from the volumetric loading term appearing in the weak form, which contributes through the domain integral in Eq.~\eqref{R_ext}. Since the equilibrium equations are expressed on the reference configuration $\Omega_{\boldsymbol{X}}$, the geometric mapping associated with $\boldsymbol{U}_0$ introduces the Jacobian factor $J_0$, resulting in an implicit dependence of the external residual on the current estimate of the unloaded configuration. The discrete balance of linear momentum for interior nodes and nodes with prescribed traction, therefore, reads

\begin{equation}
    \boldsymbol{R}_{\mathrm{ext}}(\boldsymbol{b},\,\bar{\boldsymbol{T}},\,\boldsymbol{U}_{0D},\,\boldsymbol{U}_{0F})
    -
    \boldsymbol{R}_{\mathrm{int}}(\boldsymbol{U}_D,\,\boldsymbol{U}_F,\,\boldsymbol{U}_{0D},\,\boldsymbol{U}_{0F},\,\boldsymbol{\mu},\,\boldsymbol{\kappa})
    = \boldsymbol{0}.
    \label{eq:discrete_momentum}
\end{equation}

For a given problem, the loading $\boldsymbol{b}$ and $\bar{\boldsymbol{T}}$, and the Dirichlet data $\boldsymbol{U}_D$ and $\boldsymbol{U}_{0D}$, are known. The unknowns are the free nodal arrays $\boldsymbol{U}_F$ and $\boldsymbol{U}_{0F}$, together with the material parameters $\mu$ and $\kappa$. The problem formulation that determines these unknowns is presented next. 

\subsection{Inverse Problem Formulation and Optimization Framework}
\label{sec:inverse_problem}

Eq.~\eqref{eq:discrete_momentum} specifies the discrete equilibrium equations and makes explicit the dependence on the degrees of freedom and material parameter vectors. As discussed in Sections~\ref{sec:problem_formulation} and~\ref{sec:fem_discretization}, equilibrium is governed by the body forces $\boldsymbol{b}$, the boundary tractions $\bar{\boldsymbol{T}}$, the displacement boundary conditions encoded in $\boldsymbol{U}_D$ and $\boldsymbol{U}_{0D}$, the material parameter vectors $\boldsymbol{\mu}$ and $\boldsymbol{\kappa}$, the free displacement degrees of freedom $\boldsymbol{U}_F$, and the free degrees of freedom $\boldsymbol{U}_{0F}$ of the field mapping the reference to the stress-free configuration. Together with the reference geometry $\Omega_{\boldsymbol{X}}$, these quantities completely specify the discrete equilibrium problem.

In the forward problem, the loading, boundary conditions, material parameter vectors, and the field $\boldsymbol{u}_0$ are all assumed to be known. The forward solver then determines $\boldsymbol{U}_F$, i.e., the deformed configuration coordinate $\boldsymbol{x}$ satisfying Eq.~\eqref{eq:discrete_momentum}. In the present work, the forward problem is solved using JAX-FEM \cite{xue_jax-fem_2023}.

The problem of interest here is an inverse problem. Rather than predicting $\boldsymbol{U}_F$ for given $(\boldsymbol{b}, \bar{\boldsymbol{T}}, \boldsymbol{U}_D,$ $ \boldsymbol{U}_{0D}, \boldsymbol{U}_{0F}, \boldsymbol{\mu}, \boldsymbol{\kappa})$, we seek to recover the material parameter vectors and the stress-free geometry from observations of the deformed state. Specifically, the observed configuration coordinate $\hat{\boldsymbol{x}}$ and the loading and boundary conditions $(\boldsymbol{b}, \bar{\boldsymbol{T}}, \boldsymbol{U}_D, \boldsymbol{U}_{0D})$ are assumed to be known, and the unknowns are $\boldsymbol{\mu}$, $\boldsymbol{\kappa}$, and $\boldsymbol{U}_{0F}$, the last of which encodes the stress-free geometry $\Omega_{\tilde{\boldsymbol{X}}}$. Since Eq.~\eqref{eq:discrete_momentum} cannot be readily inverted for these unknowns, we instead formulate an iterative optimization algorithm. Given a current estimate of $(\boldsymbol{U}_{0F}, \boldsymbol{\mu}, \boldsymbol{\kappa})$, the forward problem is solved with JAX-FEM to obtain the predicted configuration coordinate $\boldsymbol{x}$, which is then compared with the observed configuration coordinate $\hat{\boldsymbol{x}}$. The discrepancy is used to update the estimate, and the process is repeated until $\boldsymbol{x} \approx \hat{\boldsymbol{x}}$ according to a convergence criterion specified later.

To make the optimizable parameters explicit, we introduce the parameterization
\begin{equation}
    \boldsymbol{\theta}_g \equiv \boldsymbol{U}_{0F},
    \qquad
    \boldsymbol{\theta}_\mu \equiv \texttt{softplus}^{-1}(\boldsymbol{\mu}),
    \qquad
    \boldsymbol{\theta}_\kappa \equiv \texttt{softplus}^{-1}(\boldsymbol{\kappa}),
    \label{eq:theta_params}
\end{equation}
where the \texttt{softplus} reparameterization enforces positivity of the material 
parameters throughout the optimization. We further account for observations of 
multiple deformed states under different loading conditions. We denote by 
$\hat{\boldsymbol{x}}^i$ the observed configuration coordinate corresponding to the $i$-th 
loading $\boldsymbol{b}^i$, and assume for simplicity that traction loading is 
absent, i.e., $\bar{\boldsymbol{T}}^i = \boldsymbol{0}$ for all $i$.

The primary objective measures the mismatch between the predicted and observed configurations at the nodes. 
The normalized nodal mismatch for loading $i$ is
\begin{equation}
\mathcal{J}^{i}_{\mathrm{p}}(\boldsymbol{\theta}_g, \boldsymbol{\theta}_{\mu}, \boldsymbol{\theta}_{\kappa}) 
=
\frac{
    \displaystyle\sum_{a=1}^{n_{sh}} \left\|\hat{\boldsymbol{u}}^{i}_a\right\|^2
}{
    \displaystyle\sum_{a=1}^{n_{sh}} \left\|\hat{\boldsymbol{x}}^{i}_a\right\|^2}
=
\frac{
    \displaystyle\sum_{a=1}^{n_{sh}} \left\|\boldsymbol{x}^{i}_a(\boldsymbol{\theta}_g, \boldsymbol{\theta}_{\mu}, \boldsymbol{\theta}_{\kappa}) 
    - \hat{\boldsymbol{x}}^{i}_a\right\|^2
}{
    \displaystyle\sum_{a=1}^{n_{sh}} \left\|\hat{\boldsymbol{x}}^{i}_a\right\|^2
}
\label{eq:nodal_mismatch}
\end{equation}

While driving $\mathcal{J}^i_{\mathrm{p}}$ to zero would, in principle, solve the inverse problem, in practice, the regularity of the displacement field $\hat{\boldsymbol{u}}$ is not preserved during optimization, which can cause the forward solves to fail to converge. A necessary local condition for an orientation-preserving mapping between $\Omega_{\boldsymbol{x}}$ and $\Omega_{\hat{\boldsymbol{x}}}$ is $\hat{J} = \det\hat{\boldsymbol{F}} > 0$ everywhere. In practice, penalizing deviations of $\hat{J}$ from unity alone proved insufficient to ensure smooth optimization. Instead, we propose a penalty based on richer information from $\hat{\boldsymbol{F}}$ by introducing the isochoric part of the associated right Cauchy-Green tensor,
\begin{equation}
\hat{\bar{\boldsymbol{C}}}
=
\hat{J}^{-2/3}\,\hat{\boldsymbol{F}}^{T}\hat{\boldsymbol{F}}.
\end{equation}
The deformation gradient mismatch objective for loading $i$ is then
\begin{equation}
\mathcal{J}^{i}_{\mathrm{d}}(\boldsymbol{\theta}_g, \boldsymbol{\theta}_{\mu}, \boldsymbol{\theta}_{\kappa})
=
\sum_{q=1}^{N_{q}}\left[
    \left(\hat{J}^{i}_{q} - 1\right)^2
    +
    \left\|\hat{\bar{\boldsymbol{C}}}^{i}_{q} - \boldsymbol{I}\right\|_F^2
\right],
\label{eq:grad_mismatch}
\end{equation}
where the sum runs over all integration points $q$ of the finite element mesh, at which $\hat{\boldsymbol{F}}$ is evaluated using standard finite element machinery, and $\|\cdot\|_F$ denotes the Frobenius norm. Both terms vanish when $\hat{\boldsymbol{F}}$ is a rotation, including the special case $\hat{\boldsymbol{F}} = \boldsymbol{I}$. Consequently, this term controls local volumetric and isochoric stretch mismatch but does not detect a local rigid rotation. Exact agreement of the predicted and observed nodal positions is measured separately by $\mathcal{J}^{i}_{\mathrm{p}}$.

\begin{figure}[H]
\begin{center}
\includegraphics [width = 1\textwidth]{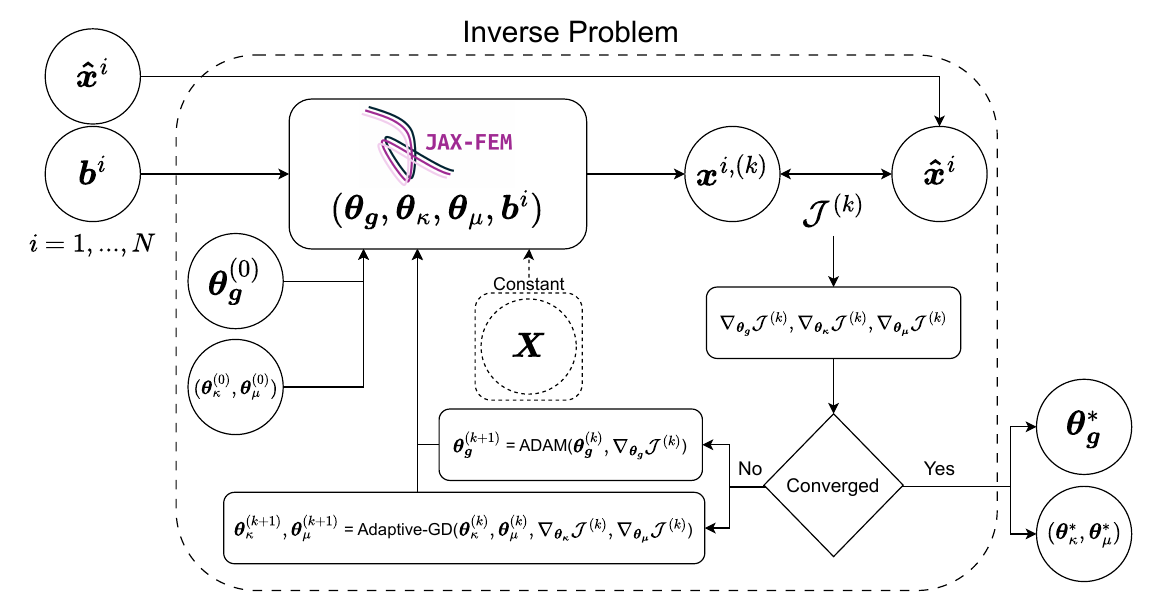} 

\caption{Schematic of the multi-loading inverse FEM framework. 
The observed deformed configurations $\hat{\boldsymbol{x}}^i$ and the corresponding loading vectors $\boldsymbol{b}^i$, $i=1,\ldots,N$, are used as input data for the inverse problem. 
The fixed reference configuration $\boldsymbol{X}$ is kept constant throughout the optimization, while the unloaded configuration $\tilde{\boldsymbol{X}}$ is represented implicitly through the geometric parameters $\boldsymbol{\theta}_g$. Material parameter vectors are denoted $\boldsymbol{\theta}_\kappa,\boldsymbol{\theta}_\mu$. 
At the $k$-th optimization iteration, the JAX-FEM forward solver computes the predicted deformed configuration $\boldsymbol{x}^{i,(k)}$ for each loading case using the current estimates of the geometric and material parameters $(\boldsymbol{\theta}_g^{(k)},\boldsymbol{\theta}_\kappa^{(k)},\boldsymbol{\theta}_\mu^{(k)})$. 
The predicted configurations are compared with the corresponding observed configurations through the objective function $\mathcal{J}^{(k)}$; see Eq.~\eqref{eq:objective_total}. 
Because the full forward FEM pipeline is differentiable, automatic differentiation is used to compute the gradients of $\mathcal{J}^{(k)}$ with respect to both the geometric and material parameters. 
The geometric parameters are updated using Adam, while the material parameters are updated using an adaptive gradient-descent scheme, enabling simultaneous recovery of the unloaded geometry and constitutive properties. After convergence, the optimized parameters $\boldsymbol{\theta}_g^*$, $\boldsymbol{\theta}_\kappa^*$, and $\boldsymbol{\theta}_\mu^*$ define the recovered unloaded geometry and material properties.}
\label{fig:JAX_FEM_framework}
\end{center}
\end{figure}

The total objective combines the nodal mismatch and the deformation gradient regularization term across all $N$ loading conditions,

\begin{equation}
\mathcal{J}(\boldsymbol{\theta}_g, \boldsymbol{\theta}_{\mu}, \boldsymbol{\theta}_{\kappa}) 
=
(100-\lambda_{\mathcal{J}})
\frac{
\displaystyle\sum_{i=1}^{N}
\mathcal{J}^{i}_{\mathrm{p}}
(\boldsymbol{\theta}_g,\boldsymbol{\theta}_{\mu},\boldsymbol{\theta}_{\kappa})
}{
\displaystyle\sum_{i=1}^{N}
\mathcal{J}^{i}_{\mathrm{p},0}
}
+
\lambda_{\mathcal{J}}
\frac{
\displaystyle\sum_{i=1}^{N}
\mathcal{J}^{i}_{\mathrm{d}}
(\boldsymbol{\theta}_g,\boldsymbol{\theta}_{\mu},\boldsymbol{\theta}_{\kappa})
}{
\displaystyle\sum_{i=1}^{N}
\mathcal{J}^{i}_{\mathrm{d},0}
},
\label{eq:objective_total}
\end{equation}

where $\mathcal{J}^{i}_{\mathrm{p},0}$ and $\mathcal{J}^{i}_{\mathrm{d},0}$ denote the values of $\mathcal{J}^{i}_{\mathrm{p}}$ and $\mathcal{J}^{i}_{\mathrm{d}}$ at the initial iterate ($0$-th iteration), respectively, and $\lambda_{\mathcal{J}} \in [0, 100]$ is a weighting parameter controlling the relative contributions of the two terms. 
Normalizing by the initial values ensures that, at the start of the optimization, the total initial objective is equal to $100$ regardless of the weighting. In this work, we set $\lambda_{\mathcal{J}} = 99$ to place greater emphasis on the deformation-gradient mismatch, which helps maintain physically admissible deformations with $\hat{J} > 0$ during the optimization. The overall optimization framework is illustrated in Fig.~\ref{fig:JAX_FEM_framework}.

\subsubsection{Optimization Algorithm}
\label{sec:optimization}

The advantage of the proposed numerical implementation using JAX-FEM is that the objective function in Eq. (\ref{eq:objective_total}) is automatically differentiable with respect to the parameters $[\boldsymbol{\theta}_g,\boldsymbol{\theta}_\mu,\boldsymbol{\theta}_\kappa]$ using JAX's built-in \texttt{grad} function. This means that the gradients $\boldsymbol{g}_{\boldsymbol{\theta}_g}:=\ \nabla_{\boldsymbol{\theta_{g}}}\mathcal{J}, \, \boldsymbol{g}_{\theta_\mu}:= \nabla_{\boldsymbol{\theta_{\mu}}}\mathcal{J} ,\,\boldsymbol{g}_{\boldsymbol{\theta}_\kappa}:= \nabla_{\boldsymbol{\theta_{\kappa}}}\mathcal{J}
$ are available after solving the forward problem(s) with JAX for a given set $[\boldsymbol{\theta}_g,\boldsymbol{\theta}_\mu,\boldsymbol{\theta}_\kappa]$. Hence, gradient-based optimization is followed to iteratively solve the inverse problem. 

The optimization comprises two coupled update mechanisms executed within the same iteration loop. The Adam optimizer is used to update the nodal parameters according to $\boldsymbol{\theta}^{(k+1)}_g \leftarrow \operatorname{AdamUpdate}\!\left(\boldsymbol{\theta}^{(k)}_g,\boldsymbol{g}^{(k)}_{\boldsymbol{\theta}_g};\alpha_0\right)$. The Adam optimizer provides robust, adaptive step sizes in this high-dimensional space \cite{kingma2014adam}. The material variables $(\boldsymbol{\theta}_\kappa,\boldsymbol{\theta}_\mu)$ are updated by a low-dimensional gradient-descent scheme with an adaptive relative step size.

Because the gradients of $\boldsymbol{\kappa}$ and $\boldsymbol{\mu}$ may differ by several orders of magnitude, the use of a fixed learning rate can lead either to instability or excessively slow convergence. To obtain robust updates that are insensitive to gradient scaling, the material variables are updated using a relative parameter change, denoted by $\texttt{rel\_step}$, such that the update magnitude is proportional to the current parameter value while the gradient sign determines the descent direction. This strategy is related to sign-based optimization methods such as resilient backpropagation (RPROP) \cite{riedmiller1993rprop}, in which the update direction depends only on the sign of the gradient and is therefore less sensitive to poorly scaled sensitivities. Similar ideas have also been explored in relative or percentage-based gradient updates \cite{abu2017percentdelta}, where parameters are updated proportionally to their current magnitude in order to ensure comparable relative evolution across variables with different scales. Motivated by these ideas, the estimates of $\boldsymbol{\kappa}$ and $\boldsymbol{\mu}$ are updated at iteration $k+1$ of the algorithm as
\begin{equation}
\boldsymbol{\theta}_\kappa^{(k+1)}
=
\boldsymbol{\theta}_\kappa^{(k)}
-
\Delta_{\boldsymbol{\theta}_\kappa}\,
\mathrm{sign}(g_{\boldsymbol{\theta}_\kappa}),
\qquad
\boldsymbol{\theta}_\mu^{(k+1)}
=
\boldsymbol{\theta}_\mu^{(k)}
-
\Delta_{\boldsymbol{\theta}_\mu}\,
\mathrm{sign}(g_{\boldsymbol{\theta}_\mu}),
\end{equation}
where 
\begin{equation}
\Delta_{\boldsymbol{\theta}_\kappa}
=
\texttt{rel\_step}\,\boldsymbol{\kappa},
\qquad
\Delta_{\boldsymbol{\theta}_\mu}
=
\texttt{rel\_step}\,\boldsymbol{\mu},
\end{equation}
and $\texttt{rel\_step}$ controls the relative change in each material parameter per iteration.

To avoid oscillations of $\boldsymbol{\theta}_\kappa$ and $\boldsymbol{\theta}_\mu$ near convergence, we monitor the recent history of the material parameters and objective value over a sliding window of fixed length. 
When the relative fluctuations of both material parameters over the sliding window fall below 
$1.5\,\texttt{min\_rel\_step}$, the objective variation falls below a prescribed objective tolerance, 
and the material relative step size has reached its minimum allowable value $\texttt{min\_rel\_step}$, 
the optimization is considered converged. 
Here, $\texttt{min\_rel\_step}$ is a user-defined hyperparameter that specifies the minimum allowable relative step size for the material-parameter updates. 
Full details of the window-based convergence criterion are provided in ~\ref{app:material_update}. 
The hyperparameters controlling the optimization procedure are summarized in ~\ref{app:hyperparameters}. 
When the convergence criteria are satisfied, the current estimates of the unknowns 
$\boldsymbol{\theta}_g,\,\boldsymbol{\theta}_\kappa$, and $\boldsymbol{\theta}_\mu$ are stored as 
$\boldsymbol{\theta}_g^\ast,\,\boldsymbol{\theta}_\kappa^\ast$, and $\boldsymbol{\theta}_\mu^\ast$.

\section{Results}
\label{sec:results}

\subsection{Homogeneous Benchmark Problem: Cube with Transverse Holes}
\label{sec:results_benchmark}

\paragraph{Geometry, Loading, and Boundary Conditions}
\label{par:benchmark_geom_bc}
\mbox{}\\

\begin{figure}[H]
\centering
\includegraphics[width=0.9\textwidth]
{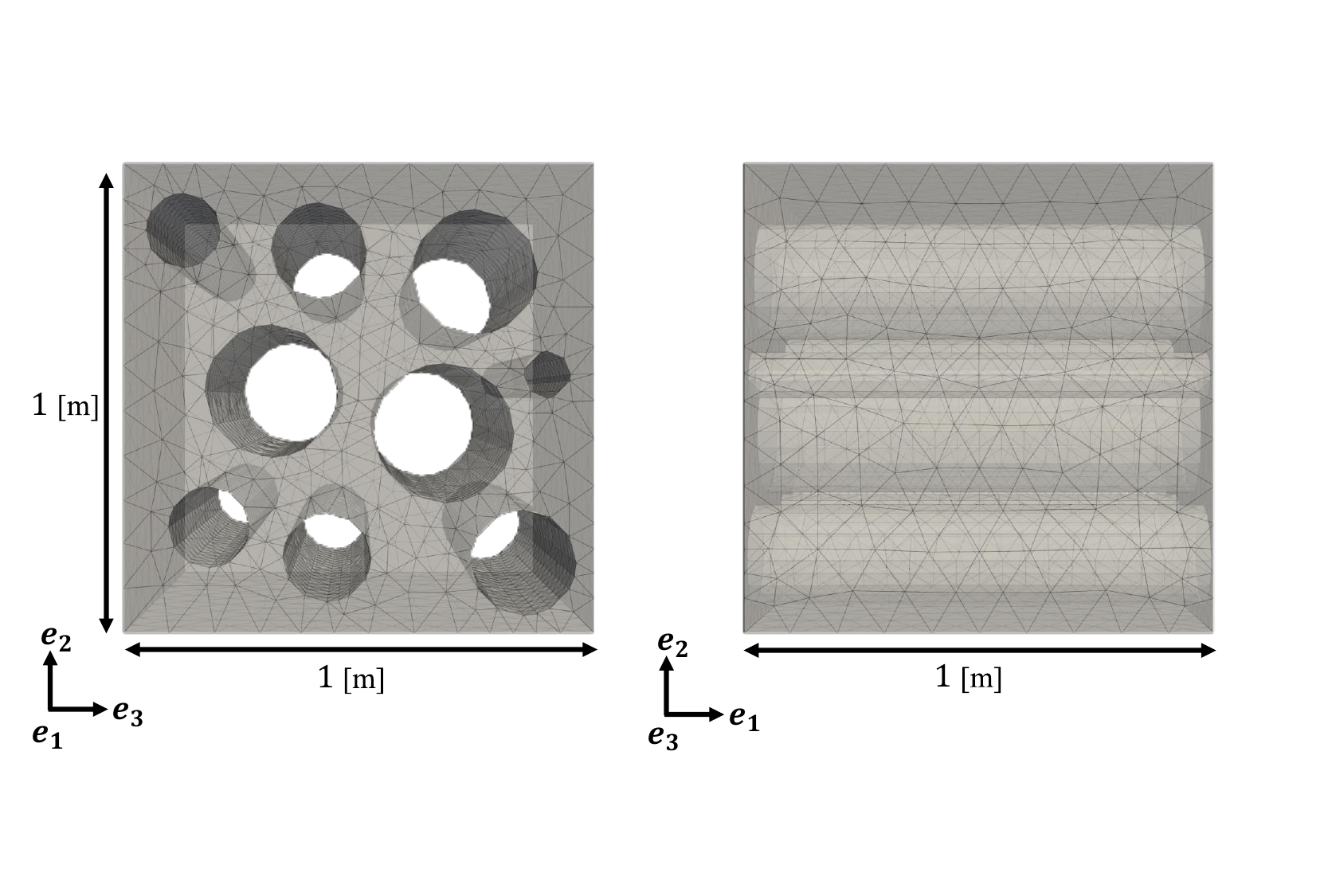}
\caption{
Finite element mesh of a unit cube containing several holes aligned along the $\boldsymbol{e}_{1}$-direction. This mesh serves as the true unloaded configuration for the benchmark-problem verification. }
\label{fig:True_sol}
\end{figure}

We exercise our optimization algorithm in a controlled benchmark where the unloaded geometry $\Omega_{\boldsymbol{\tilde{X}}}$ is known. 
The coordinates in $\Omega_{\boldsymbol{\tilde{X}}}$ are  $\boldsymbol{\tilde{X}} = \tilde{X}_{i} \boldsymbol{e}_i$, where $\{\boldsymbol{e}_j\}_{j=1,2,3}$ is the Cartesian unit basis. 
The unloaded geometry is a unit cube with several cylindrical holes aligned along the $\boldsymbol{e}_1$ direction; see Fig.~\ref{fig:True_sol}. 
The homogeneous cube is made of a compressible neo-Hookean material with shear modulus $\mu_{\mathrm{mat}} = 3.846~[\mathrm{Pa}]$ and bulk modulus $\kappa_{\mathrm{mat}} = 8.333~[\mathrm{Pa}]$, or equivalently, Young's modulus $E_{\mathrm{mat}} \approx 10~[\mathrm{Pa}]$ and Poisson's ratio $\nu_{\mathrm{mat}} \approx 0.3$. 

The geometry is discretized with $\text{17,406}$ linear tetrahedral elements. To generate synthetic observations for the inverse problem, the reference configuration is chosen to coincide with the true unloaded configuration, such that $\Omega_{\boldsymbol{X}} = \Omega_{\tilde{\boldsymbol{X}}}$ for the dataset generation. The domain is subjected to a body force $\rho \boldsymbol{b}$ representing gravity with different scaling factors and orientations. A unit density $\rho = 1\,[\mathrm{kg/m^3}]$ is assumed, while the gravitational acceleration magnitude is fixed as $g = 9.81\,[\mathrm{m/s^2}]$.

\begin{figure}[H]
\centering

\begin{minipage}[t]{0.30\textwidth}
    \centering
    \includegraphics[width=\linewidth]{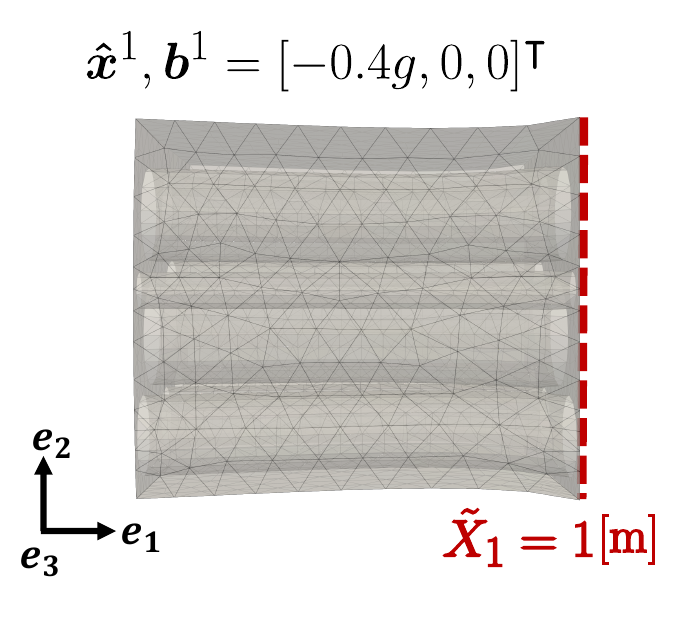}
    (a)
\end{minipage}
\hfill
\begin{minipage}[t]{0.30\textwidth}
    \centering
    
    \includegraphics[width=\linewidth]{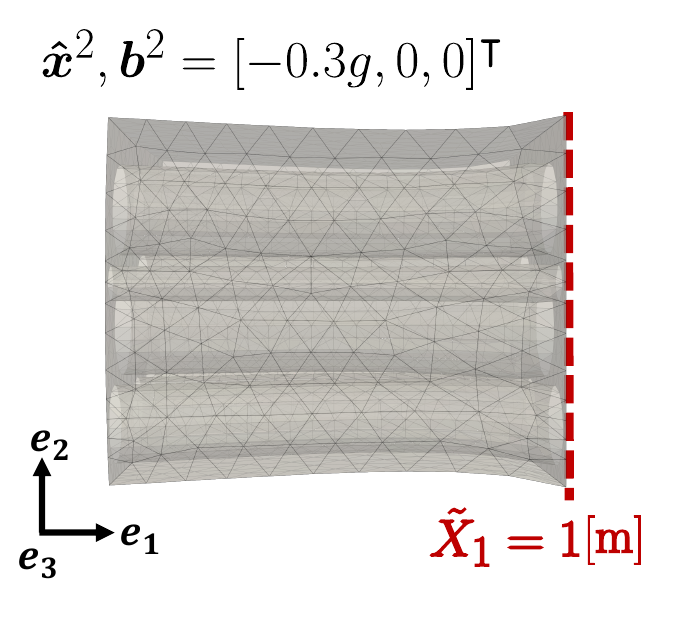}
    (b)\\[-0.1em]
\end{minipage}
\hfill
\begin{minipage}[t]{0.30\textwidth}
    \centering
    
    \includegraphics[width=\linewidth]{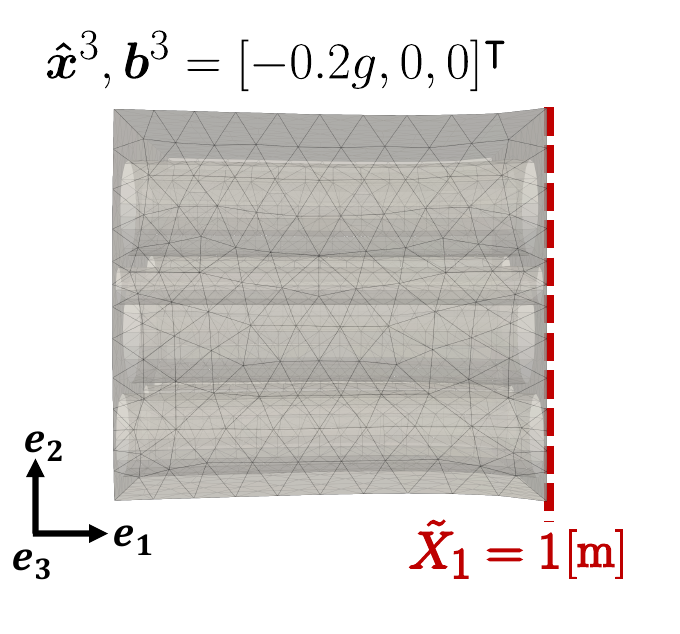}
    (c)\\[-0.1em]
\end{minipage}

\vspace{0.1em}

\begin{minipage}[t]{0.30\textwidth}
    \centering
    
    \includegraphics[width=\linewidth]{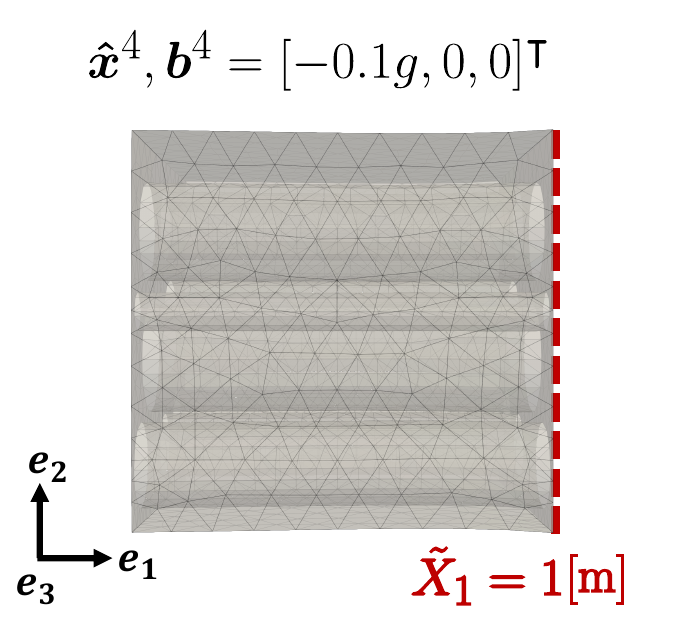}
    (d)\\[-0.1em]
\end{minipage}
\hfill
\begin{minipage}[t]{0.30\textwidth}
    \centering
    
    \includegraphics[width=\linewidth]{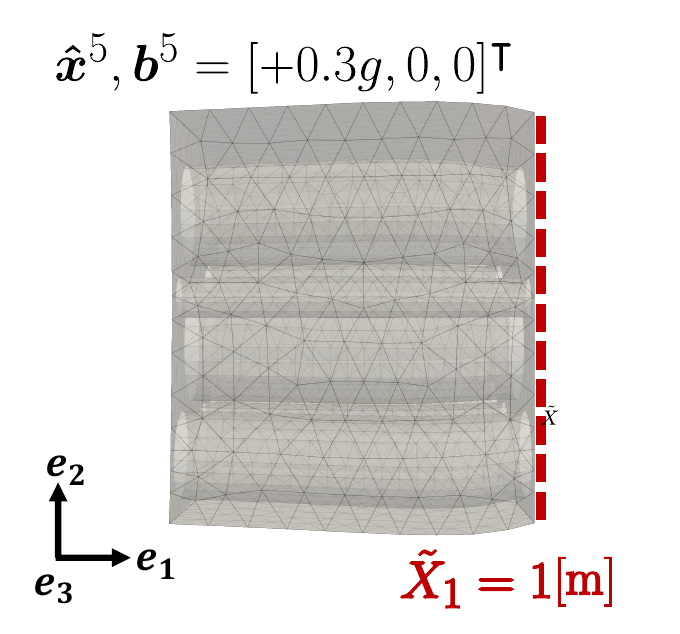}
    (e)\\[-0.1em]
\end{minipage}
\hfill
\begin{minipage}[t]{0.30\textwidth}
    \centering
    
    \includegraphics[width=\linewidth]{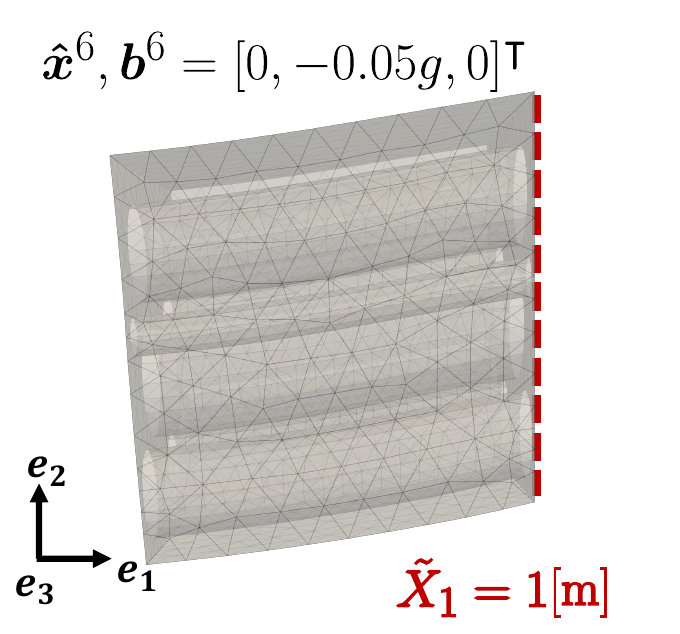}
    (f)\\[-0.1em]
\end{minipage}

\vspace{0.1em}

\begin{minipage}[t]{0.30\textwidth}
    \centering
    
    \includegraphics[width=\linewidth]{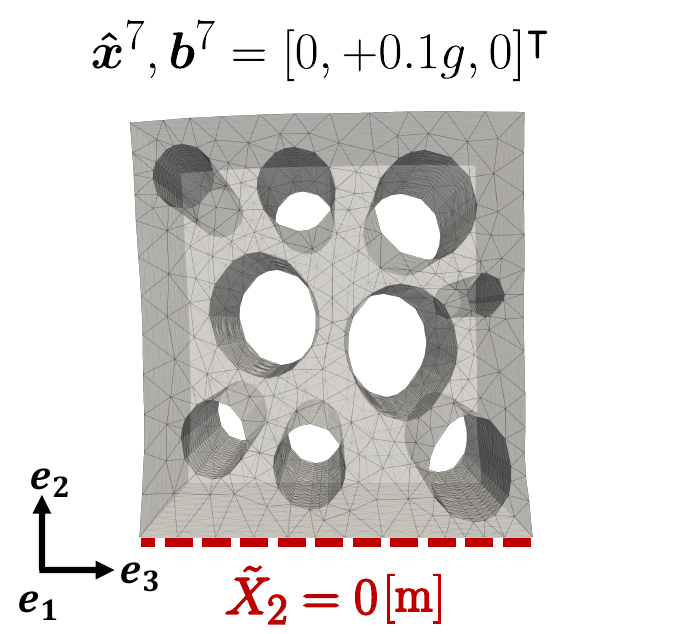}
    (g)\\[-0.1em]
\end{minipage}
\hfill
\begin{minipage}[t]{0.30\textwidth}
    \centering
    
    \includegraphics[width=\linewidth]{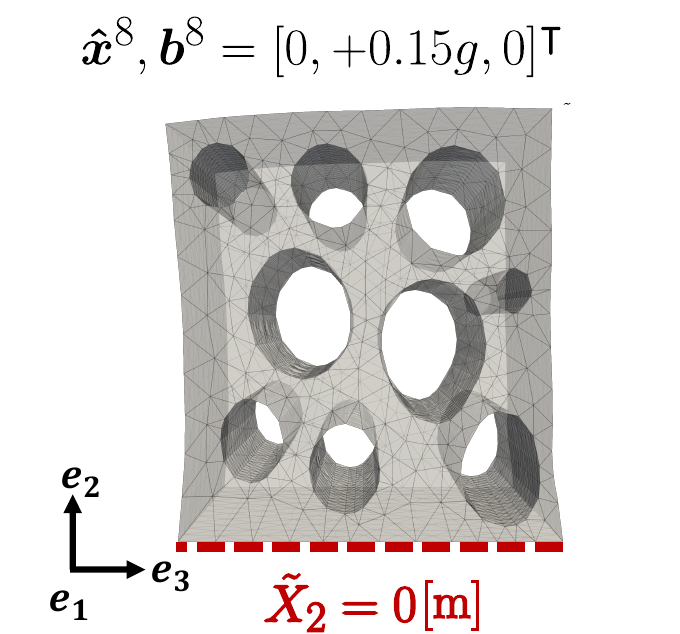}
    (h)\\[-0.1em]
\end{minipage}
\hfill
\begin{minipage}[t]{0.30\textwidth}
    \centering
    
    \includegraphics[width=\linewidth]{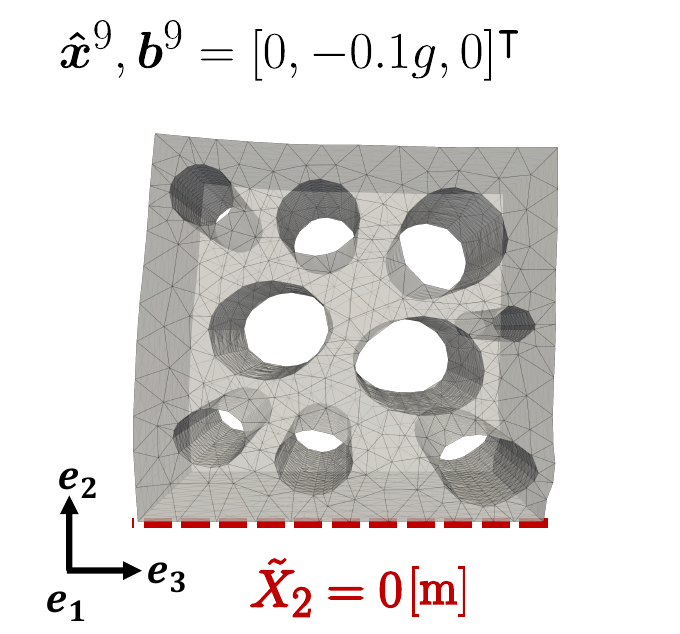}
    (i)\\[-0.1em]
\end{minipage}

\vspace{-0.1em}

\begin{minipage}[t]{0.30\textwidth}
    \centering
    
    \includegraphics[width=\linewidth]{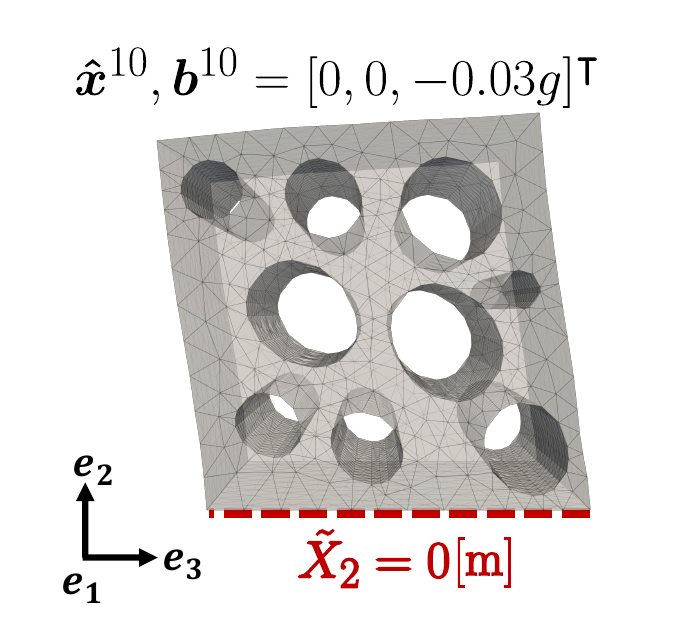}
    (j)\\[-0.1em]
\end{minipage}

\caption{
Synthetic observed configurations generated from the stress-free configuration using different body-force vectors $\boldsymbol{b}^i$ representing scaled gravity loadings. The dashed red line indicates the cube's face where a zero-displacement boundary condition is imposed.   
Each configuration $\hat{\boldsymbol{x}}^i$ is obtained by solving the forward problem with unit density $\rho = 1\,[\mathrm{kg/m^3}]$ and gravitational acceleration magnitude $g = 9.81\,[\mathrm{m/s^2}]$. 
Configuration (b) is used as the reference configuration for Tests~1--6, while configuration (g) is used as the reference configuration for Tests~7--9; see Table~\ref{tab:benchmark_tests}.
}
\label{fig:observed_configs}
\end{figure}

Multiple observed configuration coordinates $\hat{\boldsymbol{x}}^i$ are generated by prescribing different body-force vectors $\boldsymbol{b}^i$, corresponding to scaled or rotated gravity directions; see Fig.~\ref{fig:observed_configs}. For each loading case, together with its associated boundary conditions, the forward problem is solved using JAX-FEM by finding the displacement degrees of freedom $\boldsymbol{U}_F$ satisfying Eq.~\eqref{eq:discrete_momentum}, which constitute the only unknowns in this setting.

The Dirichlet boundary $\Gamma_{u}^{\boldsymbol{\tilde{X}}}$ is prescribed on different planes of the domain depending on the loading case. The boundary is defined as $\Gamma_u^{\boldsymbol{\tilde{X}}} = \{ \tilde{X}_1 = 1\,[\mathrm{m}] \}$ for cases (a-f), and $\Gamma_u^{\boldsymbol{\tilde{X}}} = \{ \tilde{X}_2 = 0\,[\mathrm{m}] \}$ for cases (g-j), as summarized in Fig.~\ref{fig:observed_configs}. We impose zero displacements on $\Gamma_u^{\boldsymbol{\tilde{X}}}$. 

The goal of the inverse problem is, as formulated in Section~\ref{sec:inverse_problem}, to reconstruct the unloaded geometry $\Omega_{\tilde{\boldsymbol{X}}}$ and the material parameters $\mu$ and $\kappa$ solely from the pairs $(\boldsymbol{b}^i, \hat{\boldsymbol{x}}^i)$.

\begin{table}
\centering
\normalsize
\caption{Summary of benchmark-problem test cases. Each test uses more than one observed configuration out of the set of observed geometries in Fig.~\ref{fig:observed_configs}. The initial values of the material parameters (at iteration $k=0$) are listed under $E_{\mathrm{mat}}^{(0)}$ and $\nu_{\mathrm{mat}}^{(0)}$. The reference configuration is set to  $\boldsymbol{\hat{x}}^{2}$ for Tests 1-6 and  $\boldsymbol{\hat{x}}^{7}$ for Tests 7-9. The initial guess for the displacement between reference and unloaded configurations is consistently set to $\boldsymbol{\theta}_g^{(0)}=\mathbf{0}$ such that, at the start of optimization, the reference configuration is effectively the initial guess for the unloaded configuration. }
\label{tab:benchmark_tests}

\begin{tabular}{c c c c c}
\hline\\[-4mm]
\textbf{Test}  & 
\textbf{$\boldsymbol{E_{\mathrm{mat}}^{(0)}}$} &
\textbf{$\boldsymbol{\nu_{\mathrm{mat}}^{(0)}}$} &
\textbf{Observed Configurations} &
\textbf{Fixed Plane} \\
\hline\\[-4mm]
1 & $1.3E_{\mathrm{mat}}$ & $1.2 \nu_{\mathrm{mat}}$ &
$\boldsymbol{\hat{x}}^{2},\boldsymbol{\hat{x}}^{1}$ & $\tilde{X}_1 = 1\,[\mathrm{m}]$ \\

2 & $1.3E_{\mathrm{mat}}$ & $1.2 \nu_{\mathrm{mat}}$ &
$\boldsymbol{\hat{x}}^{2},\boldsymbol{\hat{x}}^{5}$ & $\tilde{X}_1 = 1\,[\mathrm{m}]$ \\

3 & $0.7E_{\mathrm{mat}}$ & $0.8 \nu_{\mathrm{mat}}$ &
$\boldsymbol{\hat{x}}^{2},\boldsymbol{\hat{x}}^{1}$ & $\tilde{X}_1 = 1\,[\mathrm{m}]$ \\

4 & $1.3E_{\mathrm{mat}}$ & $1.2 \nu_{\mathrm{mat}}$ &
$\boldsymbol{\hat{x}}^{2},\boldsymbol{\hat{x}}^{6}$ & $\tilde{X}_1 = 1\,[\mathrm{m}]$\\

5 & $1.3E_{\mathrm{mat}}$ & $1.2 \nu_{\mathrm{mat}}$ &
$\boldsymbol{\hat{x}}^{2},\boldsymbol{\hat{x}}^{1},\boldsymbol{\hat{x}}^{3}$ & $\tilde{X}_1 = 1\,[\mathrm{m}]$\\

6 & $1.3E_{\mathrm{mat}}$ & $1.2 \nu_{\mathrm{mat}}$ &
$\boldsymbol{\hat{x}}^{2},\boldsymbol{\hat{x}}^{1},\boldsymbol{\hat{x}}^{3},\boldsymbol{\hat{x}}^{4}$ & $\tilde{X}_1 = 1\,[\mathrm{m}]$\\

7 & $1.3E_{\mathrm{mat}}$ & $1.2 \nu_{\mathrm{mat}}$ &
$\boldsymbol{\hat{x}}^{7},\boldsymbol{\hat{x}}^{8}$ & $\tilde{X}_2 = 0\,[\mathrm{m}]$\\

8 & $1.3E_{\mathrm{mat}}$ & $1.2 \nu_{\mathrm{mat}}$ &
$\boldsymbol{\hat{x}}^{7},\boldsymbol{\hat{x}}^{9}$ & $\tilde{X}_2 = 0\,[\mathrm{m}]$\\

9 & $1.3E_{\mathrm{mat}}$ & $1.2 \nu_{\mathrm{mat}}$ &
$\boldsymbol{\hat{x}}^{7},\boldsymbol{\hat{x}}^{10}$ & $\tilde{X}_2 = 0\,[\mathrm{m}]$\\

\hline
\end{tabular}
\end{table}

\begin{figure}[H]
\centering
\begin{minipage}{0.48\textwidth}
    \centering
    \includegraphics[width=\textwidth]{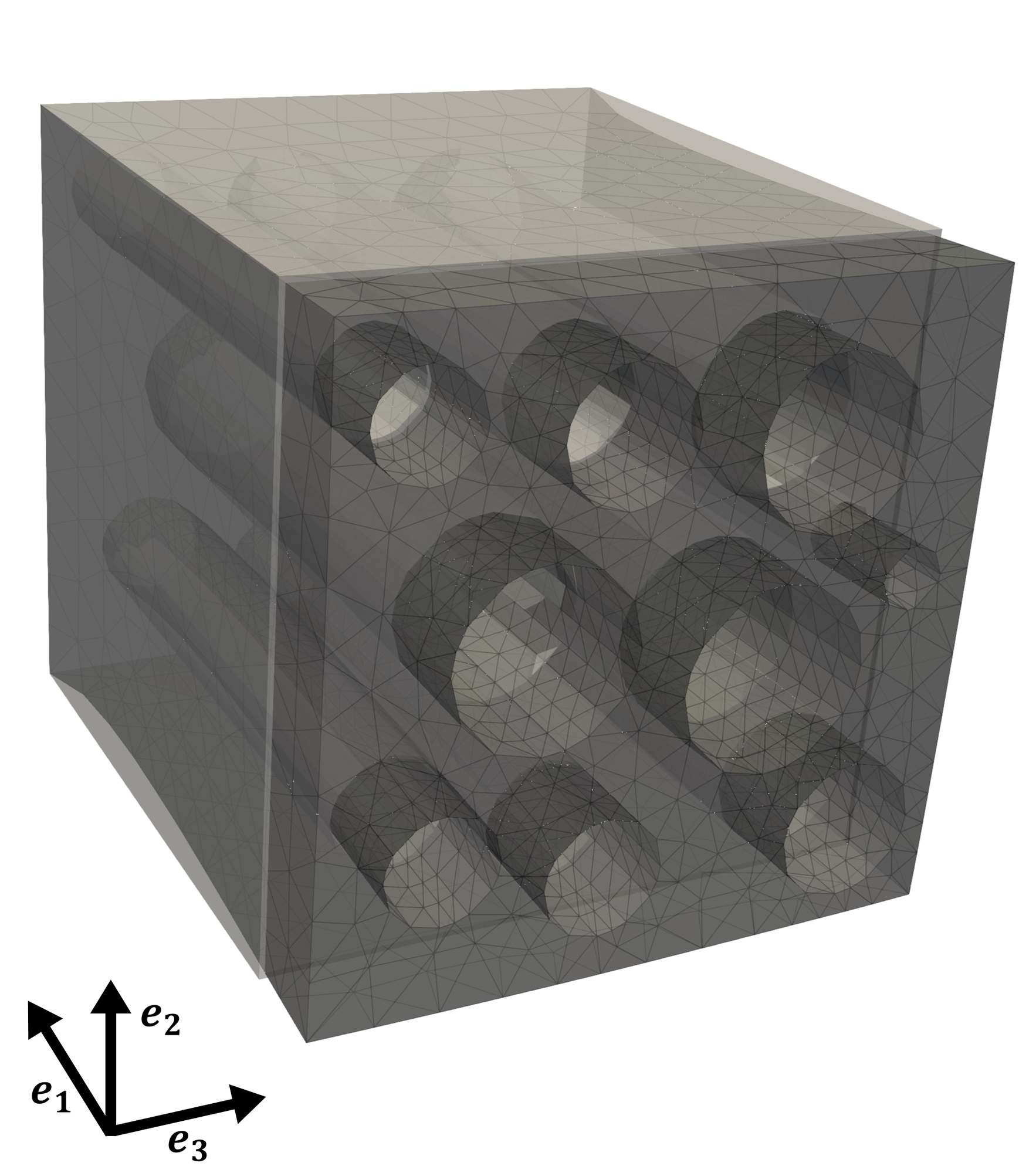}
    (a)\\
\end{minipage}
\hfill
\begin{minipage}{0.48\textwidth}
    \centering
    \includegraphics[width=\textwidth]{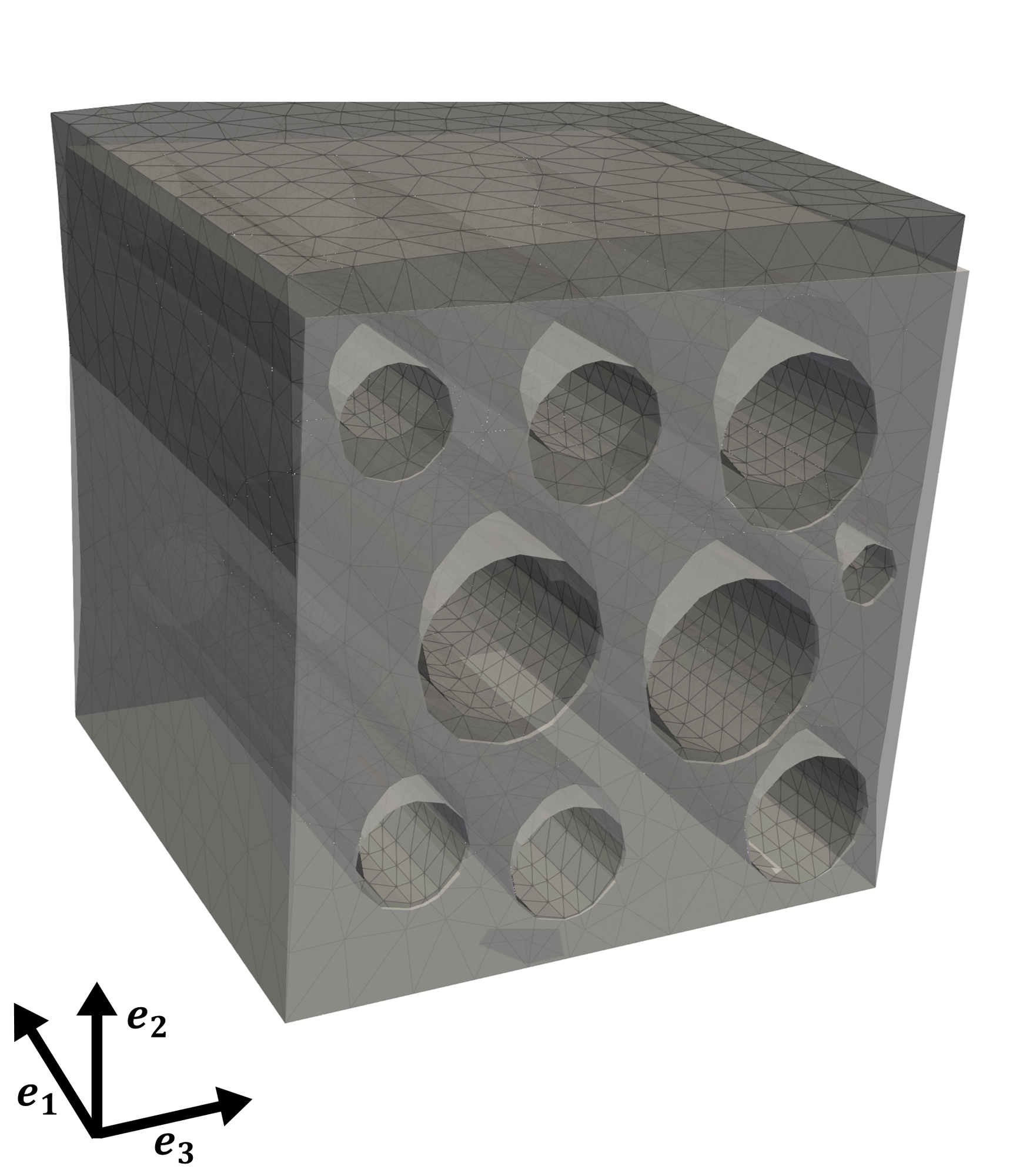}
    (b)\\
\end{minipage}
\caption{
Reference configurations, which coincide with the initial guesses of the unloaded configurations, used in the verification studies.
(a) Deformation corresponding to $\boldsymbol{b}^{2}$ is used as the reference configuration for Tests~1-6 (darker gray) overlaid with the true unloaded configuration (lighter gray).
(b) Deformation corresponding to $\boldsymbol{b}^{7}$ is used as the reference configuration for Tests~7-9 (darker gray) overlaid with the unloaded configuration (lighter gray).}
\label{figInitial_Guess}
\end{figure}

A total of nine inverse problems were solved, as summarized in Table~\ref{tab:benchmark_tests}. These tests examine the influence of the loading type, the number of observed configurations, and the initial guess on the quality of the reconstructed solution. To initialize the optimization, one must provide a reference configuration and an initial guess for the unloaded geometry, the latter encoded in $\boldsymbol{\theta}_{g}$. 
In all tests, the initial guess (iteration $k=0$) is set to $\boldsymbol{\theta}_{g}^{(0)} = \boldsymbol{0}$, i.e., the unloaded geometry is initially taken to coincide with the reference configuration. For Tests~1--6 (Table~\ref{tab:benchmark_tests}), the reference configuration is taken as the observed geometry under loading $\boldsymbol{b}^{2} = [-0.3g, 0,0]^{\intercal}$, i.e.,  $\Omega_{\boldsymbol{X}} := \Omega_{\hat{\boldsymbol{x}}^2}$. For Tests~7--9, it is taken as the observed configuration under loading $\boldsymbol{b}^{7} = [0,+0.1g,0]^{\intercal}$, i.e.,  $\Omega_{\boldsymbol{X}}:= \Omega_{\hat{\boldsymbol{x}}^7}$; see Fig.~\ref{figInitial_Guess} where the initial guesses are compared to the true unloaded configuration. By fixing the reference configuration and initial guess of the stress-free state within each group, the effect of the remaining optimization variables is isolated.

\paragraph{Effect of Loading-Mode Diversity}
\mbox{}\\

Tests 1, 2, and 4 in Table~\ref{tab:benchmark_tests} all share the same initial material guess and the same initial geometry, and differ only in the combination of observed loading cases. Test 1 includes two tensile observations (T-T); Test 2 includes one tensile and one compressive observation (T-C), and Test 4 includes one in-plane loading combined with a tensile, bending-dominated loading (T-B).

\begin{figure}[H]
    \centering
    \begin{minipage}{0.48\textwidth}
        \centering
        \includegraphics[width=\textwidth]{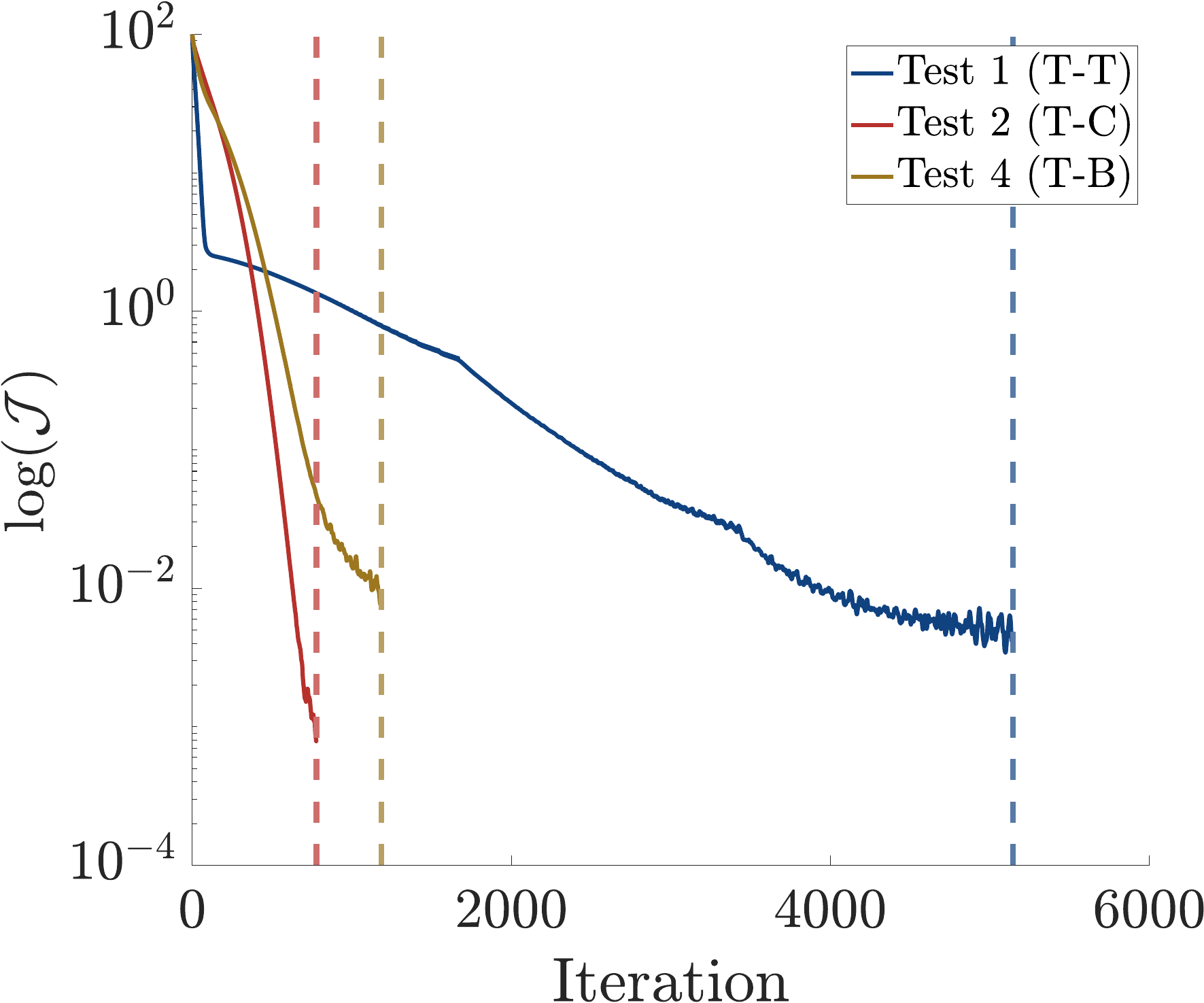}\\
        (a)
    \end{minipage}
    \par
    \vspace{0.3cm}
    \begin{minipage}{0.48\textwidth}
        \centering
        \includegraphics[width=\textwidth]{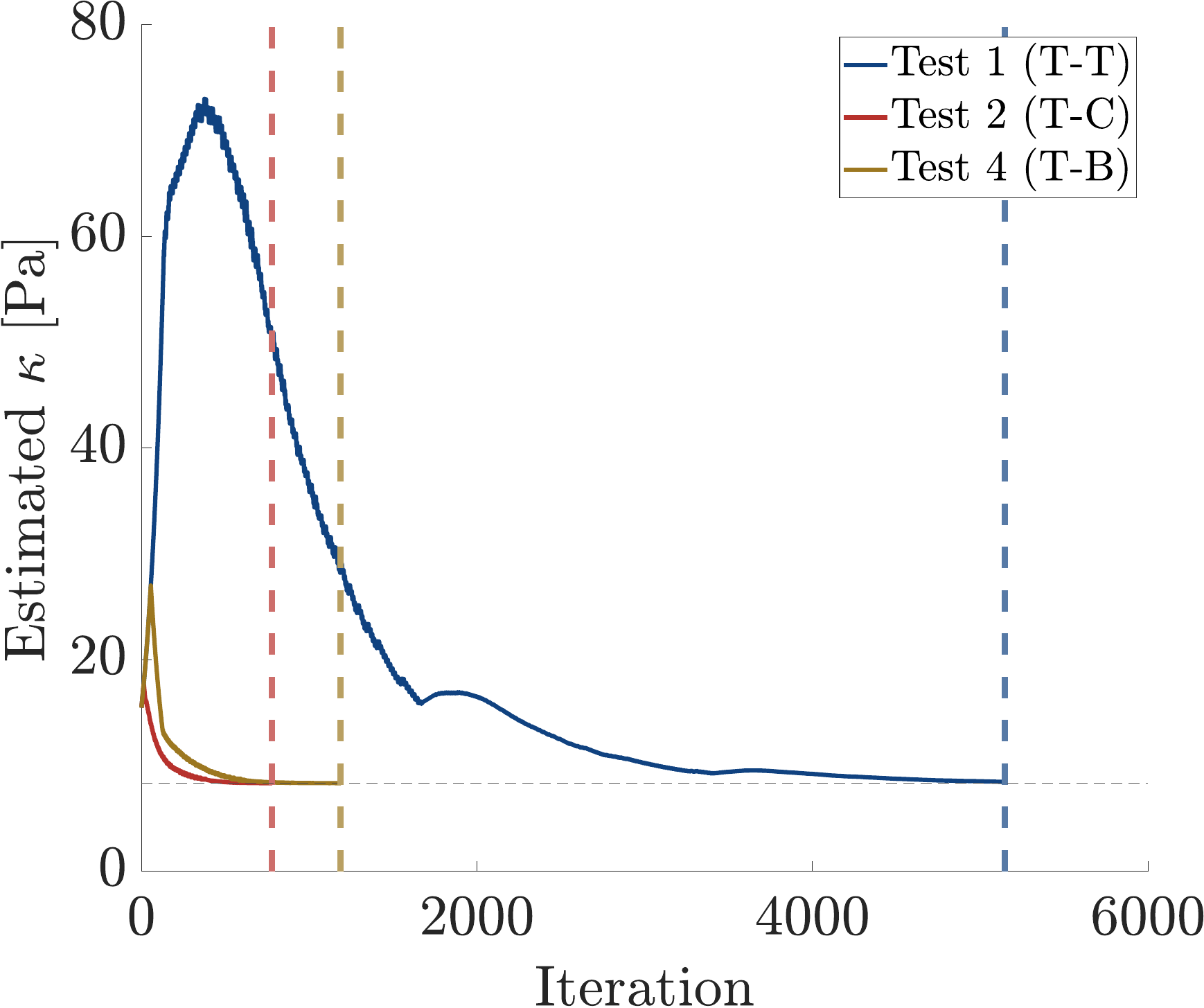}\\
        (b)
    \end{minipage}
    \hfill
    \begin{minipage}{0.48\textwidth}
        \centering
        \includegraphics[width=\textwidth]{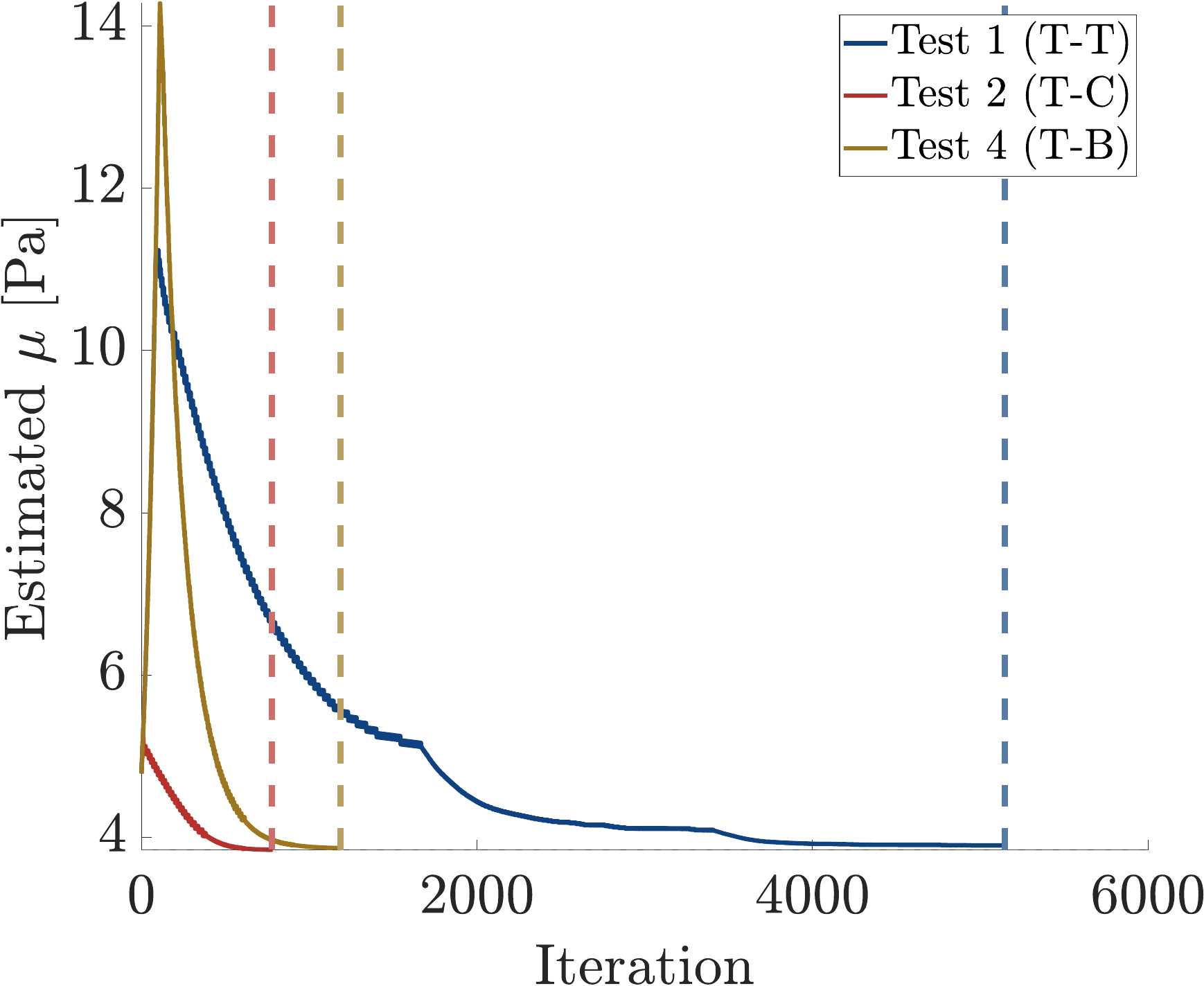}\\
        (c)
    \end{minipage}
    
    \caption{Evolution of the objective function $\mathcal{J}$ and the estimated material parameters for different combinations of loading modes. Results are reported for three test cases with loading applied in the $\mathbf{e}1$ direction: tensile–tensile (T-T), tensile–compressive (T-C), and tensile–bending (T-B). (a) Evolution of the objective function $\mathcal{J}$. (b) Evolution of the bulk modulus $\kappa{\mathrm{mat}}$. (c) Evolution of the shear modulus $\mu_{\mathrm{mat}}$. Horizontal dashed lines in (b) and (c) indicate the corresponding ground-truth parameter values. Vertical dashed lines indicate the iteration at which each test first satisfies the convergence criterion.} 
    \label{fig:J_param_Loading-Mode}
\end{figure}

Fig.~\ref{fig:J_param_Loading-Mode} shows the evolution of the objective function $\mathcal{J}$ and the estimated material parameters $\kappa_{\mathrm{mat}}$ and $\mu_{\mathrm{mat}}$ for the three loading-mode combinations. All cases satisfy the convergence criterion, suggesting that the inverse problem can be solved even when the observations are limited to a single loading direction ($\mathbf{e}_1$ in these tests). However, the convergence behavior differs substantially across the loading combinations. 

The T-T case converges the slowest and exhibits a prolonged plateau, whereas introducing complementary deformation modes accelerates convergence. The T-B case converges faster than T-T, while the T-C case reaches the convergence criterion in the fewest iterations. The material-parameter histories exhibit the same overall trend: the T-C case approaches the ground-truth values most rapidly, followed by T-B, while T-T shows slower convergence and larger transient oscillations. Despite these differences in convergence behavior, all three cases ultimately recover the material parameters with high accuracy. These results indicate that observations probing distinct mechanical responses provide richer information about the coupled unloaded geometry and material properties, thereby improving the convergence of the inverse problem.

In addition to the recovery of the material parameters, we quantify the accuracy of the reconstructed unloaded geometry using two geometry-based error measures. To quantify the improvement of the optimized unloaded configuration coordinate $\tilde{\boldsymbol{X}}^{\ast}$ relative to the initial guess $\tilde{\boldsymbol{X}}^{(0)}$, we define the \emph{relative squared error reduction} (RSER) as 
\begin{equation}
\mathrm{RSER} = 
\frac{\displaystyle\sum_{a=1}^{n_{sh}} \left \|\boldsymbol{\tilde{X}}_{a}^{\ast}-\boldsymbol{\tilde{X}}_{a} \right \|^2}{\displaystyle\sum_{a=1}^{n_{sh}} \left \|\boldsymbol{X}_{a}^{(0)}-\boldsymbol{\tilde{X}}_{a}\right \|^2}
\label{eq:RSER}
\end{equation}
The RSER values for Tests 1, 2, and 4 are $2.03\times10^{-4}$, $1.44\times10^{-6}$, and $3.45\times10^{-4}$, respectively, indicating that the converged unloaded configurations are significantly improved relative to their initial guesses. 

In addition, the \emph{normalized squared reconstruction error} (NSRE), defined as:
\begin{equation}
\mathrm{NSRE} = \frac{\displaystyle\sum_{a=1}^{n_{sh}} \left \|\tilde{\boldsymbol{X}}^{\ast}_{a}-\boldsymbol{\tilde{X}}_{a}\right \|^2}{\displaystyle\sum_{a=1}^{n_{sh}} \left \| \boldsymbol{\tilde{X}}_{a} \right \|^2}
\label{eq:NSRE}
\end{equation}
is $3.10\times10^{-6}$, $2.20\times10^{-8}$, and $5.27\times10^{-6}$ for Tests 1, 2, and 4, respectively. These results demonstrate that the reconstructed unloaded configurations closely match the true unloaded geometry. A quantitative summary of the errors, including the relative error of the recovered material parameters, is provided in Appendix~\ref{tab:summary_all_tests}. 

\paragraph{Effect of Loading Orientation}
\mbox{}\\

In the previous set of tests, the loading direction investigated was $\boldsymbol{e}_1$, with fixed boundary condition orthogonal to that plane, i.e., the plane $\tilde{X}_1=1\,[\mathrm{m}]$ was fixed. Because the geometric features are aligned with the $\boldsymbol{e}_1$ direction, here we perform a test analogous to the previous one, but with loading orientation $\boldsymbol{e}_2$. To consider a similar combination of tension and compression modes, we use observations where $\tilde{X}_2=0\,[\mathrm{m}]$ is the fixed boundary plane. The three tests considered here are Tests 7, 8, and 9 in Table~\ref{tab:benchmark_tests}.

\begin{figure}[H]
    \centering
    \includegraphics[width=0.48\textwidth]{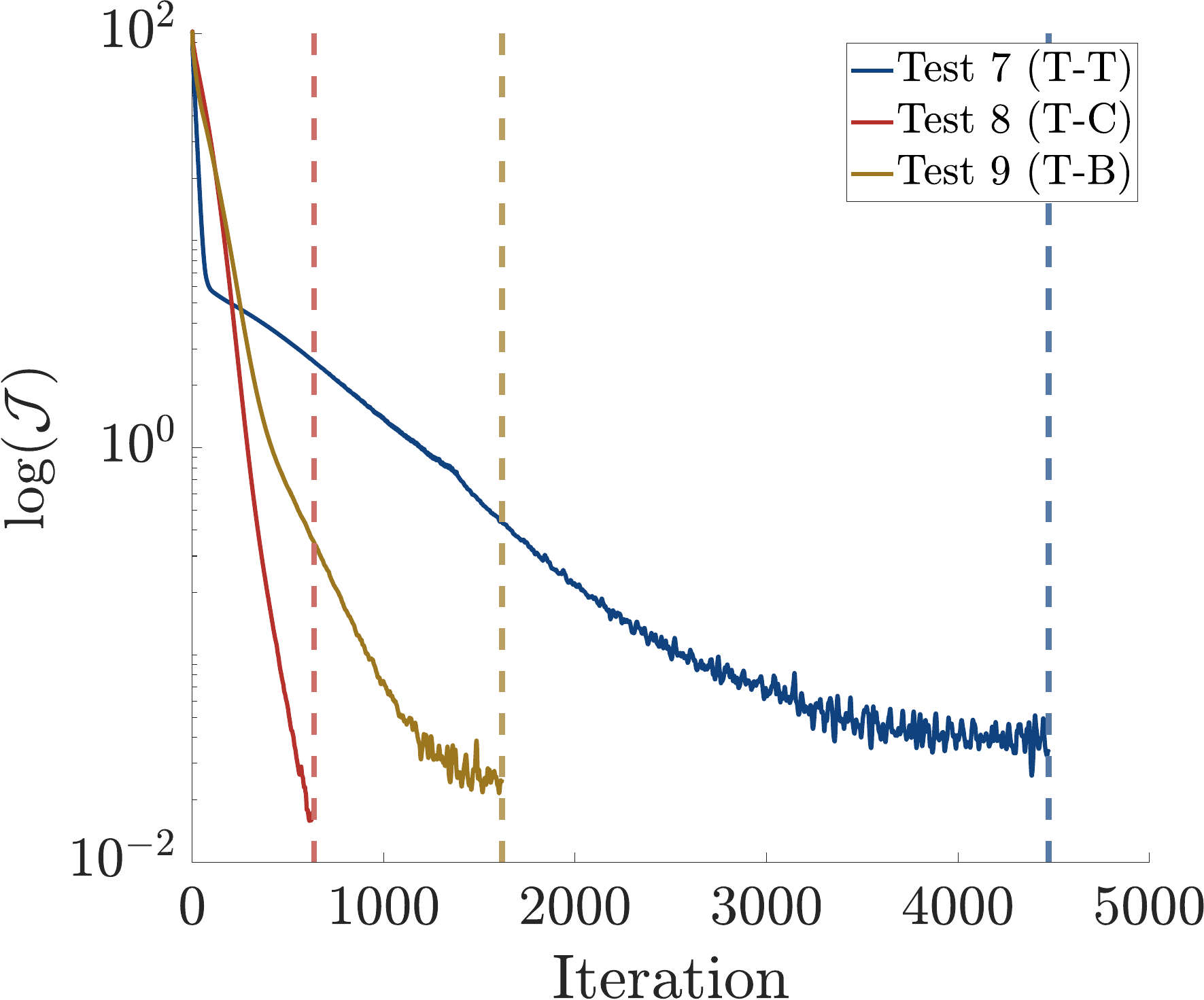}
    \caption{Evolution of the objective function $\mathcal{J}$ for different combinations of loading modes under a vertically applied load (in the direction $\mathbf{e}_2$). Three test cases are considered using a fixed plane at $\tilde{X}_2=0\,[\mathrm{m}]$: 
    tensile-tensile (T-T), tensile-compressive (T-C), and tensile-bending (T-B). Vertical dashed lines indicate the iteration at which each test first satisfies the convergence criterion.}
    \label{fig:J_benchmark_combined}
\end{figure}

Fig.~\ref{fig:J_benchmark_combined} shows the evolution of the objective function for these tests. The same qualitative trends observed in the previous subsection persist: mixed-mode observations converge more rapidly and more smoothly than single-mode cases. This suggests that the benefit of loading mode diversity is not specific to a particular loading orientation, but rather reflects a fundamental identifiability property of the inverse problem.

Quantitatively, the RSER values for Tests~7, 8, and~9 are $1.32\times10^{-4}$, $2.74\times10^{-4}$, and $1.10\times10^{-4}$, respectively, indicating substantial improvement of the reconstructed unloaded configurations relative to their initial guesses. 
In addition, the corresponding NSRE values are $3.99\times10^{-7}$, $8.29\times10^{-7}$, and $3.32\times10^{-7}$ for Tests~7, 8, and~9, respectively, confirming that the converged configurations closely approximate the true unloaded geometries. 
Although the evolutions of $\kappa$ and $\mu$ are not shown for these tests, both material parameters were correctly identified, with relative errors within the prescribed convergence tolerance; the corresponding material-parameter errors are reported in Appendix~\ref{tab:summary_all_tests}.

\paragraph{Effect of the Number of Observed Configurations}
\mbox{}\\

Here, we investigate the influence of the number of observed configurations on convergence behavior. Three tests are compared, all using different force scales of tensile loading in the same direction but differing in how many observed configurations are provided: two observations, three observations, and four observations (Tests 1, 5, and 6 in Table~\ref{tab:benchmark_tests}).

\begin{figure}[H]
    \centering
    \includegraphics[width=0.48\textwidth]{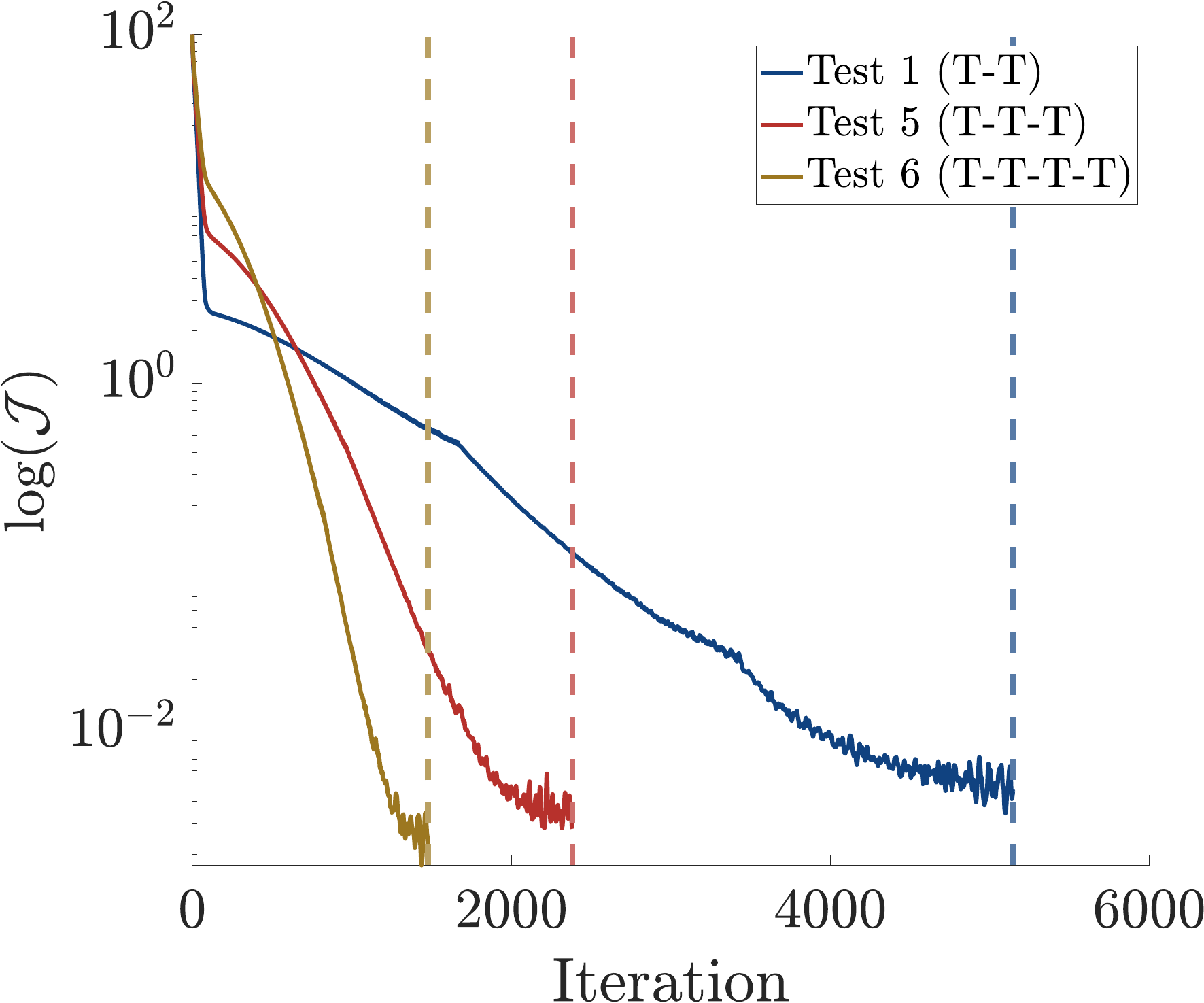}

    \caption{Evolution of the objective function $\mathcal{J}$ for different numbers of observed configurations. Three test cases are considered using tensile loading in the same direction with different force scales, while varying the number of observed configurations: two, three, and four observations (Tests~1, 5, and~6). Vertical dashed lines indicate the iteration at which each test first satisfies the convergence criterion.}
    \label{fig:J_num_obs}
\end{figure}

Fig.~\ref{fig:J_num_obs} shows the evolution of the objective $\mathcal{J}$ for all three cases. Increasing the number of observed configurations systematically accelerates convergence and reduces the duration of the transient plateau. The two-observation case converges the slowest, while the four-observation case achieves the fastest and most stable reduction in the objective $\mathcal{J}$.

The RSER values for Tests~1, 5, and~6 are $2.03\times10^{-4}$, $1.35\times10^{-5}$, and $4.83\times10^{-6}$, respectively, confirming a substantial reduction in error relative to the initial guesses. In addition, the corresponding NSRE values are $3.10\times10^{-6}$, $2.07\times10^{-7}$, and $7.38\times10^{-8}$ for Tests~1, 5, and~6, respectively, indicating that the reconstructed unloaded configurations closely match the true unloaded geometry. The recovered material parameters show the same trend, converging close to their true values as summarized in Appendix~\ref{tab:summary_all_tests}.

\paragraph{Sensitivity to Initial Guess of the Material Properties}
\mbox{}\\

Finally, the sensitivity of the convergence speed to the initial guess of the material properties is examined using two representative tests (Tests 1 and 3 in Table~\ref{tab:benchmark_tests}). Test 1 starts from an overestimated stiffness, while Test 3 starts from an underestimated stiffness; both use the same pair of observed configurations.

\begin{figure}[H]
    \centering
    \begin{minipage}{0.48\textwidth}
        \centering
        \includegraphics[width=\textwidth]{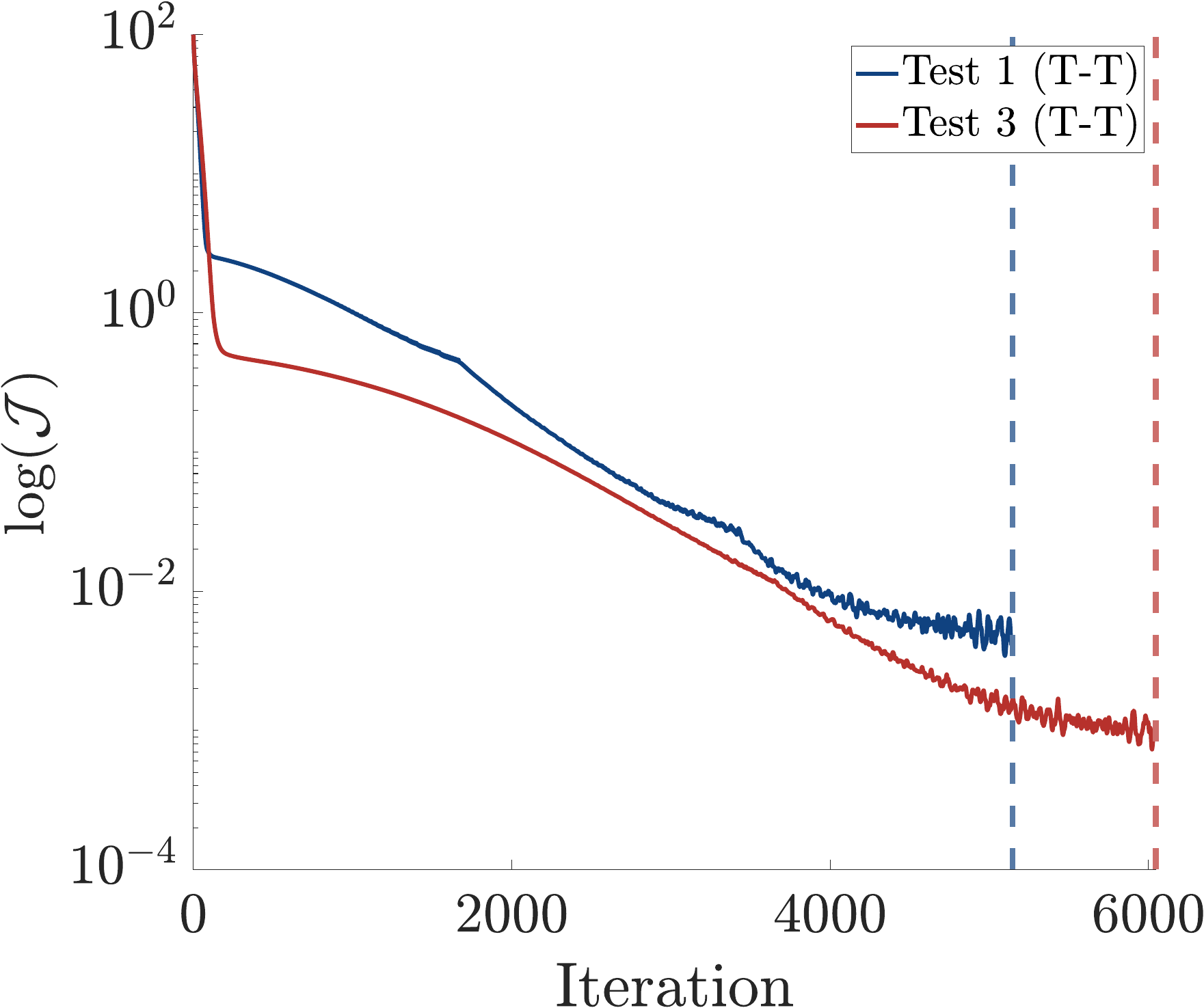}\\
        (a)
    \end{minipage}
    \par
    \vspace{0.3cm}
    \begin{minipage}{0.48\textwidth}
        \centering
        \includegraphics[width=\textwidth]{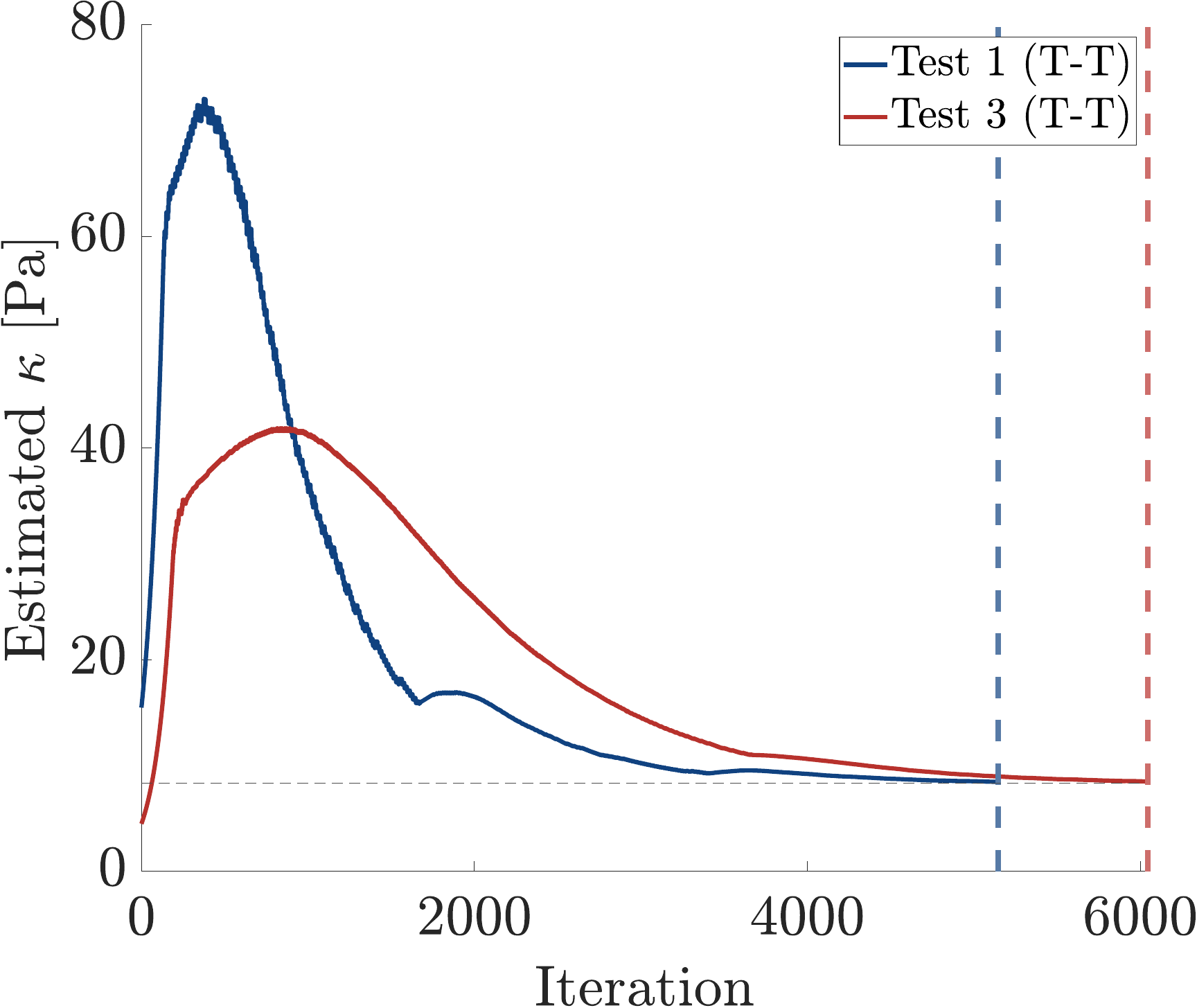}\\
        (b)
    \end{minipage}
    \hfill
    \begin{minipage}{0.48\textwidth}
        \centering
        \includegraphics[width=\textwidth]{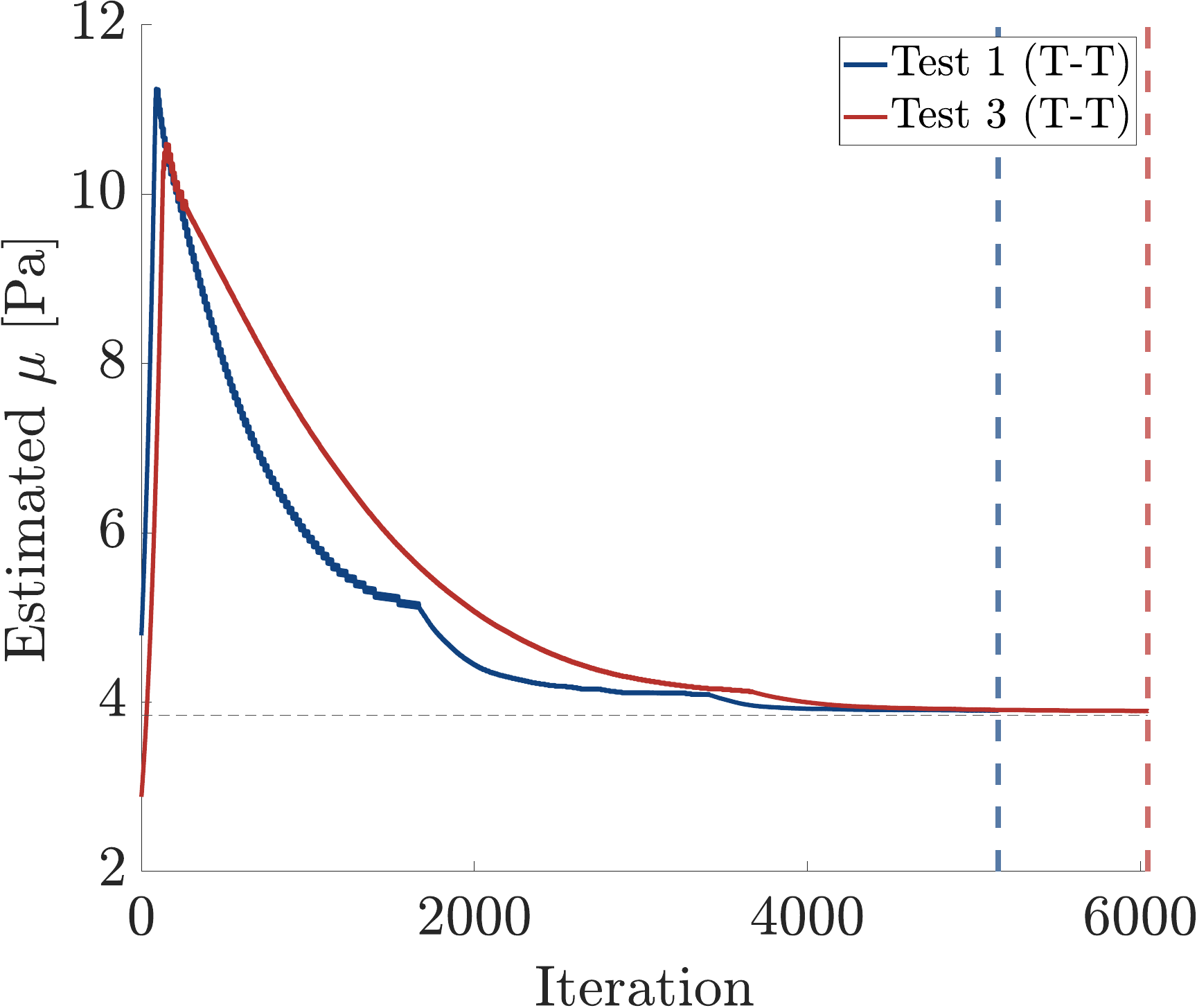}\\
        (c)
    \end{minipage}
    
    \caption{Evolution of the objective function $\mathcal{J}$ and the estimated material parameters for different initial guesses of the material parameters. Test 1 starts from an overestimated stiffness and Test 3 starts from an underestimated stiffness. Both tests use the same pair of observed configurations. (a) Evolution of the objective function $\mathcal{J}$. (b) Evolution of the bulk modulus $\kappa_{\mathrm{mat}}$. (c) Evolution of the shear modulus $\mu_{\mathrm{mat}}$. Horizontal dashed lines in (b) and (c) indicate the corresponding ground-truth parameter values. Vertical dashed lines indicate the iteration at which each test first satisfies the convergence criterion.}
    \label{fig:J_init_guess_param}
\end{figure}

Figure~\ref{fig:J_init_guess_param} shows the evolution of the objective function $\mathcal{J}$ and the estimated material parameters $\kappa_{\mathrm{mat}}$ and $\mu_{\mathrm{mat}}$ for the two different initial material guesses. Although the early-stage objective histories differ, both tests converge to similar residual levels within a comparable number of iterations. Likewise, despite starting from significantly different initial values, both material parameters converge to nearly identical final estimates. The case with initially overestimated stiffness exhibits a larger transient overshoot, whereas the case with initially underestimated stiffness converges more smoothly but slightly more slowly. These results indicate that the proposed inverse FEM framework is robust to moderate variations in the initial material guess, provided that sufficiently informative observed configurations are available.

Consistent with this observation, the RSER values for Tests~1 and~3 are $2.03\times10^{-4}$ and $1.96\times10^{-4}$, respectively, confirming a substantial reduction in error relative to the initial guesses. In addition, the corresponding NSRE values are $3.10\times10^{-6}$ and $3.00\times10^{-6}$ for Tests~1 and~3, respectively, indicating that the reconstructed unloaded configurations closely match the true unloaded geometry as summarized in Appendix~\ref{tab:summary_all_tests}.

The synthetic benchmark problem for the homogeneous material case provides a controlled setting to evaluate the robustness of the proposed method for recovering both the unloaded geometry and material parameters under large deformation and complex geometry. This distinguishes the present approach from direct inverse formulations~\cite{gee_computational_2010, rajagopal_determining_2007, peirlinck_modular_2018} and fixed-point unloading methods~\cite{bols_computational_2013, carter_biomechanical_nodate, mira_biomechanical_2018, rausch_augmented_2017, sellier_iterative_2011}, which primarily focus on recovering the unloaded geometry under prescribed material properties. Recent work, such as ~\cite{parikh_biomechanical_2023,hajhashemkhani_identification_2021}, has started to tackle the identification of stress-free configuration as well as material parameters, but with very simplified parameterizations of the geometry (hundreds of nodes of simple geometries such as a hemisphere), and for homogeneous materials. In contrast, the present formulation directly optimizes all the free nodal coordinates (on the order of thousands) as well as material parameters within a single differentiable inverse problem.

The benchmark further clarifies the role of data informativeness in the present differentiable formulation. Consistent with the observations of Hajhashemkhani et al.~\cite{hajhashemkhani_identification_2021}, mixed loading modes and additional observed configurations improve the constraint of the coupled geometry-material inverse problem. In our setting, these results show that diverse deformation states accelerate convergence and reduce the ambiguity between geometry and stiffness when optimizing volumetric nodal coordinates and material parameters simultaneously.

\subsection{Heterogeneous Benchmark Problem: Cube with  Inclusions}
\label{sec:results_benchmark_Hetero}

The heterogeneous benchmark extends the homogeneous tests by considering a material field with known region labels but unknown region-wise properties. This setting is motivated by image-based biomechanical models in which heterogeneous regions, such as inclusions, tumors, or anatomical subdomains, may be obtained from segmentation or mesh labeling \cite{guo_image_2005,schnabel_validation_2003,sivaramakrishna_3d_2005,amjad_computationally_2020}. Unlike heterogeneous material identification studies that assume the reference configuration is prescribed \cite{klinge_inverse_2015,mei_mechanics_2017,mei_comparative_2018,liu_resolving_2024}, the present benchmark treats the unloaded configuration as an unknown while also identifying the material parameters of each region.

\paragraph{Geometry, Loading, Material Parameters and Boundary Conditions}
\label{par:benchmark_geom_bc_Hetero}
\mbox{}\\

In this section, we showcase our algorithm using a synthetic unit cube $\Omega_{\boldsymbol{\tilde{X}}}$ representing a particle-fiber reinforced composite material. The cube consists of a soft matrix region $\Omega_{\mathrm{mat}}$, two spherical inclusions $\Omega_{\mathrm{in}_1}$ representing particle reinforcements, and three cylindrical inclusions $\Omega_{\mathrm{in}_2}$ representing fiber reinforcements. The two spherical inclusions are assigned one set of material properties, and the three cylindrical inclusions are assigned another set of properties. The shear moduli are prescribed as $\mu_{\mathrm{mat}} = 3.846 ~[\mathrm{Pa}]$, $\mu_{\mathrm{in}_1} = 5.556 ~[\mathrm{Pa}]$, and $\mu_{\mathrm{in}_2} = 7.407 ~[\mathrm{Pa}]$ for the matrix, particle inclusions, and fiber inclusions, respectively. The corresponding bulk moduli are $\kappa_{\mathrm{mat}} = 8.333 ~[\mathrm{Pa}]$, $\kappa_{\mathrm{in}_1} = 16.667 ~[\mathrm{Pa}]$, and $\kappa_{\mathrm{in}_2} = 22.222 ~[\mathrm{Pa}]$. Equivalently, these parameters correspond to Young's moduli $E_{\mathrm{mat}} \approx 10 ~[\mathrm{Pa}]$, $E_{\mathrm{in}_1} \approx 15 ~[\mathrm{Pa}]$, and $E_{\mathrm{in}_2} \approx 20 ~[\mathrm{Pa}]$, with Poisson's ratios $\nu_{\mathrm{mat}} \approx 0.30$, $\nu_{\mathrm{in}_1} \approx 0.35$, and $\nu_{\mathrm{in}_2} \approx 0.35$.

\begin{figure}[H]
    \centering
    \includegraphics[width=1\textwidth]{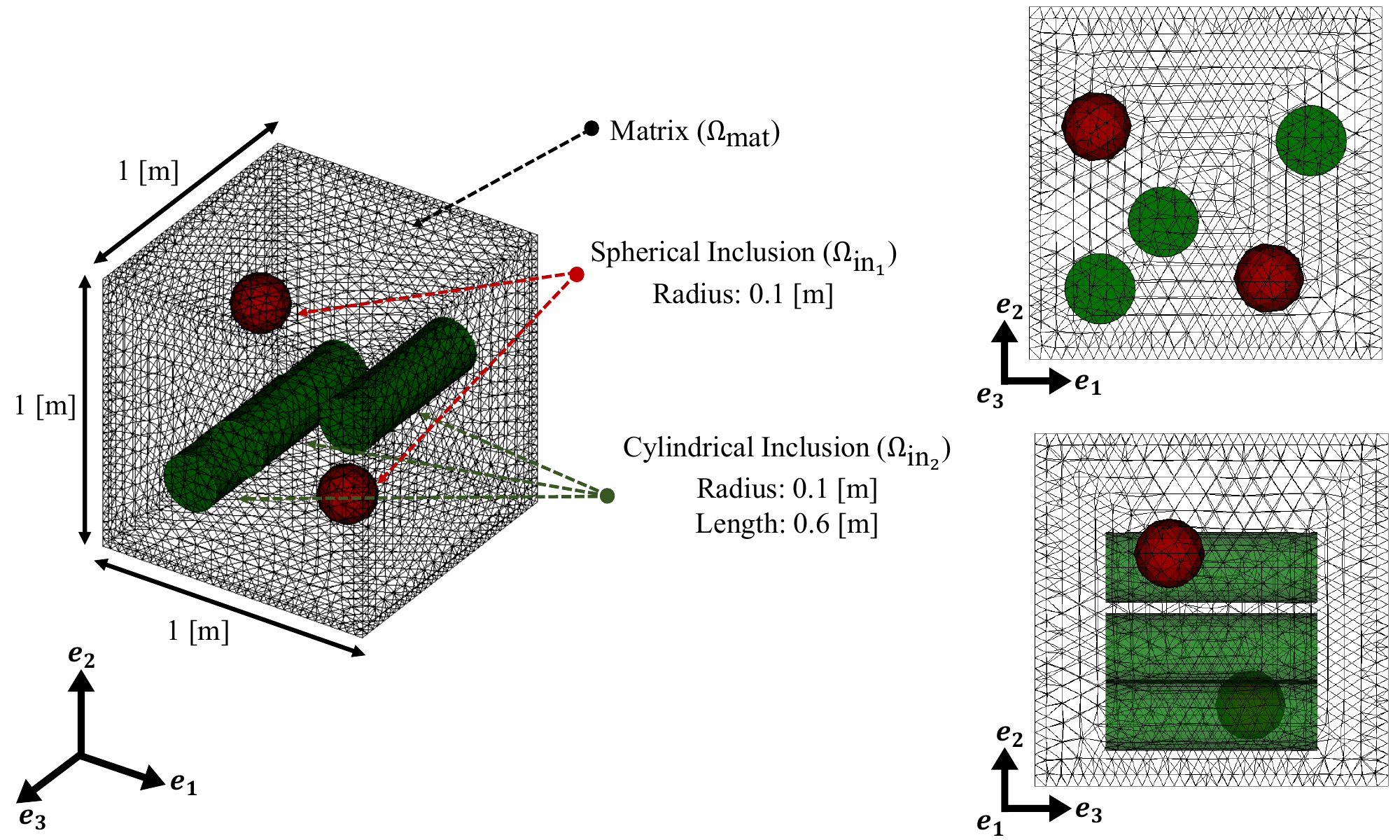}
    \caption{Discretized geometry of the heterogeneous cube benchmark. The unit cube contains a matrix region $\Omega_{\mathrm{mat}}$, two spherical inclusions $\Omega_{\mathrm{in}_1}$ with radius $0.1\,[\mathrm{m}]$, and three cylindrical inclusions $\Omega_{\mathrm{in}_2}$ with radius $0.1\,[\mathrm{m}]$ and length $0.6\,[\mathrm{m}]$ oriented along the $\boldsymbol{e}_3$-direction. The domain is discretized using 29,331 linear tetrahedral elements.} 

    \label{fig:Hetero_cube_geom}
\end{figure}

Fig.~\ref{fig:Hetero_cube_geom} shows the discretized heterogeneous cube geometry of $\Omega_{\tilde{\boldsymbol{X}}}$, which is meshed with 29,331 linear tetrahedral elements. The unit cube consists of a matrix region, $\Omega_{\mathrm{mat}}$, two spherical inclusions, $\Omega_{\mathrm{in}_1}$, each with radius $0.1\,[\mathrm{m}]$, and three cylindrical inclusions, $\Omega_{\mathrm{in}_2}$, each with radius $0.1\,[\mathrm{m}]$ and length $0.6\,[\mathrm{m}]$, oriented along the $\boldsymbol{e}_3$-direction.

\begin{figure}[H]
\centering
\begin{minipage}{0.4\textwidth}
    \centering
    \includegraphics[width=\textwidth]{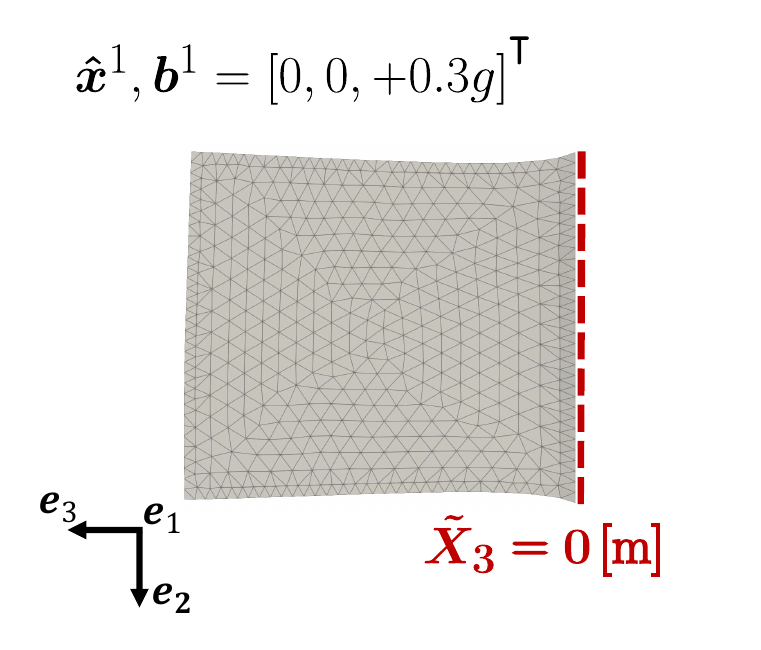}
    (a)\\
\end{minipage}
\hfill
\begin{minipage}{0.4\textwidth}
    \centering
    \includegraphics[width=\textwidth]{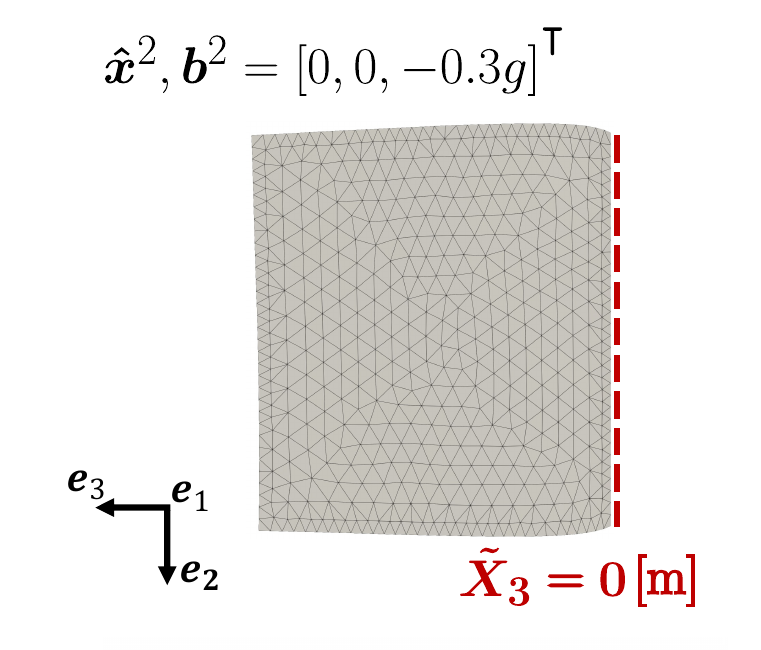}
    (b)\\
\end{minipage}
\caption{
Synthetic observed configurations for the heterogeneous composite cube generated under two body-force vectors, (a) $\boldsymbol{b}^1$ and (b) $\boldsymbol{b}^2$, corresponding to scaled or rotated gravity loadings. The dashed red line denotes the plane where zero displacement is prescribed. Configuration (a) is used as the reference configuration for the heterogeneous benchmark test.
}
\label{fig:observed_hetero}
\end{figure}

Following the loading setup and forward-solving procedure introduced in Sec.~\ref{sec:results_benchmark}, the synthetic composite cube is loaded by a gravitational body force. In this benchmark, we use a normalized density of $\rho = 1~[\mathrm{kg/m^3}]$ and prescribe the gravitational acceleration magnitude as $g = 9.81~[\mathrm{m/s^2}]$. The observed configurations shown in Fig.~\ref{fig:observed_hetero} are obtained from two loading cases defined by the scaled or rotated body-force vectors $\boldsymbol{b}^1$ and $\boldsymbol{b}^2$, resulting in $\hat{\boldsymbol{x}}^1$ and $\hat{\boldsymbol{x}}^2$, respectively. For both loading cases, the Dirichlet boundary is assigned to the plane $\Gamma_u^{\boldsymbol{\tilde{X}}} = \{ \tilde{X}_3 = 0\,[\mathrm{m}] \}$, where all displacement components are fixed to zero.

\paragraph{Effect of Material Heterogeneity on the Optimization}
\label{sec:Hetero_benchmark}
\mbox{}\\

We conduct a heterogeneous inverse test using two observed configurations. The first observed configuration, Fig.~\ref{fig:observed_hetero}(a), is used as the geometric initial guess. The initial material parameters are prescribed uniformly across all material regions based on the matrix properties, with $E_{\mathrm{mat}}^{(0)} = E_{\mathrm{in}_1}^{(0)} = E_{\mathrm{in}_2}^{(0)} = 0.8E_{\mathrm{mat}}$ and $\nu_{\mathrm{mat}}^{(0)} = \nu_{\mathrm{in}_1}^{(0)} = \nu_{\mathrm{in}_2}^{(0)} = 0.8\nu_{\mathrm{mat}}$.

\begin{figure}[H]
\centering
\begin{minipage}{0.48\textwidth}
    \centering
    \includegraphics[width=\textwidth]{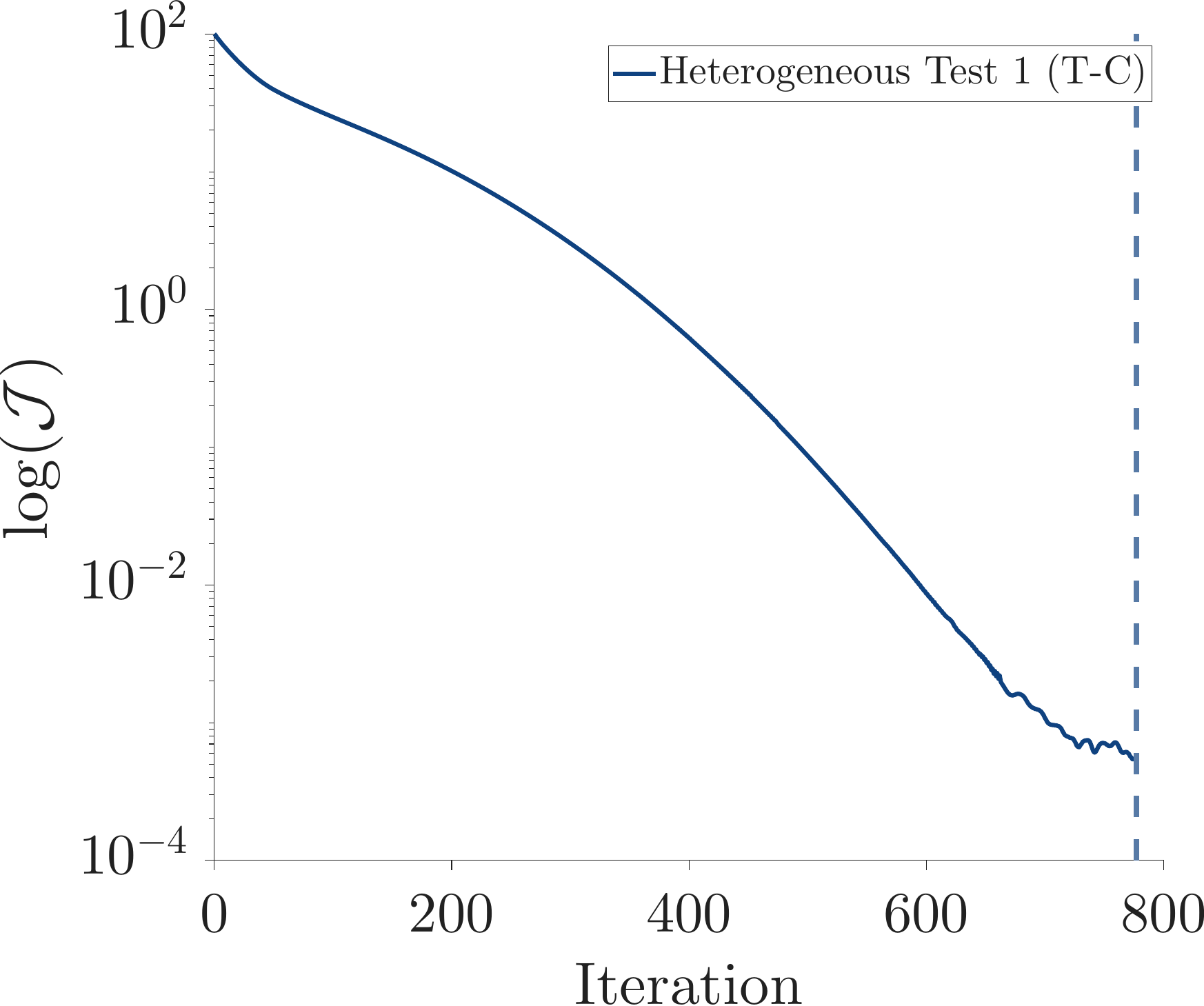}\\
    (a)
\end{minipage}
\par
\vspace{0.3cm}
\begin{minipage}{0.48\textwidth}
    \centering
    \includegraphics[width=\textwidth]{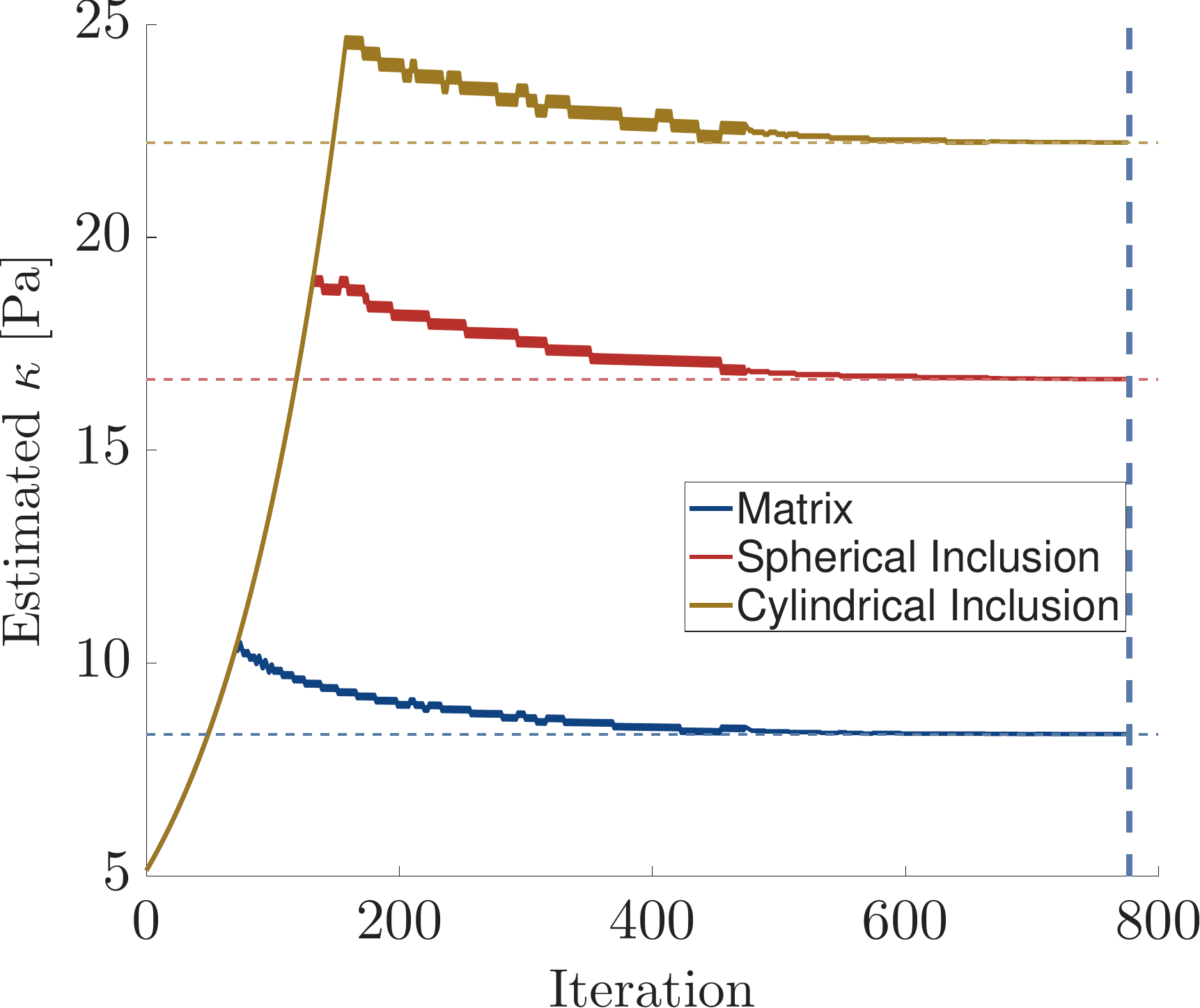}\\
    (b)
\end{minipage}
\hfill
\begin{minipage}{0.48\textwidth}
    \centering
    \includegraphics[width=\textwidth]{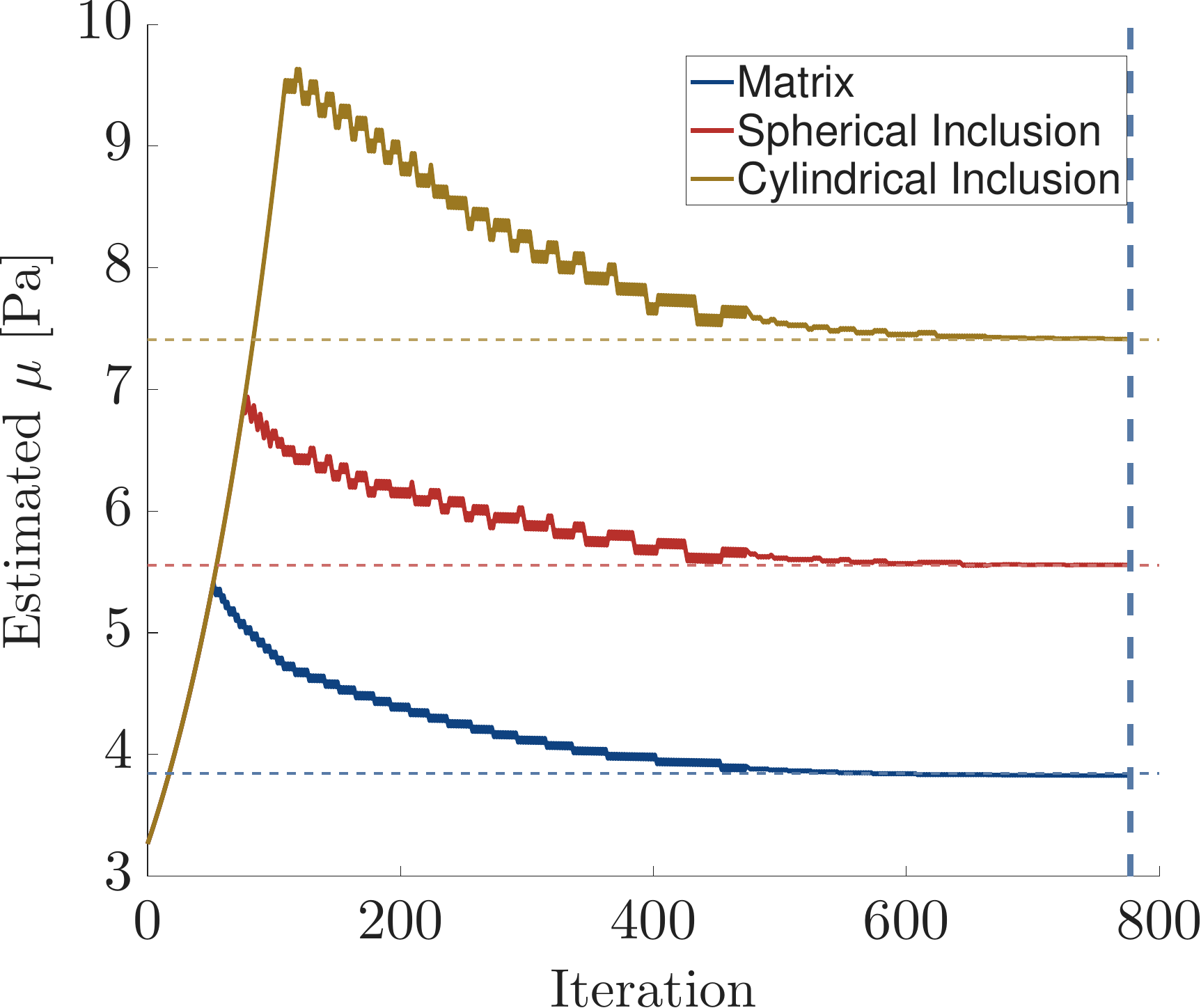}\\
    (c)
\end{minipage}

\caption{Evolution of the objective function $\mathcal{J}$ and the estimated region-wise material parameters for the heterogeneous cube test using the tensile--compressive observed configuration pair. (a) Evolution of the objective function $\mathcal{J}$. (b) Evolution of the bulk modulus $\kappa$ for the matrix $\kappa_{\mathrm{mat}}$, spherical inclusions $\kappa_{\mathrm{in}_1}$, and cylindrical inclusions $\kappa_{\mathrm{in}_2}$. (c) Evolution of the shear modulus $\mu$ for the matrix $\mu_{\mathrm{mat}}$, spherical inclusions $\mu_{\mathrm{in}_1}$, and cylindrical inclusions $\mu_{\mathrm{in}_2}$. Horizontal dashed lines in (b) and (c) indicate the corresponding ground-truth parameter values. Vertical dashed lines indicate the iteration at which the convergence criterion is first satisfied.}
\label{fig:J_Hetero_cost_param}
\end{figure}

Fig.~\ref{fig:J_Hetero_cost_param} shows the evolution of the objective function $\mathcal{J}$ and the estimated region-wise material parameters for the heterogeneous inverse test. The objective decreases steadily and reaches a low residual level, indicating stable convergence of the optimization. The estimated bulk and shear moduli of the matrix, spherical inclusions, and cylindrical inclusions converge toward their corresponding ground-truth values. The similar convergence behavior across the three material regions indicates that the observed configurations provide sufficient information to identify the distinct region-wise material properties simultaneously with the unloaded geometry.

When convergence is achieved, the value of RSER is $1.16\times10^{-6}$, which demonstrates a substantial reduction in error relative to the initial guess. The NSRE value at convergence is $4.24\times10^{-8}$, which confirms that the reconstructed unloaded configuration closely approximates the true unloaded geometry. The corresponding region-wise material errors, objective value, and reconstruction metrics are reported in Appendix~\ref{tab:summary_hetero_cube}.

The convergence of the matrix, spherical inclusion, and cylindrical inclusion properties to their true values shows that the proposed formulation can separate regional stiffness effects from geometric effects in this controlled heterogeneous setting. This is important because an error in the unloaded geometry can partially compensate for an error in the material properties when only loaded configurations are observed. The results therefore demonstrate that, when the heterogeneous regions are known, the same differentiable inverse framework can recover both the stress-free geometry and multiple region-dependent material parameters without introducing a separate formulation for the heterogeneous case.

\subsection{Inverse Problem in Homogenized Breast Mechanics}
\label{sec:results_breast}

\paragraph{Geometry and Material Parameters}
\label{Geometry and Material Parameters}
\mbox{}\\

\begin{figure}[H]
\centering
\begin{minipage}{0.25\textwidth}
    \centering
    \includegraphics[width=\textwidth]{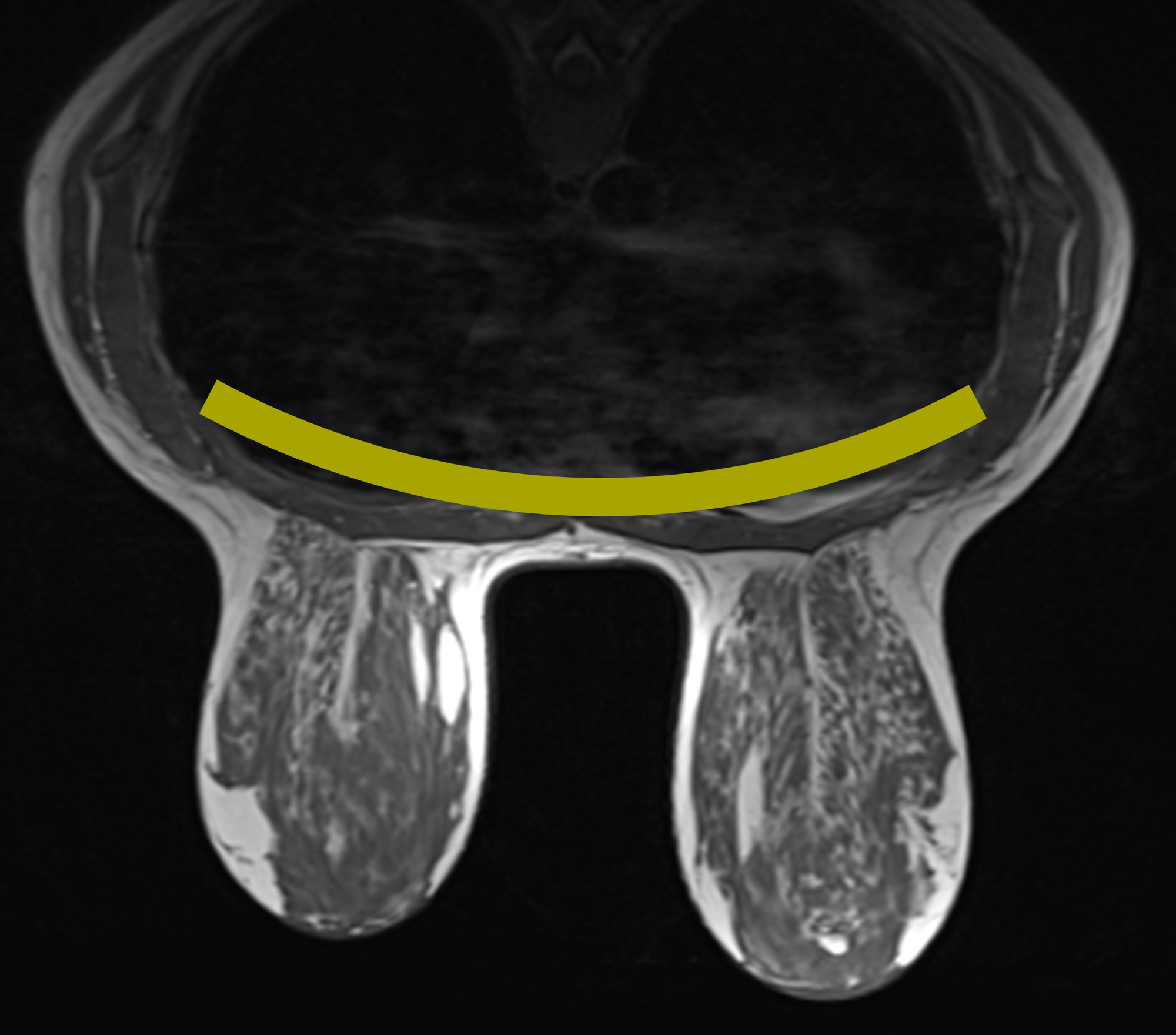}\\
    (a)
\end{minipage}
\hfill
\begin{minipage}{0.35\textwidth}
    \centering
    \includegraphics[width=\textwidth]{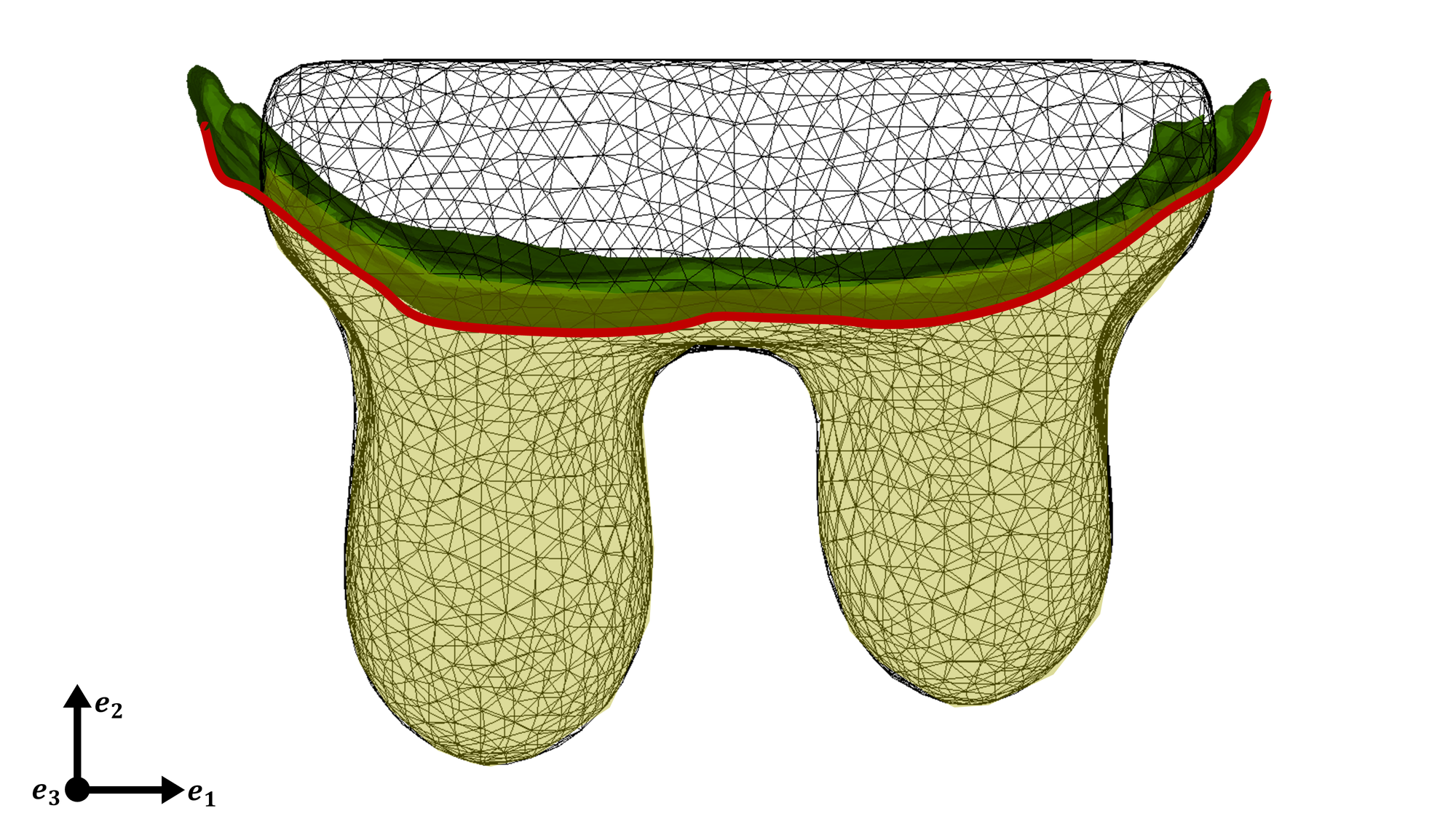}\\
    (b)
\end{minipage}
\hfill
\begin{minipage}{0.35\textwidth}
    \centering
    \includegraphics[width=\textwidth]{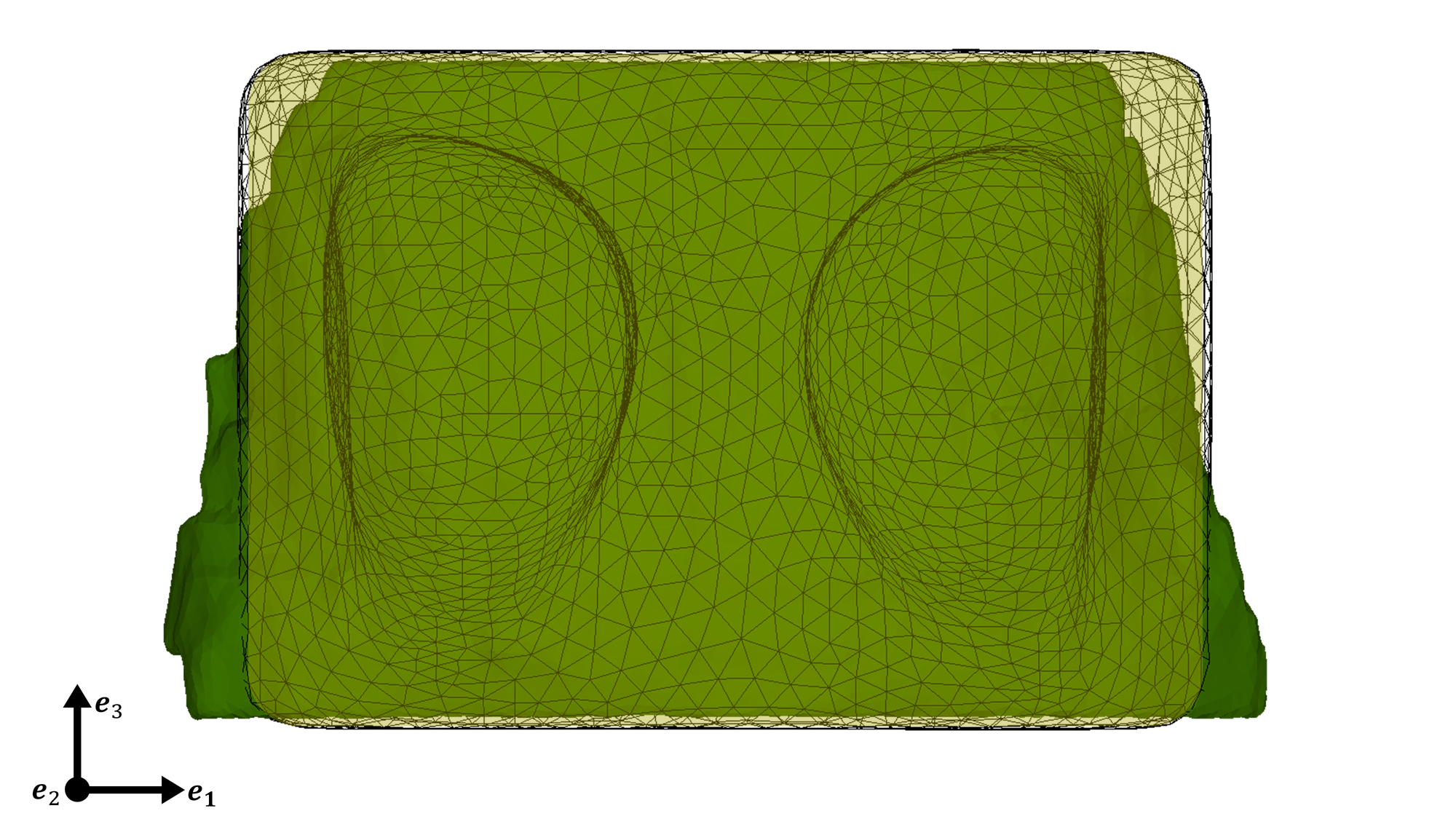}\\
    (c)
\end{minipage}

\caption{Pipeline for identification of the breast geometry. 
(a) Axial MRI slice highlighting the sternum and pectoralis major muscle (yellow curve), which serve as anatomical anchors for registration of the superior-medial region of the breast.
(b) Top view of the patient-specific breast surface mesh reconstructed from MRI (black), together with the registered sternum and pectoralis major muscle (green surface). The yellow region constitutes the computational domain with free displacement degrees of freedom. All nodes above the red surface are treated as pertaining to a zero-displacement \textit{boundary}. 
(c) Bottom view of the patient-specific breast surface mesh reconstructed from MRI (black) together with the registered sternum and pectoralis major muscle (green surface).}
\label{fig:Breast_BC_combined}
\end{figure}

The patient-specific breast geometry and the corresponding mechanical properties used here were provided by the authors of \cite{harbin_computational_2025}. Their geometrical model is constructed from a subject-specific segmentation pipeline applied to the breast MRI database in \cite{saha_machine_2018}. The geometry obtained from \cite{harbin_computational_2025} was meshed with 21,147 linear tetrahedral elements. However, the original geometry includes not only the breast, but also the surrounding tissues that are much stiffer than the breast and can be considered rigid in the simulation. To avoid generating a new geometry, we included these tissues in the simulation, but set their displacement field to zero. To determine which nodes in the mesh should receive zero-displacement constraints, the patient’s sternum and pectoralis major muscle were segmented from the MRI data using 3D Slicer~\cite{Fedorov2012Slicer}; see Fig.~\ref{fig:Breast_BC_combined}(a). The segmented structures were rigidly registered to the breast mesh [see dark green surface in Fig.~\ref{fig:Breast_BC_combined}(b) and (c)]. The region of the sternum and pectoralis major muscle surface with outer normal in the $-\bse_2$ direction (see red surface in Fig.~\ref{fig:Breast_BC_combined}(b)) was used to separate the breast tissue from the tissue considered rigid in the simulation. All nodes above the red surface were treated as pertaining to a zero-displacement \textit{boundary}. In JAX-FEM, this is possible because boundary conditions can be applied to subsets of the domain regardless of whether or not they are on the boundary. In practice, all the solver needs is for the user to specify whether a node has a fixed or free displacement degree of freedom.

Reference \cite{harbin_computational_2025} also estimated neo-Hookean material parameters using patient-specific breast densities and the rule of mixtures. For the case considered here, the estimated neo-Hookean material parameters were $\mu_{\mathrm{mat}} = 953.96~[\mathrm{Pa}]$ and $\kappa_{\mathrm{mat}} = 47379.90~[\mathrm{Pa}]$, or equivalently, Young's modulus $E_{\mathrm{mat}} \approx 2842.794~[\mathrm{Pa}]$ and Poisson's ratio $\nu_{\mathrm{mat}} \approx 0.49$. The density is $\rho = 942.82~[\mathrm{kg/m^3}]$. For numerical stability of the forward FEM solve, we adopt a slightly reduced Poisson's ratio ($\nu_{\mathrm{mat}} = 0.48$) throughout the simulations. 

\paragraph{Generation of Synthetic Observed Configurations}
\mbox{}\\

The MRI dataset provides breast geometry only in the prone position, but at least two distinct observed configurations are required to simultaneously identify the unloaded configuration and the material properties. Because acquiring multiple breast configurations from the same patient (e.g., prone and supine) is not part of the standard clinical workflow, we generate the observed geometries synthetically in this work.

\begin{figure}[H]
\centering
\begin{minipage}{0.48\textwidth}
    \centering
    \includegraphics[width=\textwidth]{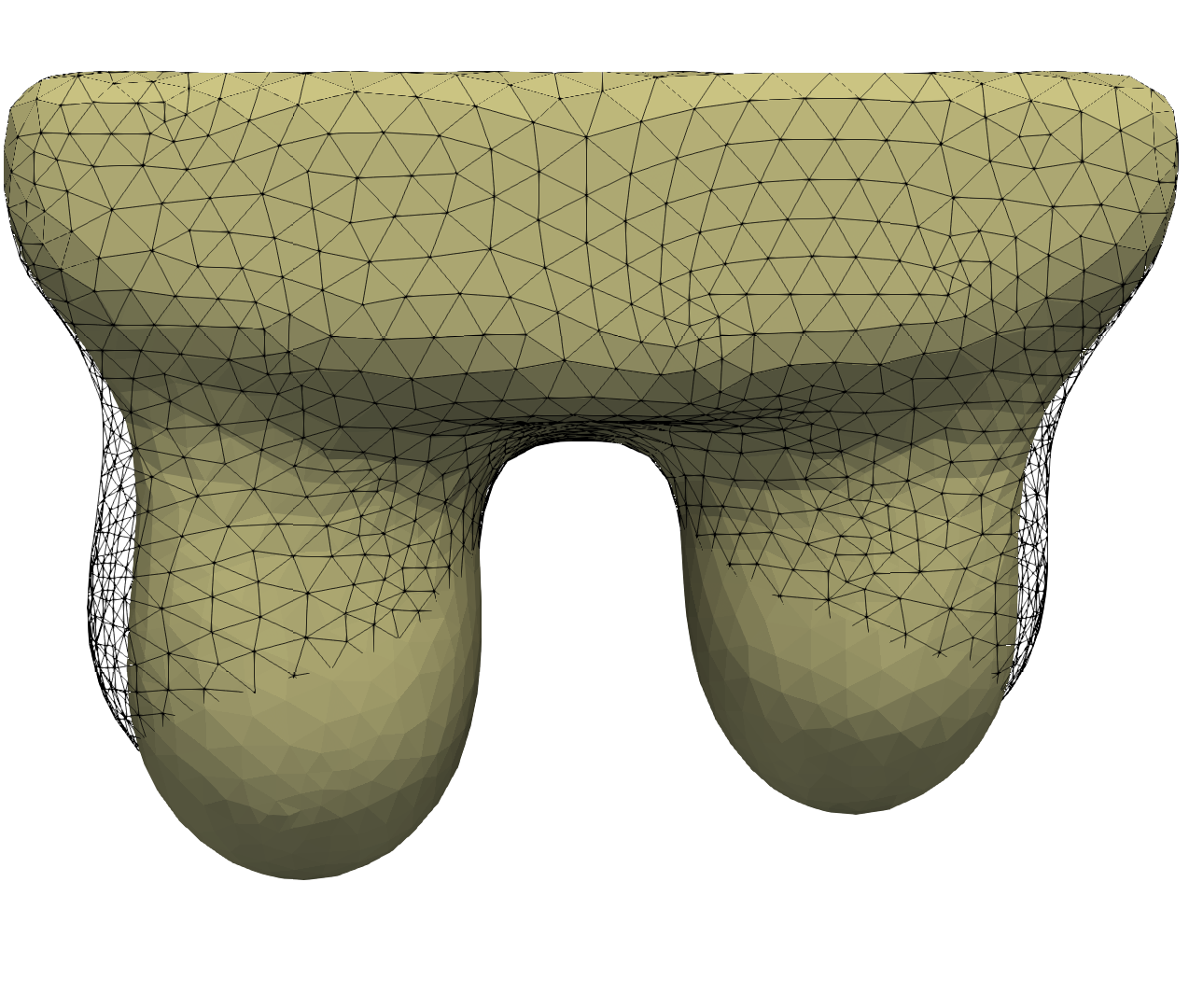}\\
    (a)
\end{minipage}
\hfill
\begin{minipage}{0.45\textwidth}
    \centering
    \includegraphics[width=\textwidth]{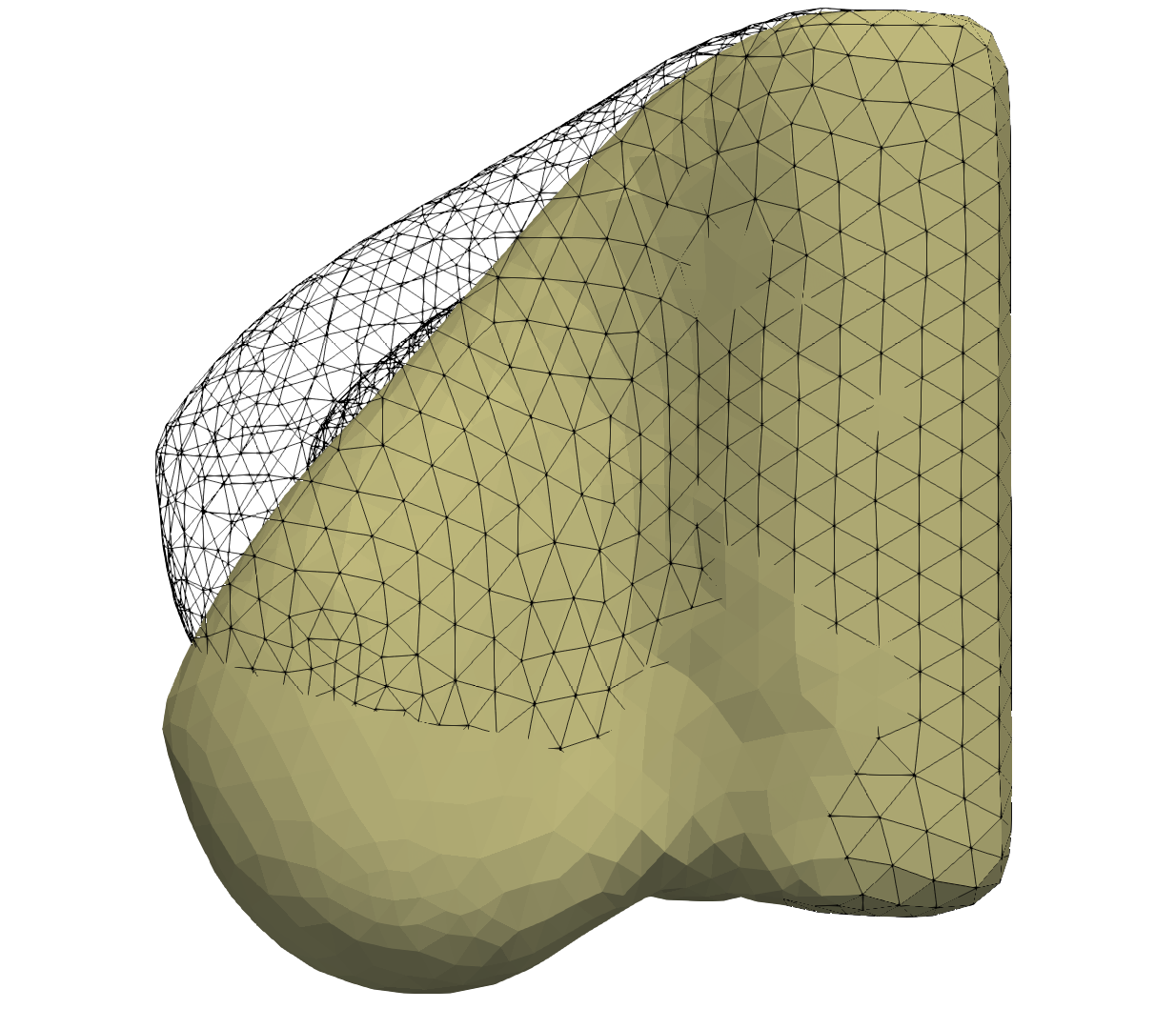}\\
    (b)
\end{minipage}
\caption{
Observed breast configurations under gravitational loading (yellow surface) overlaid on the unloaded configuration (black mesh). 
(a) Gravitational loading in prone position, $\boldsymbol{b}^{p}$, produces the observed $\Omega_{\boldsymbol{\hat{x}}^{p}}$.
(b) Gravitational loading in the standing position, $\boldsymbol{b}^{s}$, produces the observed configuration $\Omega_{\boldsymbol{\hat{x}}^{s}}$.
}
\label{fig:breast_obs}
\end{figure}

To establish a well-controlled benchmark problem with known ground truth, we first construct an approximation to the \textit{unloaded} configuration by loading the prone MRI geometry with a body force opposite to gravity and using the true mechanical parameters and density described in Section~\ref{Geometry and Material Parameters}.
This procedure does not provide the physiologic unloaded state because reversing the direction of loading is not the mathematical inverse of a hyperelastic deformation. However, we will consider this the ground truth unloaded configuration $\Omega_{\tilde{\bsX}}$ such that we are able to define an inverse problem with known solution. Starting from the configuration $\Omega_{\tilde{\bsX}}$, we generate two synthetic observed configurations: The first one is denoted $\Omega_{\boldsymbol{\hat{x}
}^{p}}$ (with coordinates $\boldsymbol{\hat{x}}^{p}$) and is obtained by applying gravitational loading corresponding to the prone position, i.e., $\boldsymbol{b}^{p} = [0,-g,0]^{\intercal}$. The second one is denoted $\Omega_{\boldsymbol{\hat{x}
}^{s}}$ (with coordinates $\boldsymbol{\hat{x}}^{s}$) and is obtained by applying gravitational loading corresponding to the standing position, i.e., $\boldsymbol{b}^{s} = [0,0,-g]^{\intercal}$. Figure~\ref{fig:breast_obs} shows the two synthetic observed configurations generated from the artificial unloaded configuration $\Omega_{\tilde{\bsX}}$ by the two loadings $\boldsymbol{b}^{p}$ and $\boldsymbol{b}^{s}$.

\paragraph{Sensitivity to Initial Material Guess}
\label{sec:breast_results}
\mbox{}\\

Here, we use the breast geometry under the prone position as the reference geometry. Because the initial guess for the displacement between the reference and unloaded configuration is set to $\boldsymbol{\theta}_g^{(0)}=\mathbf{0}$, the reference configuration is also the initial guess for the unloaded configuration (Fig.~\ref{fig:breast_obs}(a)). The inverse problem is solved with three different initial guesses for the material parameters; see Table~\ref{tab:breast_tests}. The values in Table~\ref{tab:breast_tests} are specified in terms of Young’s modulus and Poisson’s ratio, which are subsequently converted to the equivalent bulk and shear moduli prior to optimization. The inverse FEM optimization is then performed directly in terms of $\boldsymbol{\theta}_\kappa$ and $\boldsymbol{\theta}_\mu$.
\begin{table}
\centering
\normalsize
\caption{Initial guesses for the material parameters for the breast inverse problem.}
\label{tab:breast_tests}
\renewcommand{\arraystretch}{1.25}
\begin{tabular}{c c c}
\hline
\textbf{Breast Test}  & 
$E_{\mathrm{mat}}^{(0)}$ &
$\nu_{\mathrm{mat}}^{(0)}$\\
\hline
1 & $0.8E_{\mathrm{mat}}$ & $0.9 \nu_{\mathrm{mat}}$ \\

2 & $0.7E_{\mathrm{mat}}$ & $0.7 \nu_{\mathrm{mat}}$ \\

3 & $0.5E_{\mathrm{mat}}$ & $0.5 \nu_{\mathrm{mat}}$ \\
\hline
\end{tabular}
\end{table}

\begin{figure}[H]
\centering
\begin{minipage}{0.48\textwidth}
    \centering
    \includegraphics[width=\textwidth]{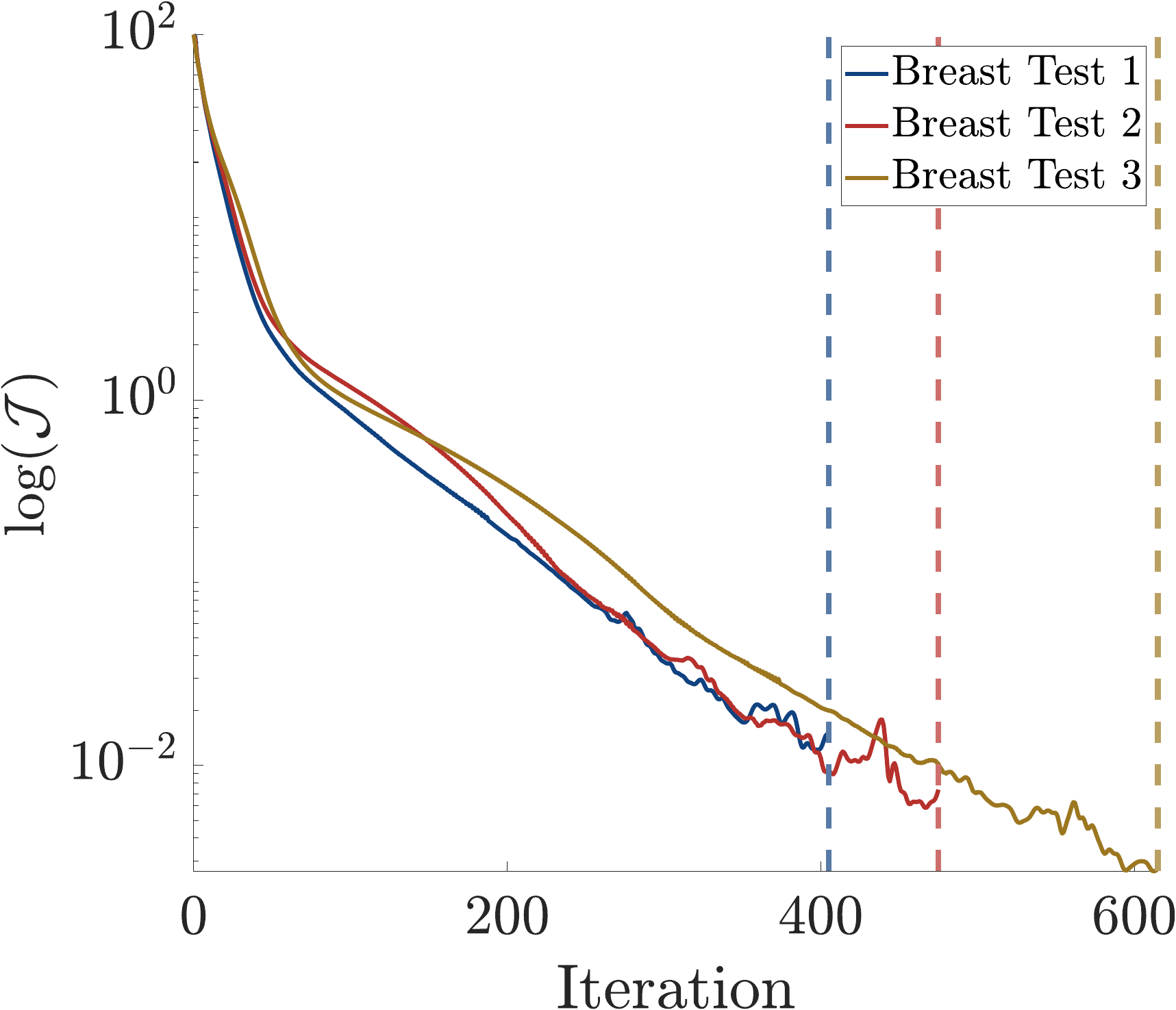}\\
    (a)
\end{minipage}
\par
\vspace{0.3cm}
\begin{minipage}{0.48\textwidth}
    \centering
    \includegraphics[width=\textwidth]{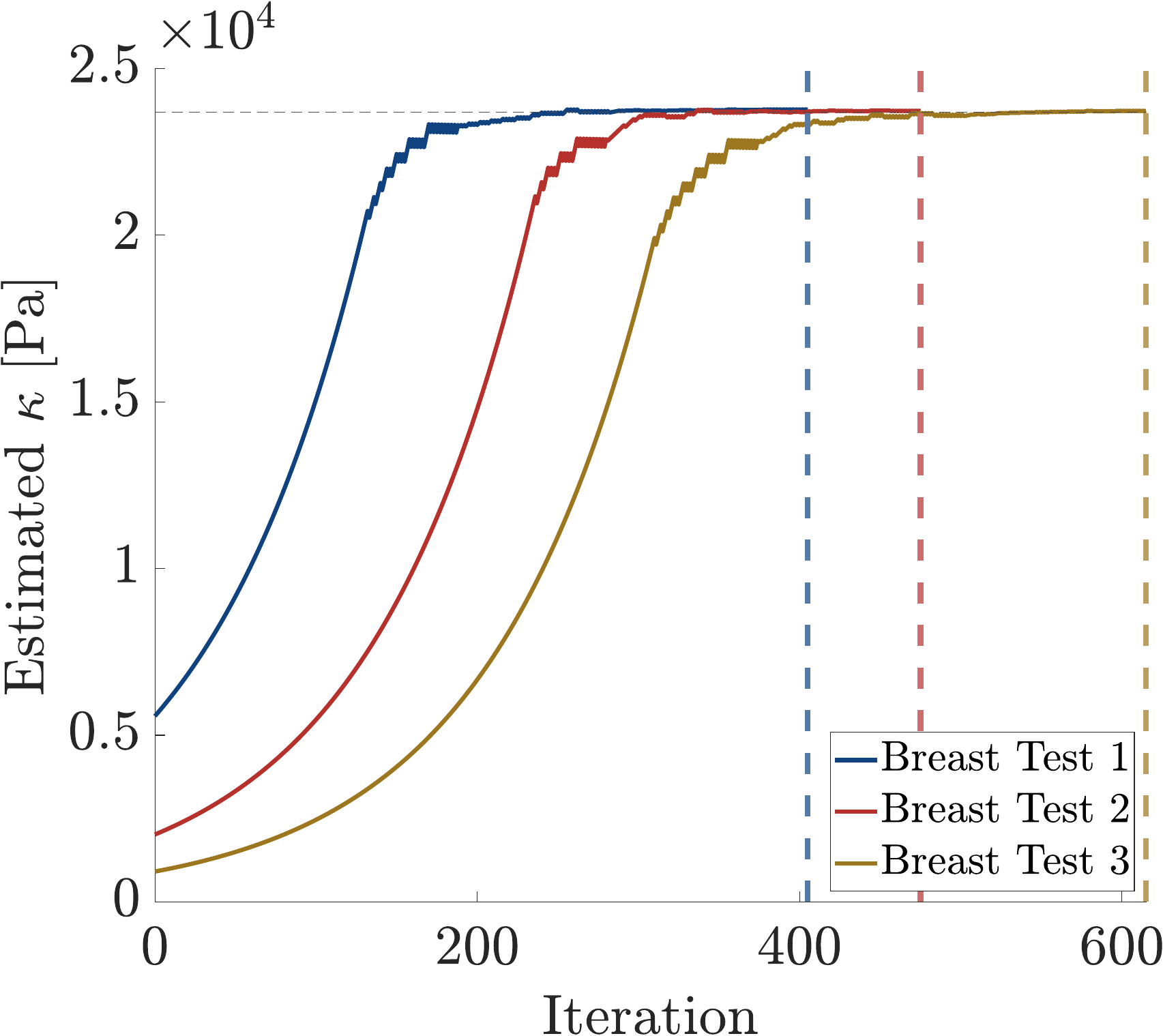}\\
    (b)
\end{minipage}
\hfill
\begin{minipage}{0.48\textwidth}
    \centering
    \includegraphics[width=\textwidth]{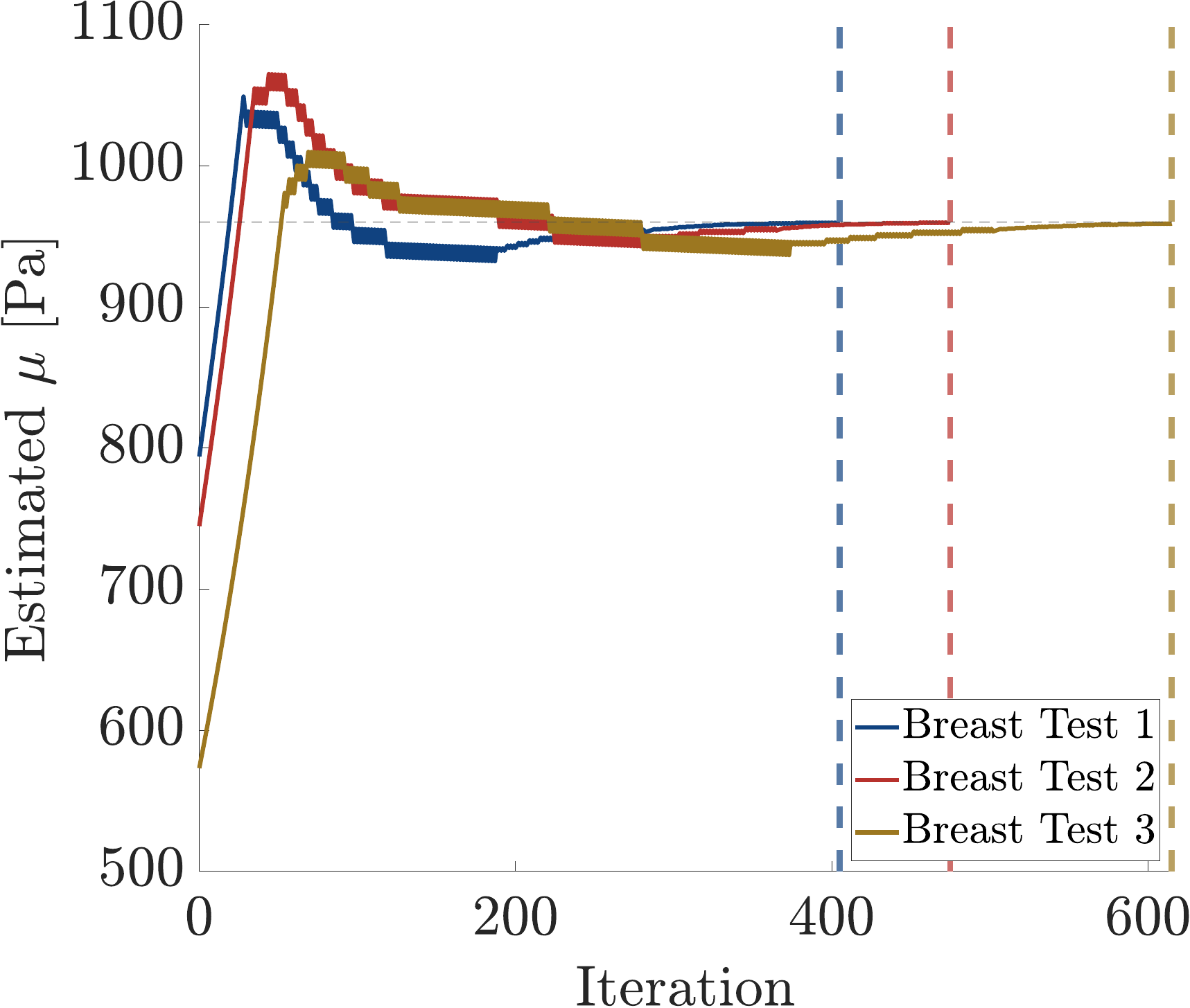}\\
    (c)
\end{minipage}

\caption{Evolution of the objective function $\mathcal{J}$ and the estimated material parameters for the patient-specific breast inverse problem with three different initial material guesses. All cases use the same pair of gravity-loaded observed configurations. (a) Evolution of the objective function $\mathcal{J}$. (b) Evolution of the bulk modulus $\kappa_{\mathrm{mat}}$. (c) Evolution of the shear modulus $\mu_{\mathrm{mat}}$. Vertical dashed lines indicate the iteration at which each test first satisfies the convergence criterion.}
\label{fig:J_breast_J_param}
\end{figure}

Figure~\ref{fig:J_breast_J_param} shows the evolution of the objective function $\mathcal{J}$ and the corresponding bulk and shear moduli for the three different initial material guesses. The objective function exhibits similar convergence trends in all three cases, although initial guesses farther from the true material parameters require additional iterations to satisfy the convergence criterion. In all tests, both $\kappa_{\mathrm{mat}}$ and $\mu_{\mathrm{mat}}$ converge toward their true values, indicating stable recovery of the material properties for this anatomically realistic geometry using only two observed loading configurations.

For less accurate initial guesses, the material parameters continue to evolve over a longer transient phase even after the objective function has reached a small value. This explains the extended convergence observed in these cases and indicates that the late-stage iterations are governed primarily by stabilization of the material parameters rather than by further reduction of the geometric mismatch.

The RSER values for Breast Tests~1, 2, and~3 are $1.69\times10^{-3}$, $1.47\times10^{-3}$, and $9.27\times10^{-4}$, respectively, confirming substantial improvement of the reconstructed unloaded configuration relative to the initial guesses. In addition, the corresponding NSRE values are $7.13\times10^{-6}$, $6.19\times10^{-6}$, and $3.90\times10^{-6}$ for Breast Tests~1, 2, and~3, respectively, indicating that the converged configurations are in close agreement with the true unloaded solution. The convergence behavior and reconstruction accuracy of all three tests are summarized in Appendix~\ref{tab:summary_breast_tests}.

The breast example demonstrates that the proposed framework can recover an unloaded soft-tissue geometry and material parameters in an image-based setting where the observed geometry is a loaded configuration. This is important because a common intuitive strategy, applying gravity or another body force in the opposite direction to the imaged configuration \cite{lee_breast_2013,del_palomar_finite_2008,sturgeon_finite-element_2016}, does not generally recover the true stress-free state \cite{carter_biomechanical_nodate}. Such reverse-loading approaches neglect the coupled effects of material uncertainty and material nonlinearity. 
This point is consistent with image-based cardiac modeling studies, where Nikou et al.~\cite{nikou_effects_2016} showed that using a numerically unloaded reference configuration instead of an early-diastolic loaded geometry affects the estimated in vivo material response. 

\subsection{Inverse Problem in Heterogeneous Breast Mechanics}
\label{sec:results_breast_Hetero}

The heterogeneous breast test evaluates whether the unloaded geometry and the region-wise material properties can be recovered simultaneously when a stiff tumor is embedded in the compliant breast matrix and only gravity-loaded configurations are observed.
\paragraph{Geometry, Loading, and Boundary Conditions}
\label{par:breast_geom_bc_Hetero}
\mbox{}\\

\begin{figure}[H]
    \centering
    \begin{minipage}{0.45\textwidth}
        \centering
        \includegraphics[width=\textwidth]{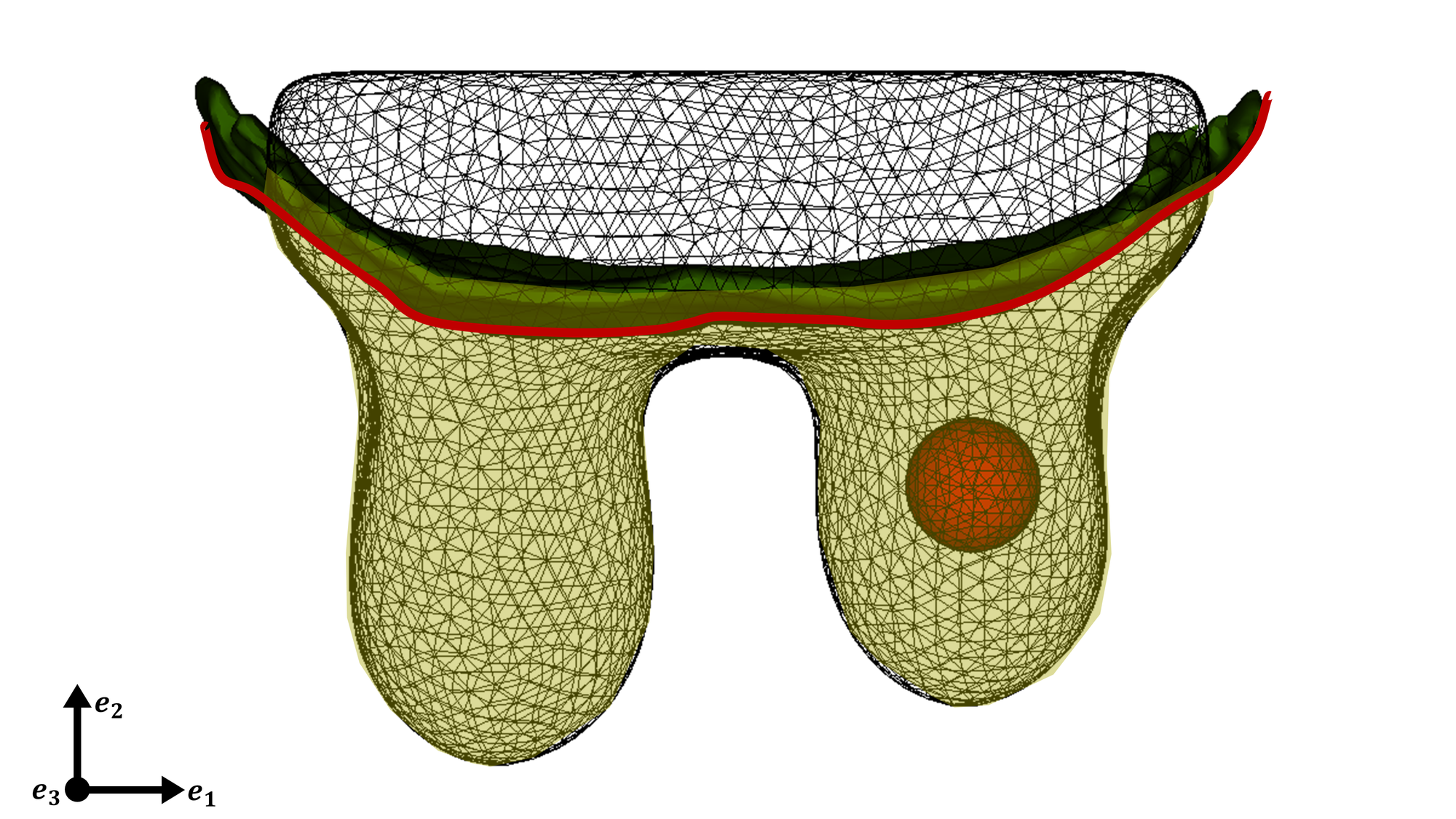}
        (a)\\
    \end{minipage}
    \hfill
    \begin{minipage}{0.45\textwidth}
        \centering
        \includegraphics[width=\textwidth]{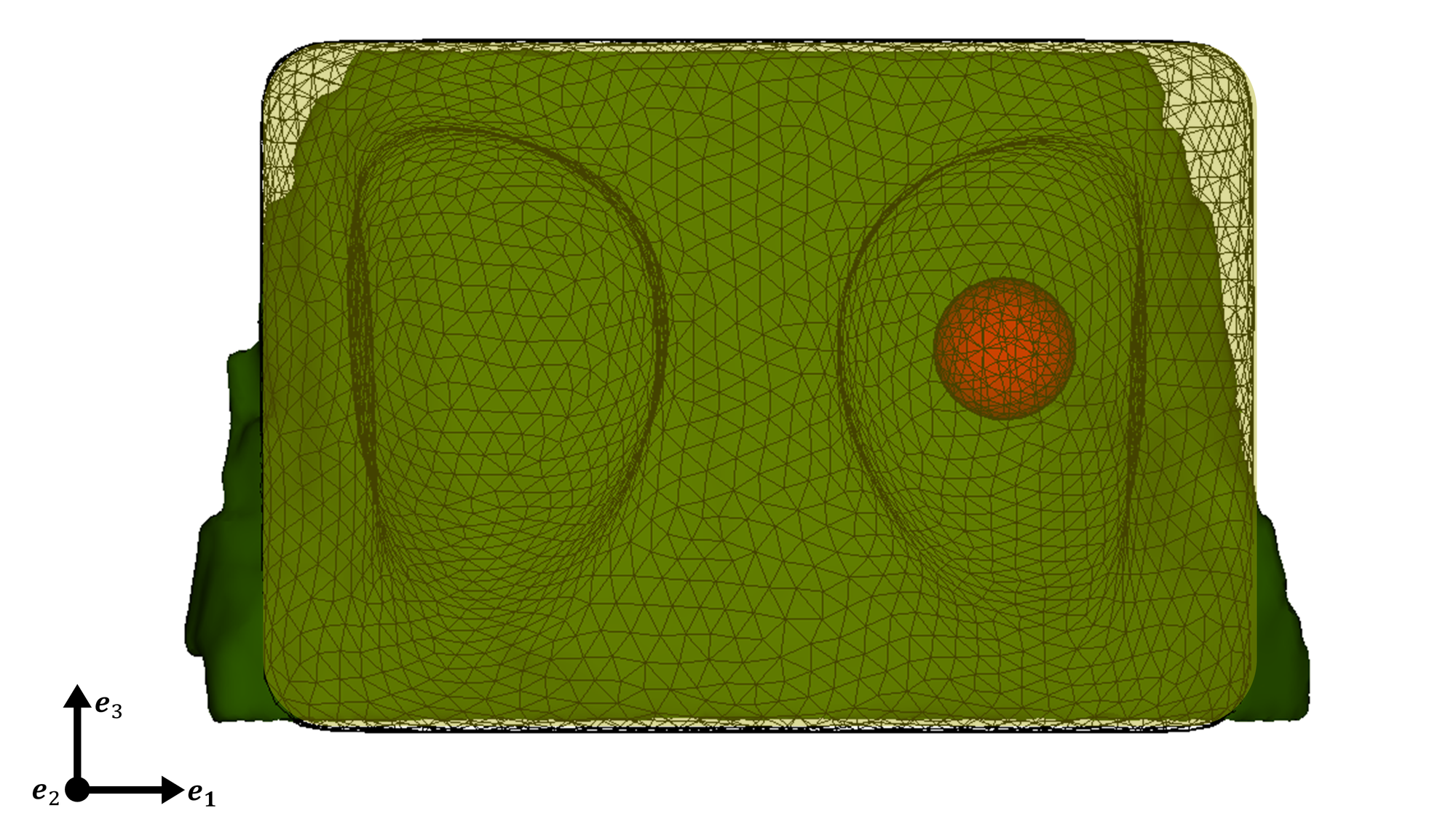}
        (b)\\
    \end{minipage}

\caption{Patient-specific heterogeneous breast geometry with a spherical tumor.
(a) Top view of the patient-specific breast surface mesh reconstructed from MRI (black), together with the registered sternum and pectoralis major muscle (green surface), the computational domain with free displacement degrees of freedom (yellow), and the spherical tumor region (red sphere). 
(b) Bottom view of the patient-specific breast surface mesh. The spherical tumor has radius $20~[\mathrm{mm}]$ and is modeled as a separate material region with Young's modulus ten times larger than that of the surrounding breast tissue. The domain contains 64,395 linear tetrahedral elements. }
\label{fig:Breast_tumor_geom}
\end{figure}

This example extends the homogenized breast mechanics problem in Sec.~\ref{sec:results_breast} by introducing a stiffer spherical region into the patient-specific breast geometry. The spherical region represents a tumor. The same breast geometry, density, loading condition, and boundary-condition treatment are used as in the homogenized case. The spherical tumor has a radius of $20~[\mathrm{mm}]$ and is embedded in the free breast tissue region, as shown in Fig.~\ref{fig:Breast_tumor_geom}. To better resolve the tumor-matrix interface, the geometry is remeshed using 64,395 linear tetrahedral elements.

The surrounding breast tissue is treated as the matrix region, $\Omega_{\mathrm{mat}}$, while the tumor is treated as a separate material region, $\Omega_{\mathrm{in}}$. The tumor is assigned a Young's modulus 10 times larger than the surrounding breast tissue, while the same Poisson's ratio is used for both regions. This stiffness contrast is motivated by the measurements of Samani et al.~\cite{samani_elastic_2007}, who reported that malignant breast tumors can be several times stiffer than normal breast tissue, with high-grade invasive ductal carcinoma being approximately 13 times stiffer than normal fibroglandular tissue. The density is kept identical to the homogenized case, $\rho = \text{942.82}$ [kg/m$^3$].

The loading condition is also kept the same as in the homogenized breast mechanics problem. Gravity is applied as a body force in the $\boldsymbol{e}_2$-direction. The zero-displacement boundary condition is prescribed using the same sternum and pectoralis-major-based segmentation procedure described in Sec.~\ref{sec:results_breast}. 

The synthetic observed configurations are generated using the same procedure as in the homogenized breast mechanics example. An artificial unloaded configuration is first constructed, followed by two synthetic observed configurations: $\Omega_{\boldsymbol{\hat{x}}^{p}}$ (with coordinates $\boldsymbol{\hat{x}}^{p}$) under the prone gravitational loading $\boldsymbol{b}^{p}$ and $\Omega_{\boldsymbol{\hat{x}}^{s}}$ (with coordinates $\boldsymbol{\hat{x}}^{s}$) under the standing gravitational loading $\boldsymbol{b}^{s}$, as defined in the homogenized breast mechanics example. The prone observed configuration, $\Omega_{\boldsymbol{\hat{x}}^{p}}$, is used as the initial guess for the unloaded configuration.

\paragraph{Effect of Tumor-Induced Material Heterogeneity}
\label{sec:Hetero_breast}
\mbox{}\\

\begin{figure}[H]
    \centering
    \includegraphics[width=0.48\textwidth]{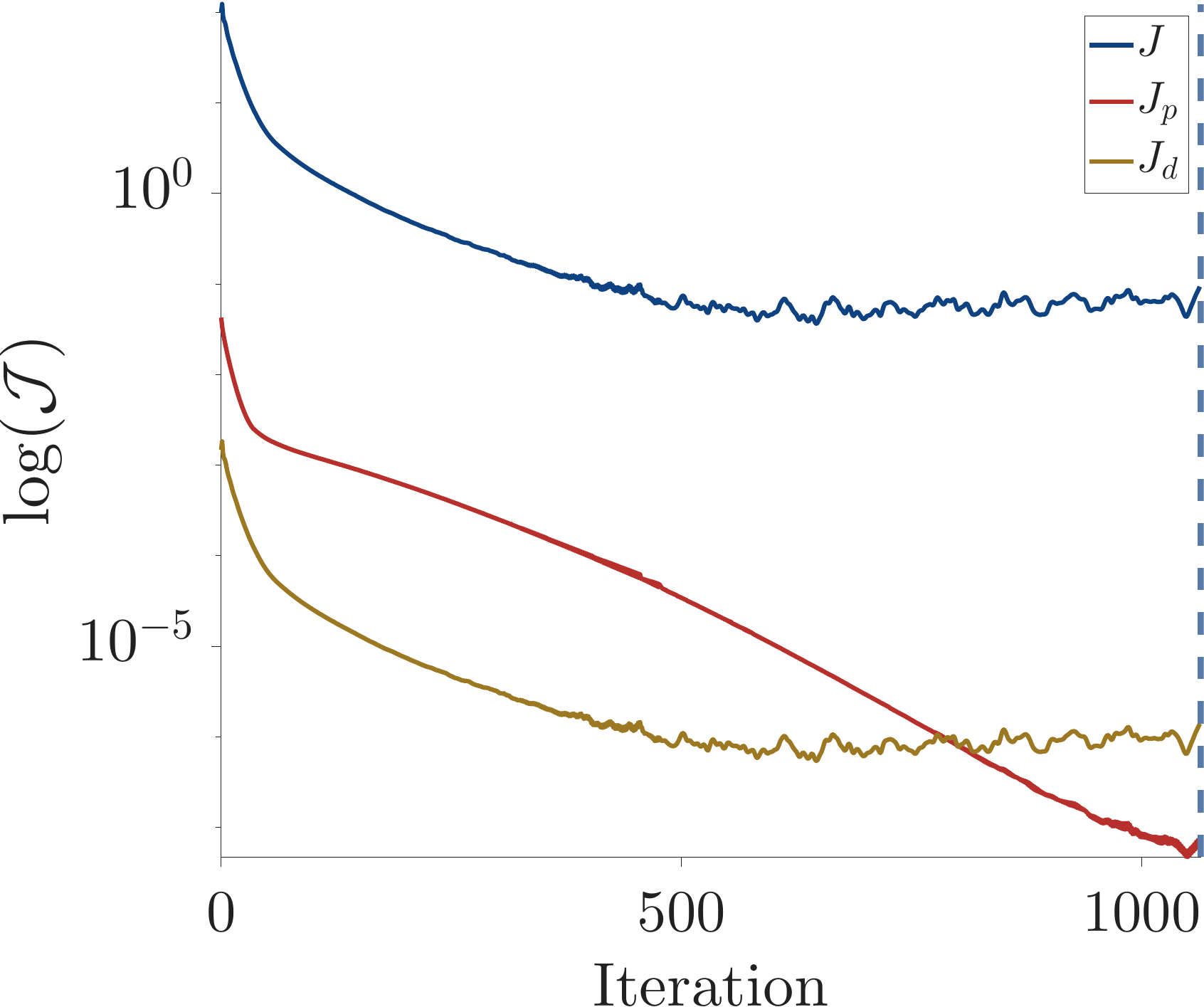}
    \caption{Optimization history for the heterogeneous patient-specific breast model with the objective weighting fixed at $\lambda_{\mathcal{J}}=99$ throughout the optimization. The figure shows the total objective function $\mathcal{J}$, the nodal-position mismatch $\mathcal{J}_{\mathrm{p}}$, and the deformation-gradient mismatch $\mathcal{J}_{\mathrm{d}}$. The total objective decreases rapidly during the early iterations but enters an oscillatory plateau after approximately 600 iterations. The component-wise histories show that this late-stage oscillatory behavior is primarily associated with $\mathcal{J}_{\mathrm{d}}$, while $\mathcal{J}_{\mathrm{p}}$ continues to decrease. The vertical dashed line indicates the manually selected stopping iteration, since the convergence criterion is not satisfied.} 
    \label{fig:J_breast_hetero_baseline}
\end{figure}

Figure~\ref{fig:J_breast_hetero_baseline} shows that the objective weighting $\lambda_{\mathcal{J}}=99$ produces a rapid initial reduction of the objective $\mathcal{J}$, but the optimization enters an oscillatory plateau after approximately 600 iterations and does not satisfy the convergence criterion. The decomposition of the objective provides further insight into this behavior. Although the nodal-position mismatch $\mathcal{J}_{\mathrm{p}}$ continues to decrease, the deformation-gradient mismatch $\mathcal{J}_{\mathrm{d}}$ ceases to exhibit sustained reduction and instead oscillates during the late iterations. Because $\lambda_{\mathcal{J}}=99$ places nearly all of the objective weight on $\mathcal{J}_{\mathrm{d}}$, this behavior prevents further sustained decrease of the total objective.

The behavior in Fig.~\ref{fig:J_breast_hetero_baseline} therefore suggests that the relative importance of the two objective terms should be changed after the nodal-position mismatch has been substantially reduced. Similar difficulties can arise in composite physics-based objectives when the component losses evolve at different rates, motivating strategies that rebalance their relative contributions during optimization \cite{bischof_multi-objective_2025}.

To investigate whether late-stage reweighting can restore convergence, the optimization is repeated using the same initial value $\lambda_{\mathcal{J}}=99$ to maintain physically admissible mappings with $\hat{J}>0$ during the early stages, but the weighting parameter is changed at iteration 600 to $\lambda_{\mathcal{J}}=0$, $10$, or $50$. These cases are compared with the baseline case in which $\lambda_{\mathcal{J}}=99$ is retained throughout the optimization.

\begin{figure}[H]
    \centering
    \begin{minipage}{0.48\textwidth}
        \centering
        \includegraphics[width=\textwidth]{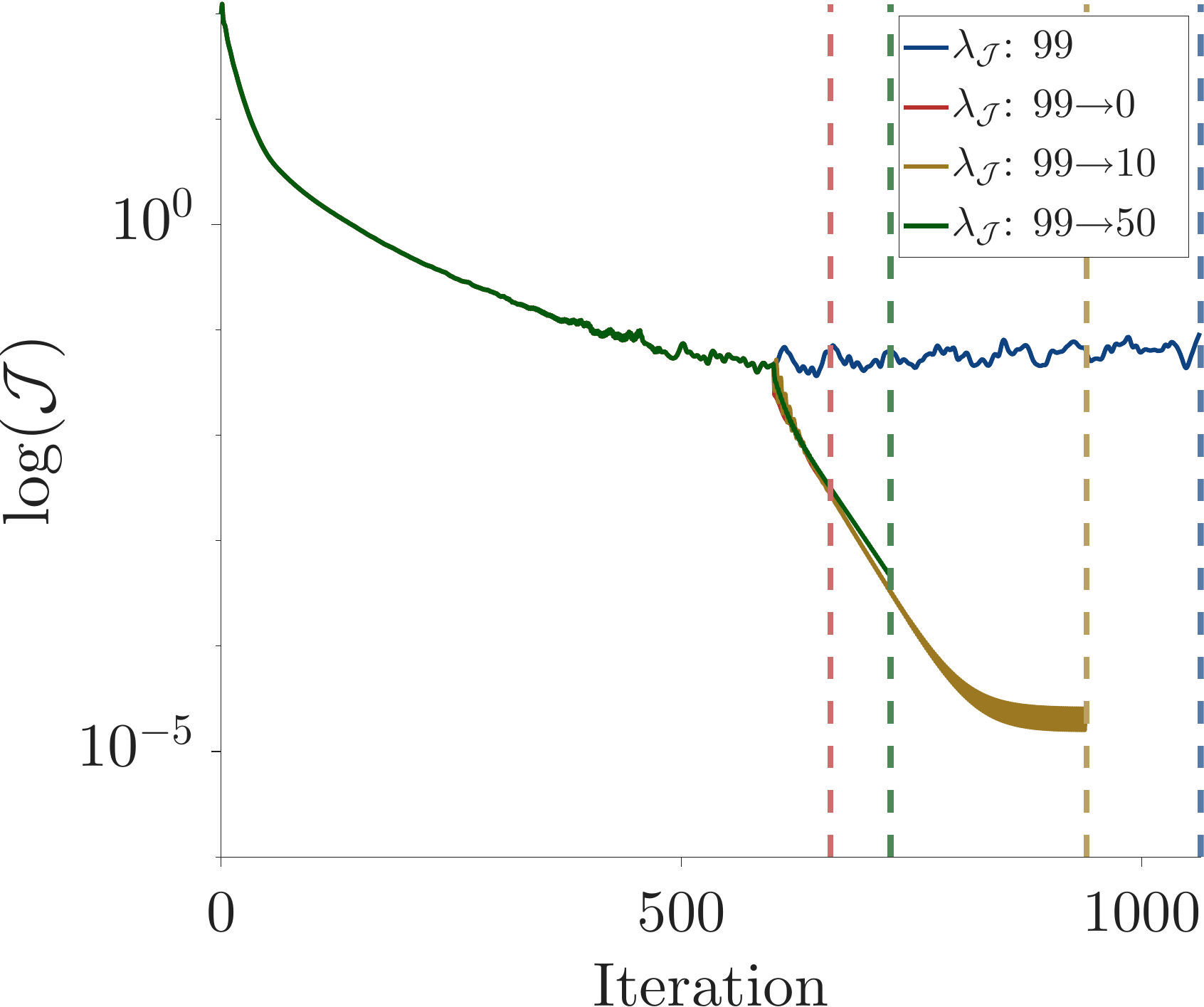}\\
        (a)
    \end{minipage}
    \par
    \vspace{0.3cm}
    \begin{minipage}{0.48\textwidth}
        \centering
        \includegraphics[width=\textwidth]{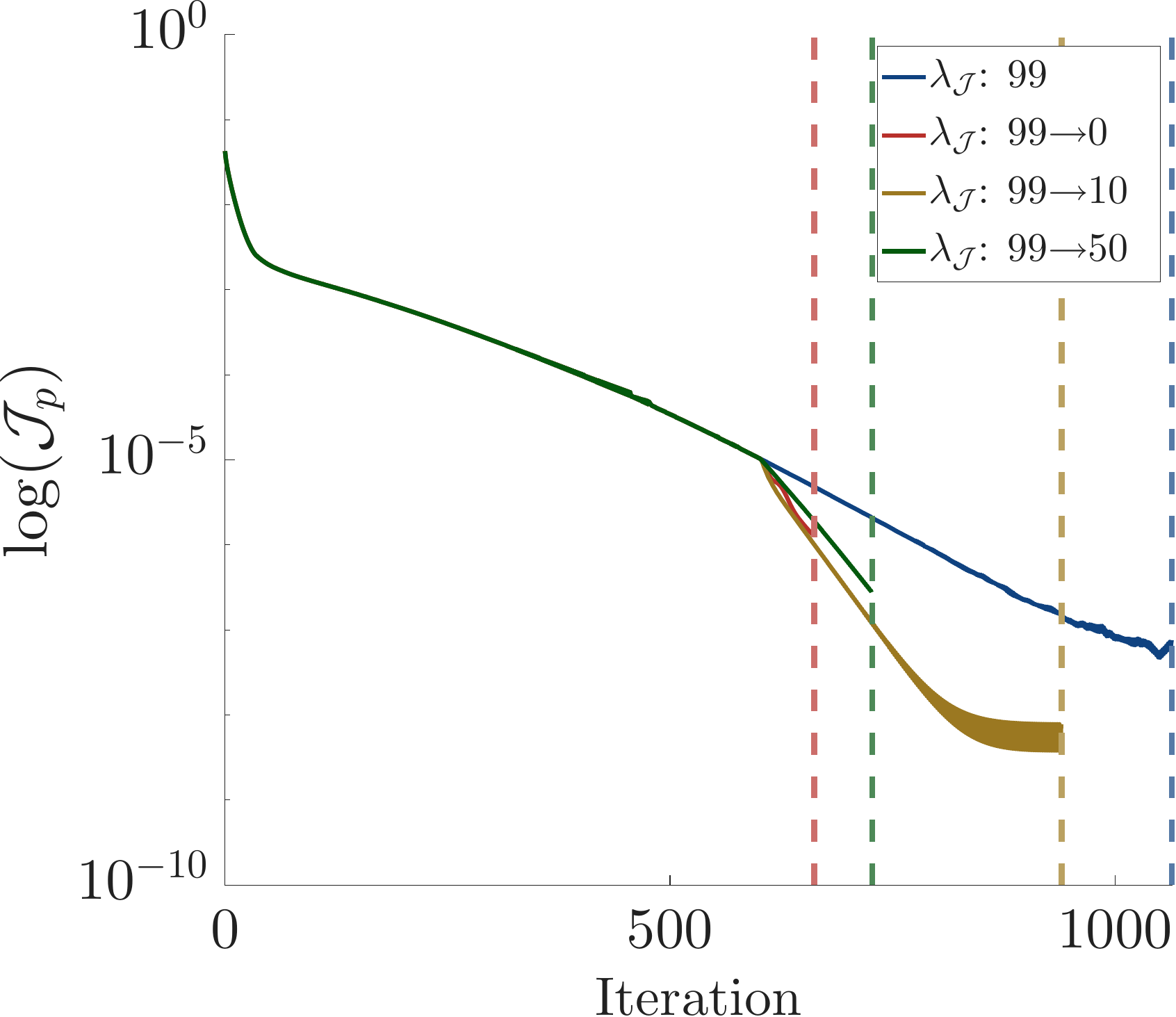}\\
        (b)
    \end{minipage}
    \hfill
    \begin{minipage}{0.48\textwidth}
        \centering
        \includegraphics[width=\textwidth]{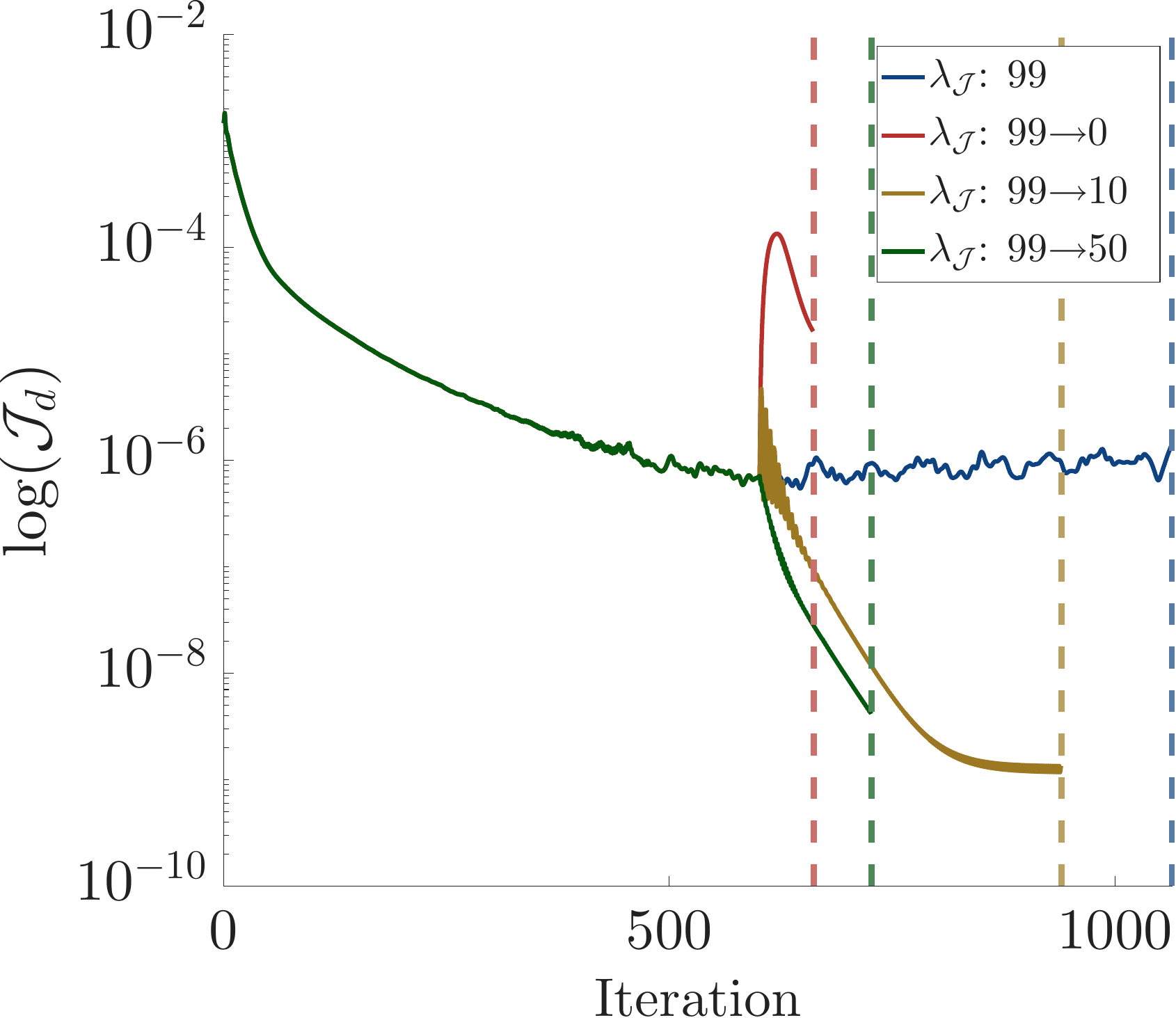}\\
        (c)
    \end{minipage}

    \caption{Effect of late-stage objective reweighting on the heterogeneous patient-specific breast inverse problem. All cases use $\lambda_{\mathcal{J}}=99$ during the first 600 iterations. At iteration 600, the weighting parameter is changed to $0$, $10$, or $50$, or retained at a fixed value of $99$. (a) Total objective function $\mathcal{J}$. (b) Nodal-position mismatch $\mathcal{J}_{\mathrm{p}}$. (c) Deformation-gradient mismatch $\mathcal{J}_{\mathrm{d}}$. Vertical dashed lines indicate the iterations at which the reweighted cases first satisfy the convergence criterion. For the fixed-$99$ case, the vertical dashed line indicates the manually selected stopping iteration because the convergence criterion is not satisfied. Since the weighting of the composite objective changes at iteration 600 for the reweighted cases, the total-objective histories in (a) are interpreted together with the individual objective terms in (b) and (c).}
    \label{fig:J_breast_hetero_reweight}
\end{figure}

Figure~\ref{fig:J_breast_hetero_reweight} shows that reducing $\lambda_{\mathcal{J}}$ after the onset of the oscillatory plateau restores sustained optimization progress. In the fixed-weight case ($\lambda_{\mathcal{J}}=99$), $\mathcal{J}_{\mathrm{d}}$ remains oscillatory and the convergence criterion is not satisfied. In contrast, when $\lambda_{\mathcal{J}}$ is reduced to $0$, $10$, or $50$ at iteration 600, the late-stage evolution of the individual objective terms changes and all three reweighted cases satisfy the convergence criterion. These results indicate that the large deformation-gradient weight is beneficial during the early stage of the optimization, whereas reducing the relative emphasis on $\mathcal{J}_{\mathrm{d}}$ is advantageous for late-stage refinement in the heterogeneous breast problem.

The converged iterations, region-wise material errors, objective values, and reconstruction metrics for all three reweighted cases are summarized in Appendix~\ref{tab:summary_hetero_breast}.


These results extend the homogeneous breast example by demonstrating that, for the present synthetic loading conditions and known tumor segmentation, the unloaded geometry and separate matrix and tumor material properties can be recovered simultaneously. This is a more challenging inverse problem because errors in the tumor stiffness can be partially compensated by changes in the reconstructed unloaded geometry. The heterogeneous breast example also exhibits greater sensitivity to the relative weighting of the objective terms, consistent with stronger coupling between the unloaded geometry and the regional material properties. With the tumor location supplied by the mesh labels, the differentiable formulation recovers distinct matrix and tumor properties without modifying the forward FEM formulation.

Previous heterogeneous breast models have demonstrated the importance of regional tissue properties for predicting large deformations and co-locating tumors across imaging positions \cite{han_development_2011,gamage_modelling_2012}, while posture-based inverse analyses have estimated tumor-related elasticity distributions from prone and supine configurations \cite{hasse_systematic_2016}. The present formulation simultaneously optimizes the full unloaded nodal geometry and distinct material parameters for the breast matrix and tumor within a patient-specific mesh through a gradient-based optimizer built on top of end-to-end automatic differentiation of the FEM solver.

\section{Conclusion}
\label{Conclusion}

We presented a fully differentiable inverse finite element framework for simultaneously reconstructing an unloaded configuration and identifying constitutive parameters from multiple observed deformed configurations. The same formulation applies to homogeneous materials and to heterogeneous domains with known region labels but unknown region-wise properties. By differentiating through the complete nonlinear FEM solution, the framework provides exact gradients with respect to both the unloaded nodal coordinates and the material parameters without requiring a problem-specific inverse solver or an explicitly assembled sensitivity matrix.

The numerical studies reveal several consistent identifiability trends. First, loading diversity is more influential than loading magnitude: observations that combine complementary deformation modes, such as tension and compression or stretching and bending, converge more rapidly than repeated observations of the same mode. Second, increasing the number of observed configurations improves convergence even when the additional data use the same loading direction, indicating that both mechanical diversity and data redundancy help constrain the coupled geometry-material inverse problem. Third, the final reconstruction is comparatively insensitive to the initial material guess. Poorer guesses extend the transient parameter-adjustment phase, but sufficiently informative observations lead to similar unloaded geometries and material estimates.

The heterogeneous cube and tumor-containing breast examples further demonstrate that the framework can leverage the difference in observed geometry prediction coming from unloaded geometry and regional stiffness. When material-region labels are known, the method recovers the unloaded configuration together with multiple matrix, inclusion, or tumor parameters without changing the forward formulation. In the heterogeneous breast case, retaining a loss weighting $\lambda_{\mathcal{J}}=99$, which heavily relies on the deformation-gradient cost, produced an oscillatory objective plateau and did not satisfy the convergence criterion. The corresponding gradient histories indicate an imbalance between the nodal-position and deformation-gradient objective terms. The large initial weight on $\mathcal{J}_{\mathrm{d}}$ helps maintain physically admissible deformations, which is essential during the early iterations. However, our results suggest that, as problem geometric complexity and stiffness contrast increase, balancing the contributions of the composite objective may be as important as selecting informative loading configurations.

The patient-specific breast examples showcase the method in a clinically relevant application: MRI-derived geometry, gravity loading, image-based boundary constraints, near-incompressible behavior, and localized material heterogeneity. Yet, the study is not without limitations. In the present work, all observations were generated synthetically in order to properly benchmark the method, and the heterogeneous regions were assumed to be known. Future work will thus focus on experimental validation and will require accounting for registration error, segmentation uncertainty, imperfect loading and boundary conditions, measurement noise, and potentially unknown tumor geometry.

Overall, the results indicate that successful joint identification of unloaded geometry and material properties is governed primarily by the information contained in the observed deformation states and that the proposed differentiable framework provides a unified route to solving these inverse problems, with broad applicability in tissue biomechanics.

\appendix
\section{Adaptive Material Update and Convergence Criteria}
\label{app:material_update}
\vspace{0.5em}
\noindent
\textbf{Sliding-window convergence criterion.}
To detect oscillatory behavior near the optimum, we monitor the recent history of $\kappa$, $\mu$, and the objective value over a sliding window of the most recent $k_W$ iterations. Let $k$ denote the current iteration. The relative bandwidth of each material parameter is defined as
\begin{equation}
  \texttt{kappa\_band\_rel}
  =
  \frac{\max(\kappa_i) - \min(\kappa_i)}
       {\mathrm{mean}(\kappa_i)},
  \qquad i = k-k_W+1,\ldots,k,
\end{equation}
and analogously for the shear modulus,
\begin{equation}
  \texttt{mu\_band\_rel}
  =
  \frac{\max(\mu_i) - \min(\mu_i)}
       {\mathrm{mean}(\mu_i)}.
\end{equation}

These quantities measure the magnitude of parameter oscillation relative to the current modulus. A tolerance proportional to the relative step size is introduced,
\begin{equation}
  \texttt{band\_tol\_rel}
  =
  1.5 \times \texttt{rel\_step}.
\end{equation}

If both relative bandwidths fall below this tolerance and the objective variation within the window is smaller than a prescribed threshold $\varepsilon_J$, the relative update scale $\texttt{rel\_step}$ is reduced by a factor $\gamma$ until the minimum value $\texttt{min\_rel\_step}$ is reached. Convergence is declared once the objective variation remains below $\varepsilon_J$, both bandwidth measures remain below $\texttt{band\_tol\_rel}$, and $\texttt{rel\_step}=\texttt{min\_rel\_step}$.

In the numerical examples, $\texttt{rel\_step}$ is reduced from an initial value of $0.01$ (approximately a 1\% relative update per iteration) to a minimum value of $4\times10^{-4}$ (approximately a 0.04\% update), allowing progressively finer refinement near the solution.

\section{Optimization hyperparameters}
\label{app:hyperparameters}

The optimization procedure involves several hyperparameters that control the step sizes and convergence criteria for both geometry and material updates.

\begin{center}
\begin{tabularx}{\textwidth}{lXc}
\hline
Parameter & Description & Value \\
\hline

$\alpha_0$ 
& initial learning rate of the Adam optimizer for geometry variables $\bstheta_g$
& $5\times10^{-4}$ \\

$\texttt{rel\_step\_0}$ 
& initial relative update magnitude for material parameters
& $10^{-2}$ \\

$\texttt{min\_rel\_step}$ 
& minimum allowable relative update magnitude
& $4\times10^{-4}$ \\

$k_W$ 
& window length used to evaluate convergence stability
& 20 \\

$\gamma$ 
& decay factor applied to rel\_step when oscillations are detected
& 0.2 \\

$\varepsilon_J$ 
& tolerance on the standard deviation of the objective values within the most recent $k_W$ iterations
& $10^{-4}$ \\

$k_{max}$
& maximum number of iterations
& $10^4$ \\

\hline
\end{tabularx}
\end{center}

\setcounter{table}{0}

\section{Summary of convergence behavior}
\label{Summary of convergence behavior}

\begin{table}[h]
\centering
\caption{Summary of convergence behavior and reconstruction accuracy for all benchmark verification tests.}
\label{tab:summary_all_tests}
\small
\setlength{\tabcolsep}{3pt}
\renewcommand{\arraystretch}{1.15}
\begin{tabular}{c r r r r r r r r}
\toprule
\textbf{Test} &
\textbf{Iter.} &
\multicolumn{4}{c}{\textbf{Relative material error (\%)}} &
\multicolumn{3}{c}{\textbf{Objective and reconstruction metrics}} \\
\cmidrule(lr){3-6}\cmidrule(lr){7-9}
& &
\multicolumn{1}{c}{$\kappa_{\mathrm{mat}}$} &
\multicolumn{1}{c}{$\mu_{\mathrm{mat}}$} &
\multicolumn{1}{c}{$E_{\mathrm{mat}}$} &
\multicolumn{1}{c}{$\nu_{\mathrm{mat}}$} &
\multicolumn{1}{c}{$\mathcal{J}$} &
\multicolumn{1}{c}{\textbf{NSRE}} &
\multicolumn{1}{c}{\textbf{RSER}} \\
\midrule
1 & 5146 & 1.5408 & 1.4026 & 1.4210 & 0.0786 & $4.59\times10^{-3}$ & $3.10\times10^{-6}$ & $2.03\times10^{-4}$ \\
2 &  778 & 0.0164 & 0.0762 & 0.0683 & 0.0346 & $8.01\times10^{-4}$ & $2.20\times10^{-8}$ & $1.44\times10^{-6}$ \\
3 & 6046 & 2.0107 & 1.2449 & 1.3463 & 0.4342 & $1.10\times10^{-3}$ & $3.00\times10^{-6}$ & $1.96\times10^{-4}$ \\
4 & 1187 & 0.0840 & 0.5330 & 0.4729 & 0.2591 & $8.57\times10^{-3}$ & $5.27\times10^{-6}$ & $3.45\times10^{-4}$ \\
5 & 2382 & 0.6733 & 0.3235 & 0.3700 & 0.2009 & $2.80\times10^{-3}$ & $2.07\times10^{-7}$ & $1.35\times10^{-5}$ \\
6 & 1478 & 0.5697 & 0.1458 & 0.2021 & 0.2437 & $2.18\times10^{-3}$ & $7.38\times10^{-8}$ & $4.83\times10^{-6}$ \\
7 & 4475 & 1.8196 & 1.1208 & 1.2134 & 0.3969 & $3.40\times10^{-2}$ & $3.99\times10^{-7}$ & $1.32\times10^{-4}$ \\
8 &  636 & 0.3290 & 0.1045 & 0.1345 & 0.1301 & $1.71\times10^{-2}$ & $8.29\times10^{-7}$ & $2.74\times10^{-4}$ \\
9 & 1618 & 0.9582 & 0.3661 & 0.4446 & 0.3391 & $2.51\times10^{-2}$ & $3.32\times10^{-7}$ & $1.10\times10^{-4}$ \\
\bottomrule
\end{tabular}
\end{table}

\begin{table}[h]
\centering
\caption{Summary of convergence behavior and reconstruction accuracy for the patient-specific breast test cases.}
\label{tab:summary_breast_tests}
\small
\setlength{\tabcolsep}{3pt}
\renewcommand{\arraystretch}{1.15}
\begin{tabular}{c r r r r r r r r}
\toprule
\textbf{Test} &
\textbf{Iter.} &
\multicolumn{4}{c}{\textbf{Relative material error (\%)}} &
\multicolumn{3}{c}{\textbf{Objective and reconstruction metrics}} \\
\cmidrule(lr){3-6}\cmidrule(lr){7-9}
& &
\multicolumn{1}{c}{$\kappa_{\mathrm{mat}}$} &
\multicolumn{1}{c}{$\mu_{\mathrm{mat}}$} &
\multicolumn{1}{c}{$E_{\mathrm{mat}}$} &
\multicolumn{1}{c}{$\nu_{\mathrm{mat}}$} &
\multicolumn{1}{c}{$\mathcal{J}$} &
\multicolumn{1}{c}{\textbf{NSRE}} &
\multicolumn{1}{c}{\textbf{RSER}} \\
\midrule
1 & 405 & 0.0911 & 0.0168 & 0.3136 & 0.0965 & $1.46\times10^{-2}$ & $7.13\times10^{-6}$ & $1.69\times10^{-3}$ \\
2 & 475 & 0.0852 & 0.0098 & 0.1494 & 0.0883 & $7.37\times10^{-3}$ & $6.19\times10^{-6}$ & $1.47\times10^{-3}$ \\
3 & 615 & 0.1124 & 0.0124 & 0.1847 & 0.1164 & $2.71\times10^{-3}$ & $3.90\times10^{-6}$ & $9.27\times10^{-4}$ \\
\bottomrule
\end{tabular}
\end{table}

\begin{landscape}
\begin{table}[p]
\centering
\caption{Summary of convergence behavior, regional material-parameter errors, and reconstruction accuracy for the heterogeneous cube test. Relative material errors are reported in percent for the matrix, inclusions 1, and 2.}
\label{tab:summary_hetero_cube}
\scriptsize
\setlength{\tabcolsep}{3pt}
\renewcommand{\arraystretch}{1.15}
\resizebox{\linewidth}{!}{%
\begin{tabular}{r *{12}{>{\centering\arraybackslash}p{1.25cm}} r r r}
\toprule
\textbf{Iter.} &
\multicolumn{4}{c}{\textbf{Matrix relative material error (\%)}} &
\multicolumn{4}{c}{\textbf{Inclusions 1 relative material error (\%)}} &
\multicolumn{4}{c}{\textbf{Inclusions 2 relative material error (\%)}} &
\multicolumn{3}{c}{\textbf{Objective and reconstruction metrics}} \\
\cmidrule(lr){2-5}\cmidrule(lr){6-9}\cmidrule(lr){10-13}\cmidrule(lr){14-16}
& $E_{\mathrm{mat}}$ & $\nu_{\mathrm{mat}}$ & $\kappa_{\mathrm{mat}}$ & $\mu_{\mathrm{mat}}$ & $E_{\mathrm{in}_1}$ & $\nu_{\mathrm{in}_1}$ & $\kappa_{\mathrm{in}_1}$ & $\mu_{\mathrm{in}_1}$ & $E_{\mathrm{in}_2}$ & $\nu_{\mathrm{in}_2}$ & $\kappa_{\mathrm{in}_2}$ & $\mu_{\mathrm{in}_2}$ & $\mathcal{J}$ & \textbf{NSRE} & \textbf{RSER} \\
\midrule
777 & 0.4610 & 0.3139 & 0.0099 & 0.5330 & 0.0289 & 0.0110 & 0.0546 & 0.0260 & 0.0704 & 0.0270 & 0.0073 & 0.0774 & $5.46\times10^{-4}$ & $4.24\times10^{-8}$ & $1.16\times10^{-6}$ \\
\bottomrule
\end{tabular}%
}
\end{table}
\end{landscape}

\begin{landscape}
\begin{table}[p]
\centering
\caption{Summary of convergence behavior, regional material-parameter errors, and reconstruction accuracy for the heterogeneous patient-specific breast model. Relative material errors are reported in percent for the breast matrix and tumor. Each $\lambda_{\mathcal{J}}$ reweighting schedule changes the weighting parameter at iteration 600.}
\label{tab:summary_hetero_breast}
\scriptsize
\setlength{\tabcolsep}{3pt}
\renewcommand{\arraystretch}{1.15}
\resizebox{\linewidth}{!}{%
\begin{tabular}{l r *{8}{>{\centering\arraybackslash}p{1.25cm}} r r r}
\toprule
\textbf{$\lambda_{\mathcal{J}}$ schedule} &
\textbf{Iter.} &
\multicolumn{4}{c}{\textbf{Matrix relative material error (\%)}} &
\multicolumn{4}{c}{\textbf{Tumor relative material error (\%)}} &
\multicolumn{3}{c}{\textbf{Objective and reconstruction metrics}} \\
\cmidrule(lr){3-6}\cmidrule(lr){7-10}\cmidrule(lr){11-13}
& & $E_{\mathrm{mat}}$ & $\nu_{\mathrm{mat}}$ & $\kappa_{\mathrm{mat}}$ & $\mu_{\mathrm{mat}}$ & $E_{\mathrm{tumor}}$ & $\nu_{\mathrm{tumor}}$ & $\kappa_{\mathrm{tumor}}$ & $\mu_{\mathrm{tumor}}$ & $\mathcal{J}$ & \textbf{NSRE} & \textbf{RSER} \\
\midrule
$99\rightarrow0$ & 661 & 0.1981 & 0.0292 & 0.5066 & 0.2076 & 0.4410 & 0.0498 & 0.7645 & 0.4571 & $3.03\times10^{-3}$ & $4.56\times10^{-4}$ & $1.18\times10^{-1}$ \\
$99\rightarrow10$ & 939 & 0.1940 & 0.0488 & 0.9879 & 0.2098 & 0.2865 & 0.0293 & 0.4193 & 0.2960 & $2.55\times10^{-5}$ & $4.71\times10^{-4}$ & $1.22\times10^{-1}$ \\
$99\rightarrow50$ & 726 & 0.0809 & 0.0227 & 0.4659 & 0.0883 & 0.0949 & 0.0101 & 0.1481 & 0.0982 & $4.65\times10^{-4}$ & $4.67\times10^{-4}$ & $1.20\times10^{-1}$ \\
\bottomrule
\end{tabular}%
}
\end{table}
\end{landscape}

\newpage
\newpage

\bibliographystyle{elsarticle-num} 
\bibliography{Papers}

\end{document}